\documentclass{aa}  

\usepackage{txfonts}
\usepackage{natbib}
\usepackage{array}
\usepackage{txfonts}
\usepackage{subfigure}
\usepackage{enumitem}
\usepackage{hyperref}
\usepackage{comment}
\usepackage{threeparttable,lscape}
\usepackage{multirow}
\usepackage{multicol}
\usepackage{float}
\usepackage{amsfonts}
\usepackage{amsmath}
\usepackage[title]{appendix}
\usepackage{booktabs,multirow}
\usepackage{appendix}
\usepackage{comment}
\usepackage{xcolor}
\usepackage{pifont}
\usepackage{pdflscape}
\usepackage{afterpage}

\usepackage{caption}

\begin{document} 

   \title{Testing the magnetic flux paradigm for AGN radio loudness with a radio intermediate quasar}
      
    \titlerunning{Testing the magnetic flux paradigm for radio loudness of AGN}
    \authorrunning{Chamani et al.}
   
   
 \author{Wara Chamani,
         \inst{1,2}\fnmsep\thanks{E-mail address: wara.chamani@aalto.fi}
         Tuomas Savolainen,\inst{1,2,4}\fnmsep\thanks{E-mail address: tuomas.k.savolainen@aalto.fi}
         Kazuhiro Hada\inst{3} 
         \and
         Ming H. Xu\inst{1,2} 
         }

\institute{Aalto University Mets\"ahovi Radio Observatory,                           Mets\"ahovintie 114, FI-02540 Kylm\"al\"a, Finland 
           \and
            Aalto University Department of Electronics and Nanoengineering, PO Box 15500, FI-00076 Aalto, Finland 
           \and 
            Mizusawa VLBI Observatory, National Astronomical Observatory of Japan, 2-21-1 Owasa, Mitaka, Tokyo 181-8588, Japan
        \and
            Max-Planck-Institut f\"ur Radioastronomie,
            Auf dem H\"ugel 69, DE-53121 Bonn, Germany
        }

   \date{Received February 26, 2021; accepted June 1, 2021}

 
  \abstract
 {For understanding the diversity of jetted active galactic nuclei (AGN) and especially the puzzling wide range in their radio-loudness, it is important to understand what role the magnetic fields play in setting the power of relativistic jets in AGN.  We have performed VLBA phase-referencing observations of the radio-intermediate quasar III\,Zw\,2 to estimate jet magnetic flux by measuring the core-shift effect. Multi-frequency observations at 4\,GHz, 8\,GHz, 15\,GHz and 24\,GHz were made using three nearby calibrators as reference sources. By combining the self-referencing core-shift of each calibrator with the phase-referencing core-shifts, we obtained an upper limit of 0.16\,mas for the core-shift between 4 and 24\,GHz in III\,Zw\,2. By assuming equipartition between magnetic and particle energy densities and adopting the flux-freezing approximation, we further estimated the upper limit for both magnetic field strength and poloidal magnetic flux threading the black hole. We find that the upper limit to the measured magnetic flux is smaller by at least a factor of five compared to the value predicted by the magnetically arrested disk (MAD) model. An alternative way to derive the jet magnetic field strength from the turnover of the synchrotron spectrum leads to an even smaller upper limit. Hence, the central engine of III\,Zw\,2 has not reached the MAD state, which could explain why it has failed to develop a powerful jet, even though the source harbours a fast-spinning black hole. However, it generates an intermittent jet, which is possibly triggered by small scale magnetic field fluctuations as predicted by the magnetic flux paradigm of \citet{Sikora2013}. We propose here that combining black hole spin measurements with magnetic field measurements from the VLBI core-shift observations of AGN over a range of jet powers could provide a strong test for the dominant factor setting the jet power relative to the accretion power available.}
 


   \maketitle

%

\section{Introduction}

One of the major enigmas regarding AGN involves their jet production efficiency i.e., the ratio of jet power $P_\mathrm{j}$ to accretion power, $\dot{M}c^2$, where $\dot{M}$ is the mass accretion rate and $c$ is the speed of light. It is not well understood how supermassive black holes (SMBH) generate relativistic jets with a wide range of jet powers in sources with similar accretion powers. This issue is tangible in the observed spread (originally even dichotomy) of the radio-loudness parameter, defined as the ratio of radio luminosity of an AGN to its optical luminosity at some specific frequencies \citep{Strittmatter1980, Kellermann1989}. Radio-loud (RLQ) and radio-quiet (RQQ) quasars can have a difference of at least four orders of magnitude in radio-loudness --- which is considered a proxy for the jet power --- but show similar black hole masses and accretion rates \citep{Sikora2007}. We should note here that "radio-quiet" sources are by no means "radio-silent" and AGN can dominate the radio emission in RQQ host galaxies at $z \sim 1$ \citep{White2017}. The radio emission related to accretion in RQQ could originate in small scale, 'aborted' jets \citep{Ghisellini2004}, in quasar-driven winds \citep{Zakamska2014}, or in a hot corona above the accretion disk \citep{Raginski2016}.

Possible explanations for the wide radio-loudness distribution include differences solely in black hole spins \citep[so-called "spin paradigm"; see e.g.,][]{Moderski1998, Sikora2007, Tchekhovskoy2010}, or in the magnetic flux threading the black hole, combined with variation in black hole spins \citep[so-called "magnetic flux paradigm";][]{Sikora2013}. These explanations are motivated by the fact that the currently favoured jet launching model uses magnetic fields to extract rotational energy from the spinning black hole \citep{Blandford1977}. Attributing the radio-loudness distribution solely to a BH spin distribution and its evolution seems to be, however, in a contradiction with the expected cosmological evolution of spin in massive black holes \citep{Volonteri2013}. Moreover, it appears that not only radio-loud AGN harbour high spinning black holes but so do radio-quiet \citep{Reynolds2013,Reynolds2020}.

Although accretion onto the SMBH is required for building up the launching jet in the first place, the level of the disk magnetization in different environments might also set the different jet production efficiency of RLQs and RQQs \citep[see][]{Blandford2019}. By highly efficient jet production we here mean jet powers as high as the accretion power or even higher \citep{RawlingsSa1991, McNamara2011, Ghisellini2014}. Such powerful relativistic jets can be ejected by the Blandford-Znajek process around a spinning black hole which is threaded by a poloidal magnetic field with a maximum strength that a given accretion rate allows. This occurs when enough large-scale poloidal magnetic flux is accumulated in the inner accretion disk so that its magnetic pressure equals the ram pressure of the accretion flow. Such state is referred to as a magnetically arrested disk \citep[MAD;][]{Narayan2003, Tchekhovskoy2011, McKinney2012}. 

The MAD scenario has been successfully tested with a sample of radio-loud AGN consisting of mostly blazars and nearby radio galaxies by \citet{Zamaninasab2014}. Their results show that for the majority of the sources, the observationally inferred jet magnetic flux is comparable to the maximum predicted magnetic flux ($\sim 50$ in their dimensionless units) in the MAD state. 

If the magnetic flux paradigm is the explanation for the observed wide range in radio-loudness, then radio-quiet AGN should have magnetic fluxes that are  generally well below the MAD level. In order to test this it is necessary to study the magnetic field strengths in sources with weak jets but high black hole spins as well as in recent RQQs which have transformed into RLQs, although with possibly short-lived/young jets \citep{Nyland2020}. Inferring the magnetic fluxes from observations for a diversity of sources can confirm whether the low jet production efficiency can indeed be attributed to the failure in the accumulation of enough magnetic flux to reach the MAD state. 

Most of the available methods to measure jet magnetic field strengths are based on resolving the jet emission regions that are partially opaque due to synchrotron self-absorption \citep{Marscher1983}. One way is to use high angular resolution, multi-frequency, radio imaging observations to measure the frequency-dependent positional shift of the synchrotron photosphere along the jet and assume an equipartition between energy densities of the radiating particles and magnetic fields in the jet \citep{Lobanov1998}. The method requires sub-milliarcsecond imaging and astrometry with very long baseline interferometry (VLBI). The above technique, commonly known as the ”core-shift” measurement, has been increasingly used to infer jet magnetic field strengths over the past decade \citep{Kovalev2008,OSullivan2009,Sokolovsky2011, Pushkarev2012, Kutkin2014, Voitsik2018, Pushkarev2018, Plavin2019}. Another, closely related method to measure the jet magnetic field strength is to use multi-frequency VLBI to resolve the size of the emission region at its synchrotron turnover frequency --- this gives the magnetic flux density without an equipartition assumption, but tends to be more difficult in practice than the core-shift measurement \citep[e.g.,][]{Savolainen2008,Hodgson2017}. Jet magnetic fields can be also inferred by modelling the broadband spectral energy distribution, especially in blazars, although such estimates are usually quite model-dependent \citep[see][and references therein]{Hovatta2019}. 

While the core-shift method has proved itself a robust way to measure magnetic field strengths in radio-loud AGN, it is challenging to make these measurements for weak jets in radio-quiet sources due to their faintness. One solution to this problem is to examine radio-intermediate quasars (RIQs) that are thought to be relativistically boosted counterparts of RQQs \citep{Falcke1996b,Falcke1996a, Blundell1998}. A good example of such an RIQ is III\,Zw\,2, also known as Mrk\,1501 and PG\,0007+106.  Its observed radio loudness ($\mathcal{R} = L_\mathrm{5GHz} / L_\mathrm{B}$, where $L_\mathrm{5GHz}$ is the radio luminosity at 5\,GHz and $L_\mathrm{B}$ optical B-band luminosity of the nucleus) ranges from 150 to 200 \citep{,Falcke1996a, Sikora2007}, and accounting for the bulk Lorentz factor places the source in or near the radio-quiet group with intrinsic radio loudness ranging from 2 to 35 \citep{Chamani2020}. This source has a compact, core-dominated, low-power jet, and it harbours a rapidly spinning black hole \citep{Chamani2020} with a mass of $1.8 \times 10^{8}$\,M$_{\odot}$ \citep{Grier2012}. The parsec scale jet was ejected towards the northwestern direction as observed at 43\,GHz in the late 1990s \citep{Brunthaler2000, Brunthaler2005}. At kiloparsec scales two-sided structure has been detected; one of the components is a weak lobe or a hot spot connected to the core with a structure resembling a jet on the southwestern side \citep{Unger1987, Brunthaler2005, Cooper2007}, whereas an even fainter lobe is seen on the northeastern side \citep{Brunthaler2005}.
In the late 1990s an episode of strong outbursts led to changes in the turnover frequency of the synchrotron spectrum, and to the ejection of a new emission feature in VLBI images moving with an apparent superluminal speed of $1.25\,\mathrm{c} \pm 0.09$\,c at 43\,GHz \citep{Brunthaler2000}. Two more ejections have been since then observed with the Very Long Baseline Array (VLBA) at 15\,GHz with apparent speeds of $1.58\,\mathrm{c} \pm 0.29$\,c and $1.358\,\mathrm{c} \pm 0.074$\,c \citep{Lister2019}.

In this work, we aim to test the magnetic flux paradigm and determine whether this particular RIQ has reached the MAD state or not. III\,Zw\,2 has a relatively low jet production efficiency yet it has a rapidly spinning black hole. Low accumulated magnetic flux would therefore be a natural expectation if the jets are launched by the Blandford-Znajek mechanism.  We demonstrate here that such a test is feasible and can be carried out for a sample of RIQ and RQQ sources in the future. We estimate the magnetic field strength of III\,Zw\,2 employing the core-shift measurements from phase-referencing observations. VLBA phase-referencing observations, although challenging, have been previously used to successfully measure the core-shift, e.g., in M\,87 \citep{Hada2011} and NGC\,4261 \citep{Haga2015}.

The paper is organized as follows: In Section~2, we present our phase-referencing multi-frequency VLBA observations, the data calibration and imaging. Section~3 is devoted to the analysis method for measuring the core-shift in III\,Zw\,2. We present in Section~4 the results of the core-shift measurements that include the self-referencing and the phase-referencing core-shifts, the upper limit estimation of the core-shift and magnetic field parameters for III\,Zw\,2. Finally, Section~5 presents the summary and discussion.

\begin{table}[t]
\centering
\begin{threeparttable}
\caption{Phase reference sources for III\,Zw\,2}
 \label{t1}
\begin{tabular}{ccc}
\hline
Source &  d\tnote{1} ($^{\circ}$) & Redshift  \\ \hline
J0006+1235 & 1.91   & 0.98\tnote{2}       \\
J0007+1027 & 0.82   & -       \\
J0008+1144 & 0.98   & -       \\ \hline
\end{tabular}
\begin{tablenotes}
\item [1] Separation from the target.
\item [2] \cite{Snellen2002}.
\end{tablenotes}
\end{threeparttable}
\end{table}

\section{Observations and Data Calibration}
\subsection{Observations}

We carried out phase-referencing observations with the VLBA in order to do quasi-simultaneous multi-band relative astrometry of the target III\,Zw\,2 with the nearby reference sources: J0006+1235, J0007+1027 and J0008+1144 (hereafter J0006, J0007 and J0008), see Table~\ref{t1}. Since III\,Zw\,2 itself is very compact and bright enough for self-calibration, we use it to calibrate the phases, which are then transferred to the reference sources. Despite this, from here on we will call the reference sources "calibrators" for the sake of brevity. Two of the calibrators (J0007 and J0008) are within 1$^\circ$ from III\,Zw\,2, while the third (J0006) is within 2$^\circ$. The bright quasar 3C\,454.3 was observed as the primary calibrator.

The observations were performed on November 8, 2017, at four frequencies: C-band (centre frequency 4.148\,GHz), X-band (7.652\,GHz), K$_\mathrm{U}$-band (15.352\,GHz; hereafter "U-band") and K-band (23.88\,GHz). The use of the wide C-band receiver allowed the simultaneous observations at 4 and 8\,GHz. The observed bandwidth was 256\,MHz at each band with eight 32\,MHz wide intermediate frequency (IF) sub-bands. The data were digitized with two bits, and dual polarization was recorded at U and K bands, whereas for C/X band only right-hand circular polarization was recorded. The total recording rate was 2\,Gbps.

The total observing time was eight hours. The observations were performed in frequency blocks of roughly 10\,min, 16\,min and 23\,min each at C/X, U and K bands, respectively. The blocks were distributed evenly over the eight hours and each block consisted of fast switching between the target and the calibrators with cycle times adjusted for the each band and source. The cycle times (T-C-T) were 165$-$205\,s at C/X band, 135$-$160\,s at U-band and 90$-$105\,s at K-band. Total on-source observing times for III\,Zw\,2 were 56\,min at K-band, 36\,min at U band and 28\,min at C/X band.

Additionally, four geodetic-VLBI style K-band observing blocks of 30\,min each were included in the schedule. Purpose of these "geo-blocks" is to estimate the slowly varying component of the residual zenith tropospheric delay at each telescope by measuring the group delay of a number of compact sources over a range of azimuths and elevations in a rapid succession \citep{Reid2009}. We used the automatic geo-block generation algorithm implemented in the VLBI scheduling software \textsc{sched} with a list of candidate sources selected to have ICRF positions accurate to better than 1\,mas and more than 100\,mJy of correlated flux on long baselines. To improve the delay measurement accuracy, we also used a geodetic-style frequency setup for the geo-blocks --- the IFs were spread out to cover the whole available 500\,MHz band with minimally redundant frequency separations. 

\subsection{Calibration}

We calibrated the data in AIPS \citep{Greisen2003} following standard procedures for a phase-referencing experiment \citep[see e.g.,][]{Reid2009}. This included \textit{a priori} corrections to the Earth Orientation Parameters and parallactic angle, corrections to dispersive ionospheric residual delays based on global maps of ionospheric total electron content derived from the Global Navigation Satellite System (GNSS) data, and calibration of instrumental single-band delays and phase offsets between IFs. We also corrected the small amplitude biases due to drifts in sampler threshold levels using autocorrelations, calibrated the complex bandpass shapes, and performed \textit{a priori} amplitude calibration using recorded system temperatures and gain curves.

\begin{figure}[ht]
\centering
   \subfigure[]   
   {
        \includegraphics[width=0.45\textwidth]{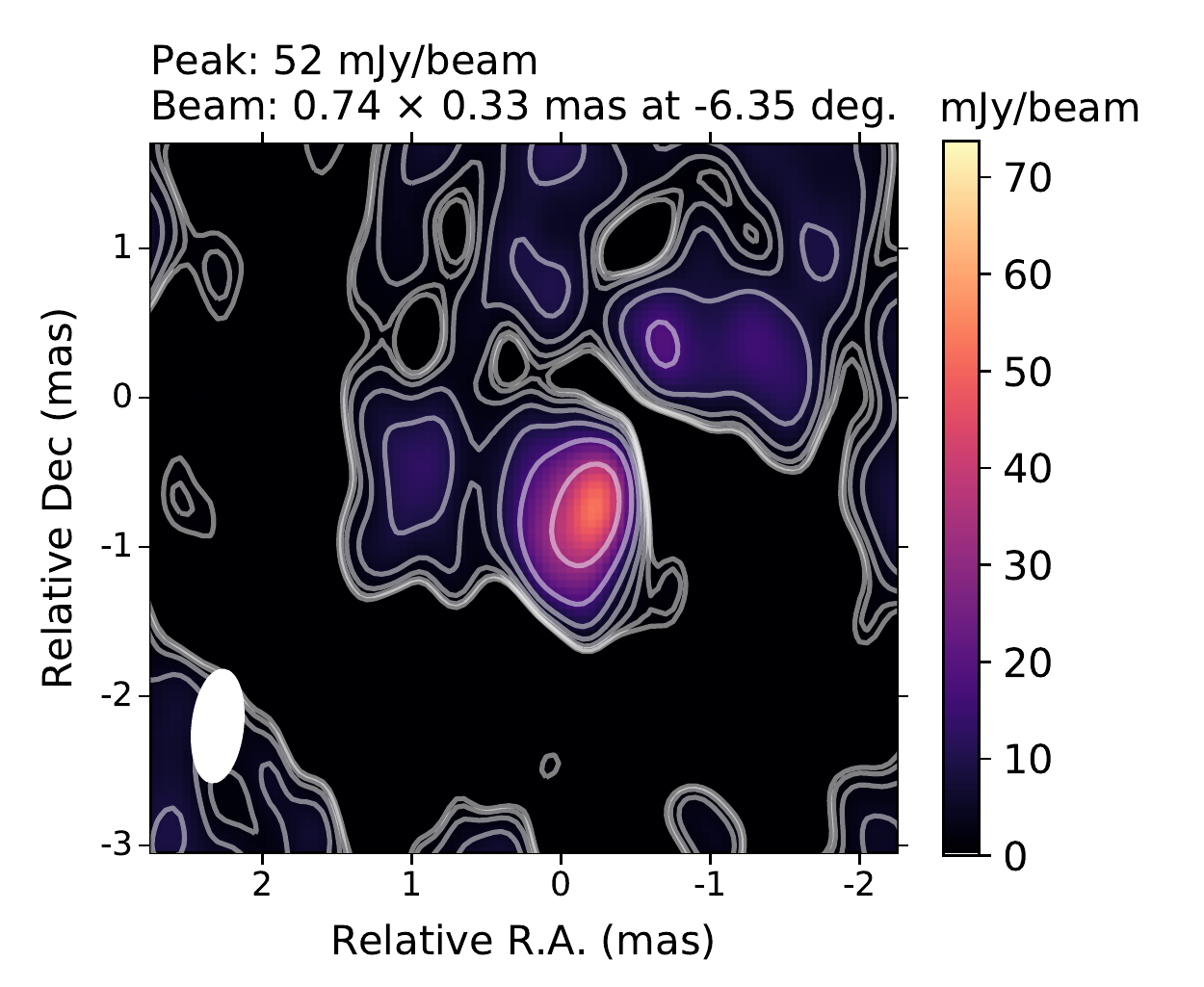}
        \label{}
    }
         \subfigure[]
    {
        \includegraphics[width=0.45\textwidth]{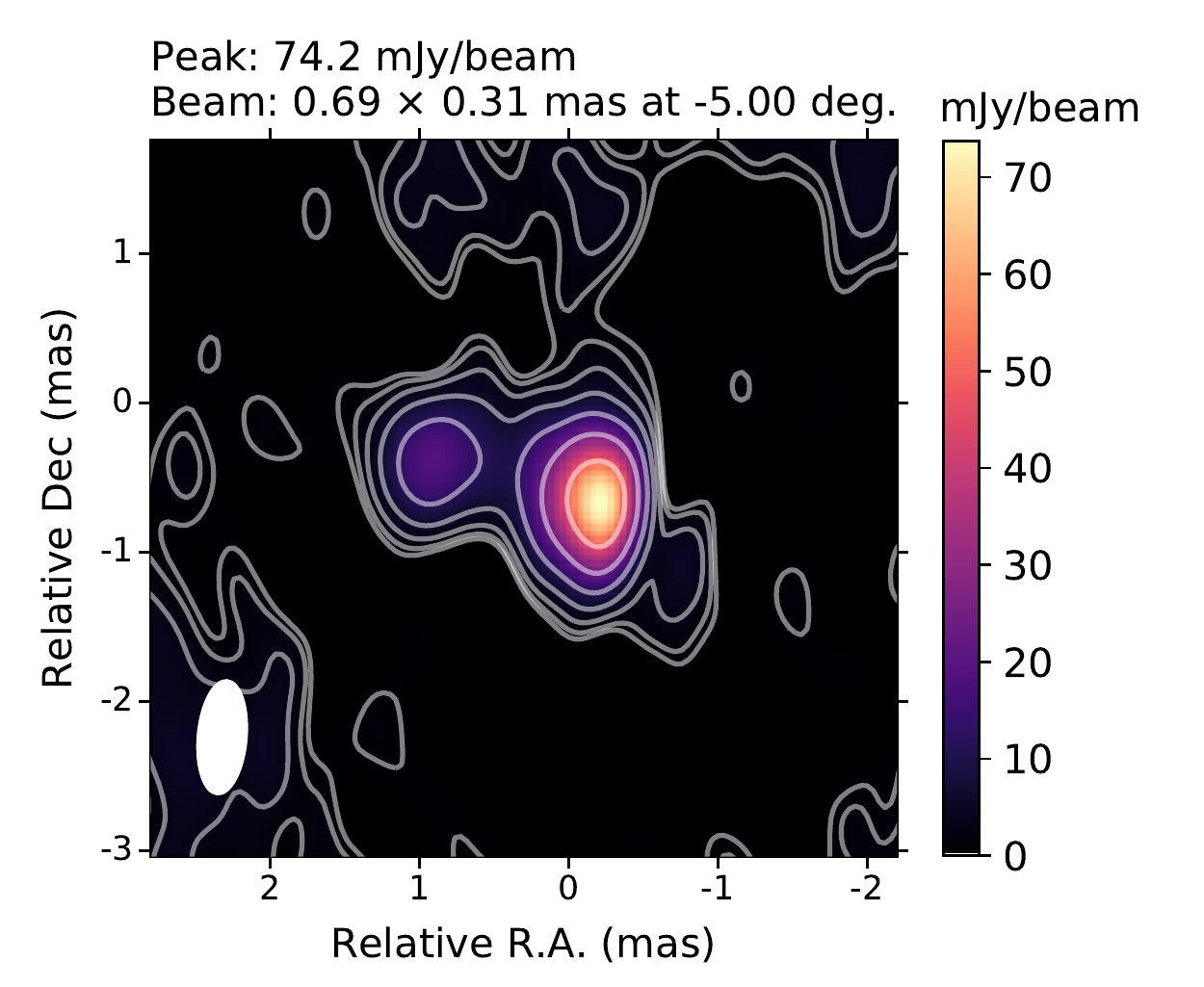}
        \label{}
    }
    \caption{Phase referenced images of J0006 observed at the K-band after a few CLEAN iterations. (a) The image without troposphere delay corrections has noise rms of 2.7\,mJy/beam and (b) the image with troposphere delay corrections has noise rms of 1.3\,mJy/beam. The interferometric beam (ellipse) is displayed on the bottom-left corner of each image. Contours represent 1\%, 2\%, 4\%, 8\%, 16\%, 32\% and 64\% of the peak intensity at each image.}
    \label{f1}
\end{figure}

For processing the geodetic block data we used the procedure presented in \citet{Mioduszewski2009}. It consists of fringe-fitting the geo-block data, using the \textsc{mbdly} task to measure the multi-band delays, and finally obtaining the zenith tropospheric delays and clock errors of the antennas from a fit to multi-band delays with the task \textsc{delzn}. These solutions were applied to each data set before the fringe-fitting. The fringe-fitting was performed with the task \textsc{fring} on the target itself, assuming a point-like source, and then the solutions were applied with \textsc{clcal} to the calibrators, and phase-referenced images were obtained for each of them. We compared phase-referenced CLEAN images with and without tropospheric delay corrections, and we found that dynamic range of the images improved significantly with the tropospheric delay calibration, especially at the higher frequencies. On average, the signal-to-noise ratio (SNR or peak over noise rms) improvements were a factor of 1.5, 1.4, 2.1, and 2.0 for the C, X, U and K bands, respectively. The most notable improvement of the SNR by a factor of 3 was obtained for the U and K bands for J0006. An illustration of this is shown in Figure~\ref{f1}, which shows the images of J0006 with and without troposphere correction at the K-band.

\begin{figure*}[!h]
\centering
   \subfigure[]
   {
        \includegraphics[width=0.4\linewidth]{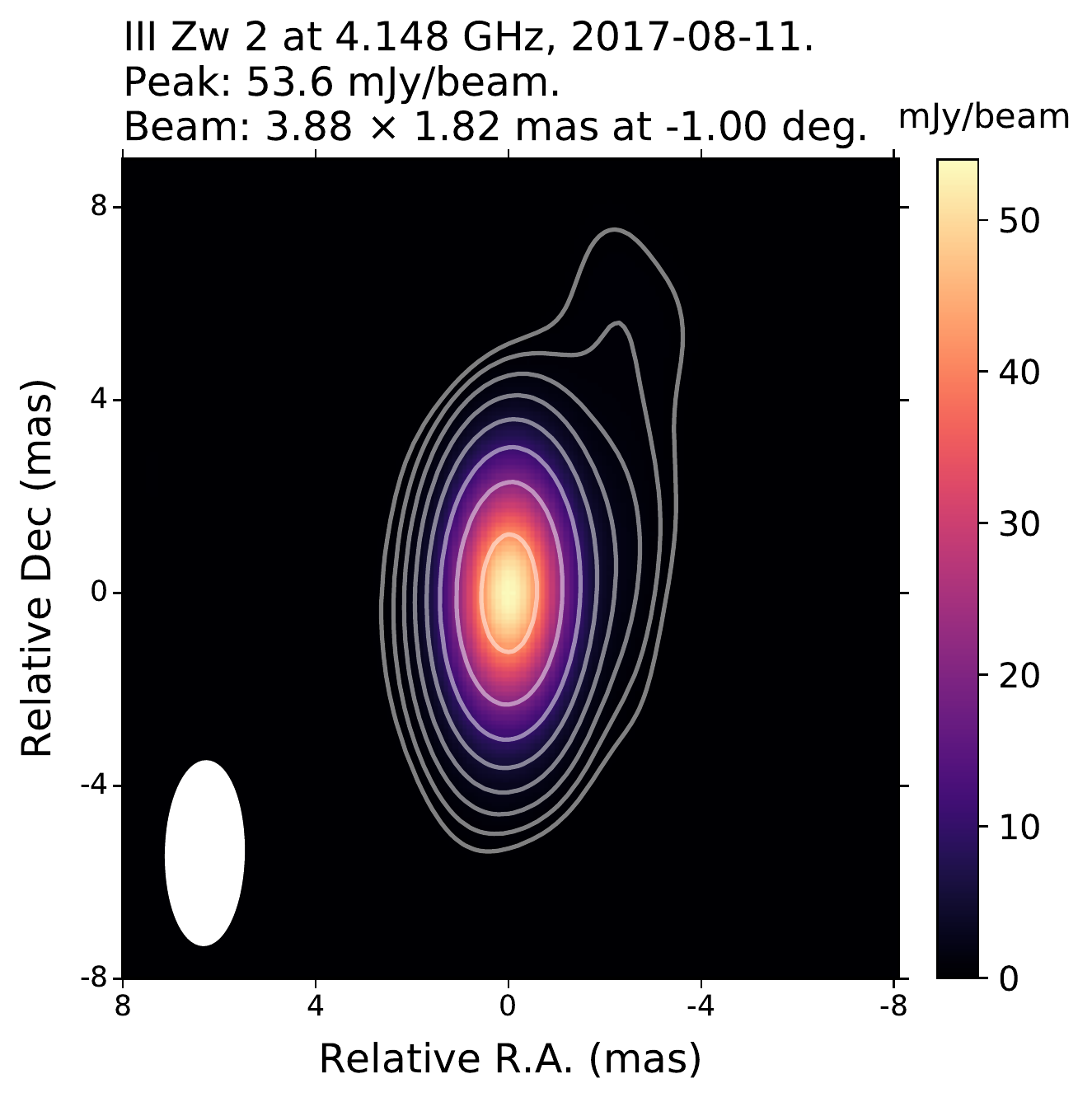}
        \label{}
    }
         \subfigure[]
    {
        \includegraphics[width=0.4\linewidth]{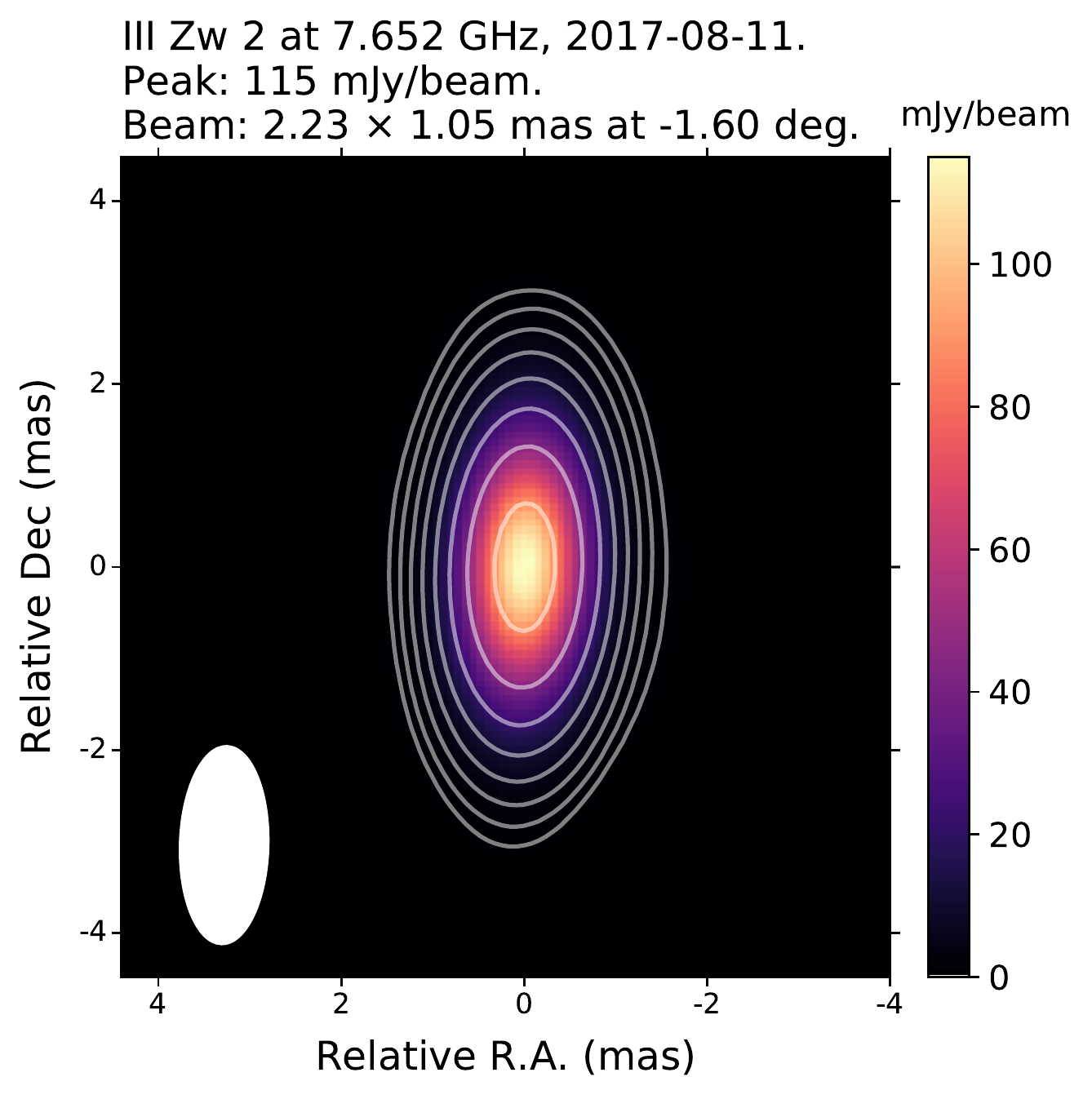}
        \label{}
    }
    \subfigure[]
    {
        \includegraphics[width=0.4\linewidth]{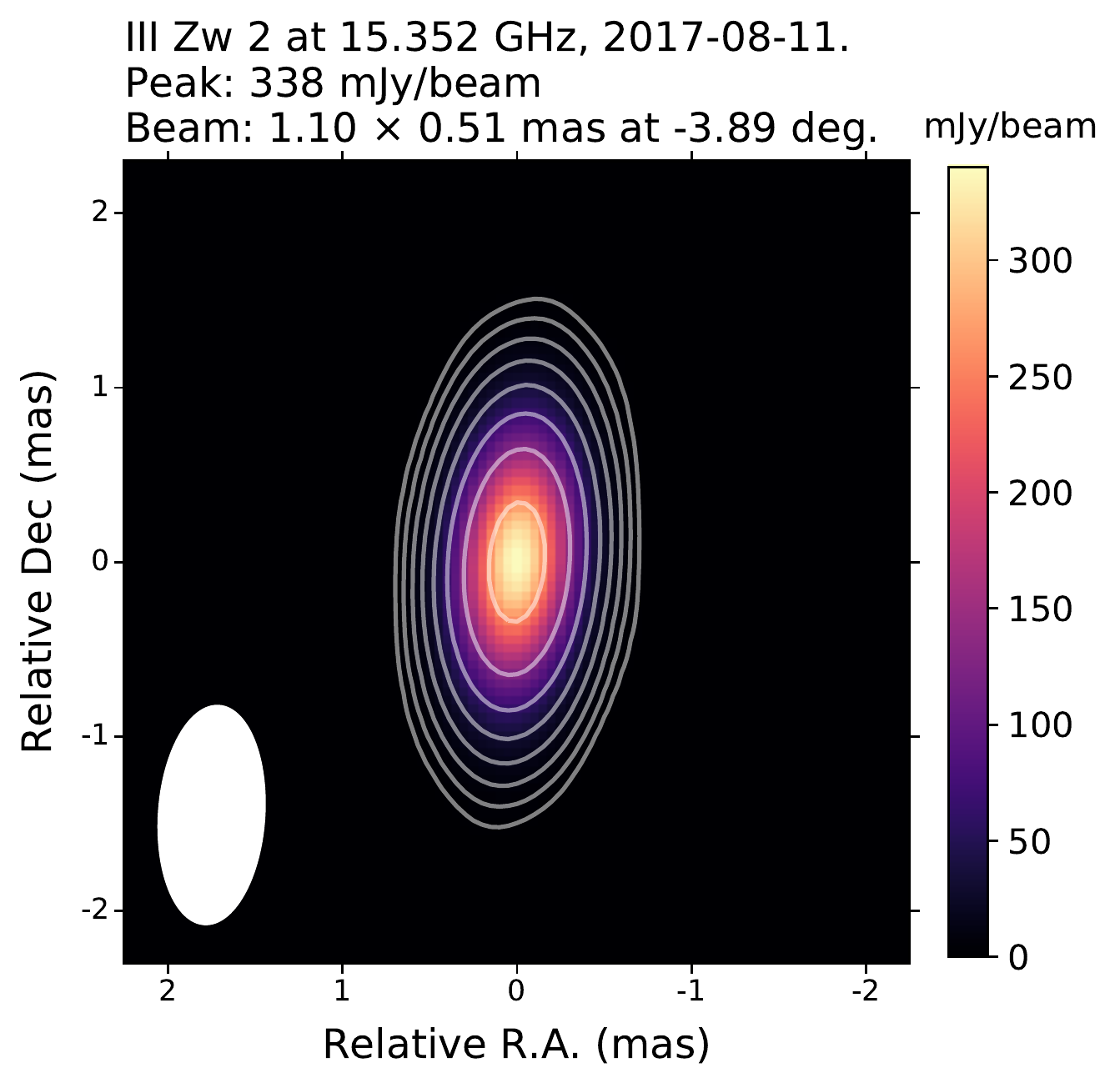}
        \label{}
    }
      \subfigure[]
    {
        \includegraphics[width=0.4\linewidth]{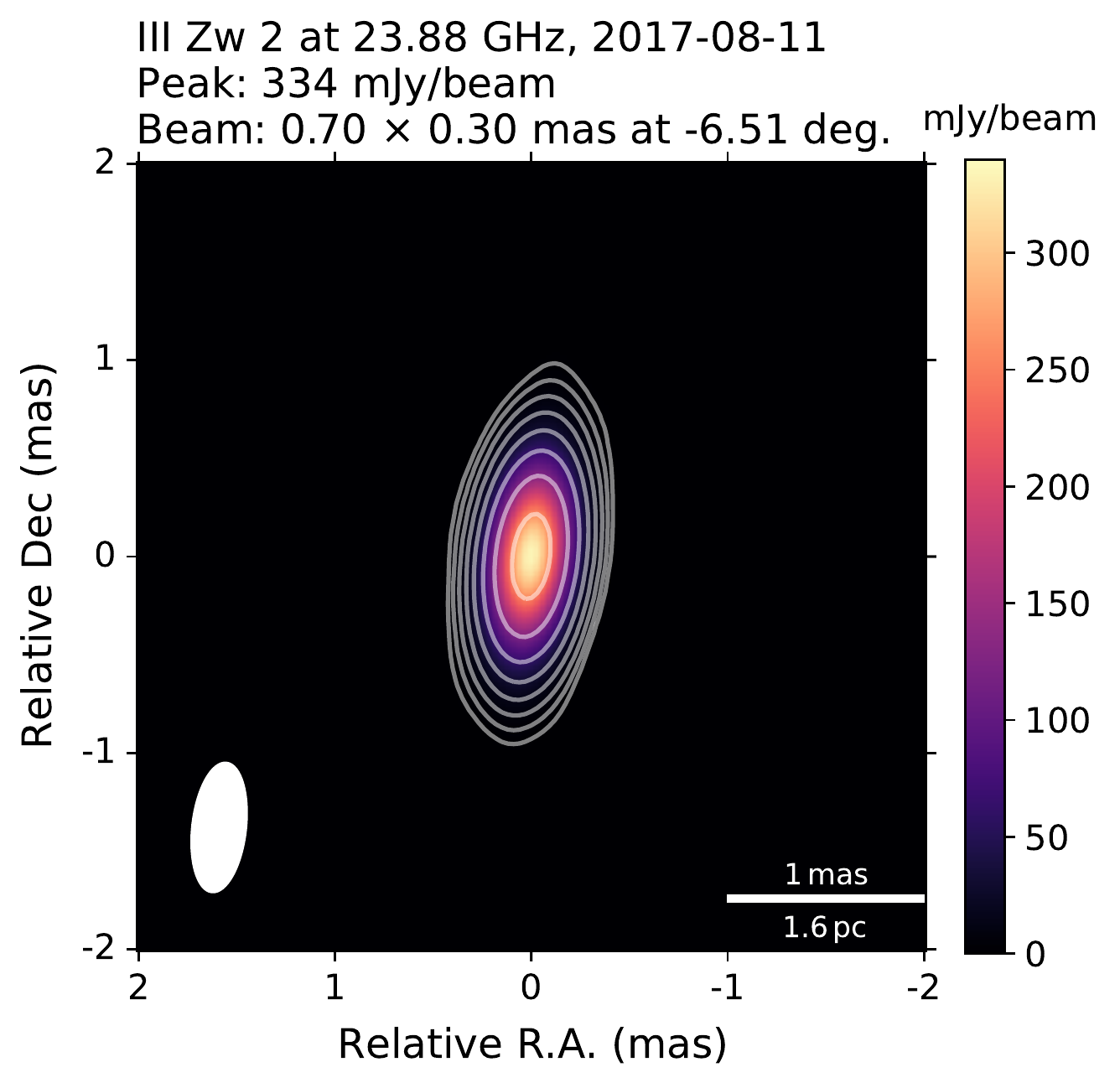}
        \label{}
    } 

    \caption{Self-calibrated images of III\,Zw\,2. The images were obtained at a) 4.148\,GHz, b) 7.652\,GHz, c) 15.352\,GHz and d) 23.88\,GHz. The rms noise level from the lower to the higher frequency is 0.1, 0.1, 0.1 and 0.2\,mJy/beam. The interferometric beam (ellipse) is displayed on the bottom-left corner of each image. Contours represent -0.6\%, 0.6\%, 1.2\%, 2.4\%, 4.8\%, 9.6\%, 19.2\%, 38.4\% and 76.8\% of the peak intensity at each image. }
    \label{f2} 
\end{figure*}

\subsection{Imaging}
We imaged the troposphere-corrected data with  {\sc Difmap} \citep{Shepherd1997} and produced both phase-referenced and self-calibrated CLEAN images with natural weighting. These are all shown in Appendices~A and B. III\,Zw\,2 has a compact structure with very little extended emission at all the bands, see Figure~\ref{f2}. However, previous observations show that the jet is oriented towards West-Southwest at kilo-parsec scales \citep{Cooper2007} and to West-Northwest at parsec scales \cite{Pushkarev2017}.

On the other hand, the CLEAN images of all the calibrators show extended jet emission at all frequencies. The most distant calibrator J0006 has a jet oriented towards East with a compact radio core seen at all frequencies and a bright component downstream which is resolved at 15 and 24\,GHz (see Figure~\ref{fB1}). The nearest calibrator J0007 has a rich and complex extended jet structure towards North with a certain degree of bending visible at all frequencies. J0007 also has a compact core and a visible bright knot located farther North (see Figure~\ref{fB2}). The second nearest calibrator J0008 has an extended jet structure oriented towards South-East with a compact core (see Figure~\ref{fB3}). We point out here that our VLBA observations have produced for the first time high-dynamic range, high-resolution images of the calibrators that show their parsec-scale radio morphology in detail.

The core spectrum of III\,Zw\,2 and each calibrator was obtained by fitting a 2D Gaussian component directly to the visibility data. The description of the target spectrum is given in Section~4.4.2. Details of the core spectra of individual calibrators are provided in Appendix~C.

\section{Analysis method }

\subsection{Core-shift measurements}

\begin{figure}[t]
 \centering
     \includegraphics[width=0.53\textwidth]{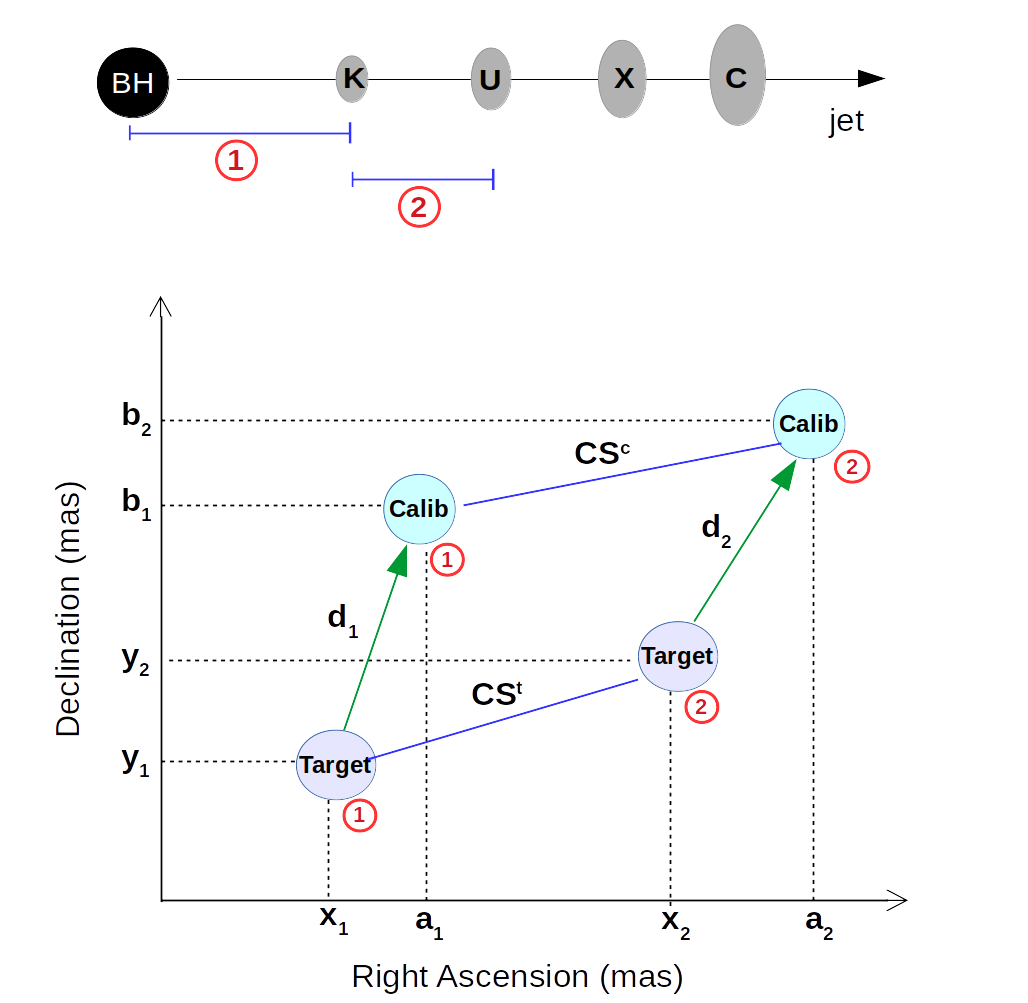}
     \caption{Figure on the top shows the black hole ('BH'; filled black circle) that launches a jet in the horizontal direction. The different core positions at each frequency are shown in  ellipses. The K-band core's distance from the black hole is shown with the red circle '1'. The distance between the adjacent frequencies K and U bands is shown with the red circle '2'.  Figure on the bottom is a schematic diagram describing the phase referencing positions ($d_{1}$, $d_{2}$) and the core-shifts of the target (CS$^{t}$) and a calibrator (CS$^{c}$). All coordinates are specified for each source.}
     \label{f3}
\end{figure}

The core is typically --- although not always --- the brightest  and the most compact region found in the VLBI images. Its position ($r$) from the central engine changes with the frequency ($\nu$) due to opacity changes, becoming visible at the distance where $\tau \sim 1$ to synchrotron self-absorption. If one assumes that the magnetic field strength and the particle number density decrease with the distance along the jet as power-laws $B=B_1(\frac{r_1}{r})^m$ and $N=N_1(\frac{r_1}{r})^n$, where $B_1$ and $N_1$ are the values at the distance of $r_1 = 1$\,pc, the dependence of the core position on the frequency follows $r\propto \nu^{-1/k_r}$ \citep{Blandford1979}. Here $k_r$ depends on $m$, $n$ and the optically thin spectral index of the emission \citep[see Eq.~1 in][]{Lobanov1998}. The model of Blandford \& Königl assumes a supersonic, narrow, conical jet in equipartition with a constant half-opening angle and Lorentz factor, which would imply $k_r = 1$. For such a jet, $m=1$ and $n=2$ are typically assumed. The measurement of the positional shift of the core as a function of frequency, together with the equipartition assumption, allows the measurement of the magnetic field strength and the particle number density. 

In our study, we measure the core-shifts of both the target and the calibrators since all calibrators are AGN with jets. Thus the measurement of phase referencing core-shift of the target is actually a combination of calibrators' own core-shifts and the target's core-shift. The latter is  what we want to investigate.\, Figure~\ref{f3} displays a diagram where the core positions and the core-shift vectors of the target and a calibrator are visualized. At position \ding{172} when both the target and the calibrator are observed at the highest frequency band 'K', their coordinates are x$_{1}$, y$_{1}$ and a$_{1}$, b$_{1}$ respectively. The distance between the target and calibrator core in the phase referencing map is denoted by $d_{1}$. At position \ding{173} both the target and the calibrator are observed at the frequency band 'U'. At this point their coordinates are x$_{2}$, y$_{2}$ and a$_{2}$, b$_{2}$ respectively. At this frequency the core position of the calibrator is $d_{2}$. The core-shifts between the K and U bands are denoted as $CS^{t} = \begin{pmatrix} C^{t}_{x}\\C^{t}_{y} \end{pmatrix}$ for the target and $CS^{c} = \begin{pmatrix}C^{c}_{x}\\C^{c}_{y} \end{pmatrix}$ for the calibrator. For the calibrator, the positions are related by
\begin{equation}
\begin{pmatrix}
    a_{1}  \\
    b_{1}
\end{pmatrix}
+
\begin{pmatrix}
    C^{c}_{x}  \\
    C^{c}_{y}
\end{pmatrix}
=
\begin{pmatrix}
    a_{2}  \\
    b_{2}
\end{pmatrix},
\end{equation}
and for the target,
\begin{equation}
\begin{pmatrix}
    x_{1}  \\
    y_{1}
\end{pmatrix}
+
\begin{pmatrix}
    C^{t}_{x}  \\
    C^{t}_{y}
\end{pmatrix}
=
\begin{pmatrix}
    x_{2}  \\
    y_{2}
\end{pmatrix}.
\end{equation}

The distances $d_{1}$ (at the high frequency) and $d_{2}$ (at the low frequency) are given by
\begin{equation}
d_{1}=
\begin{pmatrix}
    d_{1x}  \\
    d_{1y}
\end{pmatrix}
=
\begin{pmatrix}
    a_{1}  \\
    b_{1}
\end{pmatrix}
-
\begin{pmatrix}
    x_{1}  \\
    y_{1}
\end{pmatrix},
\end{equation}
and
\begin{equation}
d_{2}=
\begin{pmatrix}
    d_{2x}  \\
    d_{2y}
\end{pmatrix}
=
\begin{pmatrix}
    a_{2}  \\
    b_{2}
\end{pmatrix}
-
\begin{pmatrix}
    x_{2}  \\
    y_{2}
\end{pmatrix}.
\end{equation}

Subtracting Equations 1 and 2, plugging in 3 and 4, and re-arranging, we obtain the core-shift for the target as

\begin{equation}
 \begin{pmatrix}
    C^{t}_{x}  \\
    C^{t}_{y}
\end{pmatrix}
=
 \begin{pmatrix}
    C^{c}_{x}  \\
    C^{c}_{y}
\end{pmatrix}
+
\begin{pmatrix}
    d_{1x}  \\
    d_{1y}
\end{pmatrix}
-
\begin{pmatrix}
    d_{2x}  \\
    d_{2y}
\end{pmatrix}.
\end{equation}

Since all of our calibrators have extended jet structure, the core-shift vectors of the calibrator, CS$^{c}$, for the given frequency pair UK can be measured using the self-referencing method, i.e., aligning optically thin jet components at different frequencies. The distances $d_1$ and $d_2$ can be measured from phase-referencing observations.  Note that the method described above does not make any \textit{a priori} assumption of the jet directions of the sources involved. The measurement of $CS^{c}$, $d_1$ and $d_2$ are described in the next section.

\begin{figure*}[]
\centering 
    \subfigure[J0006: all vectors consistently point in the jet direction towards East; for a comparison see Figure~\ref{fB1}.]
    {
        \includegraphics[width=0.47\textwidth]{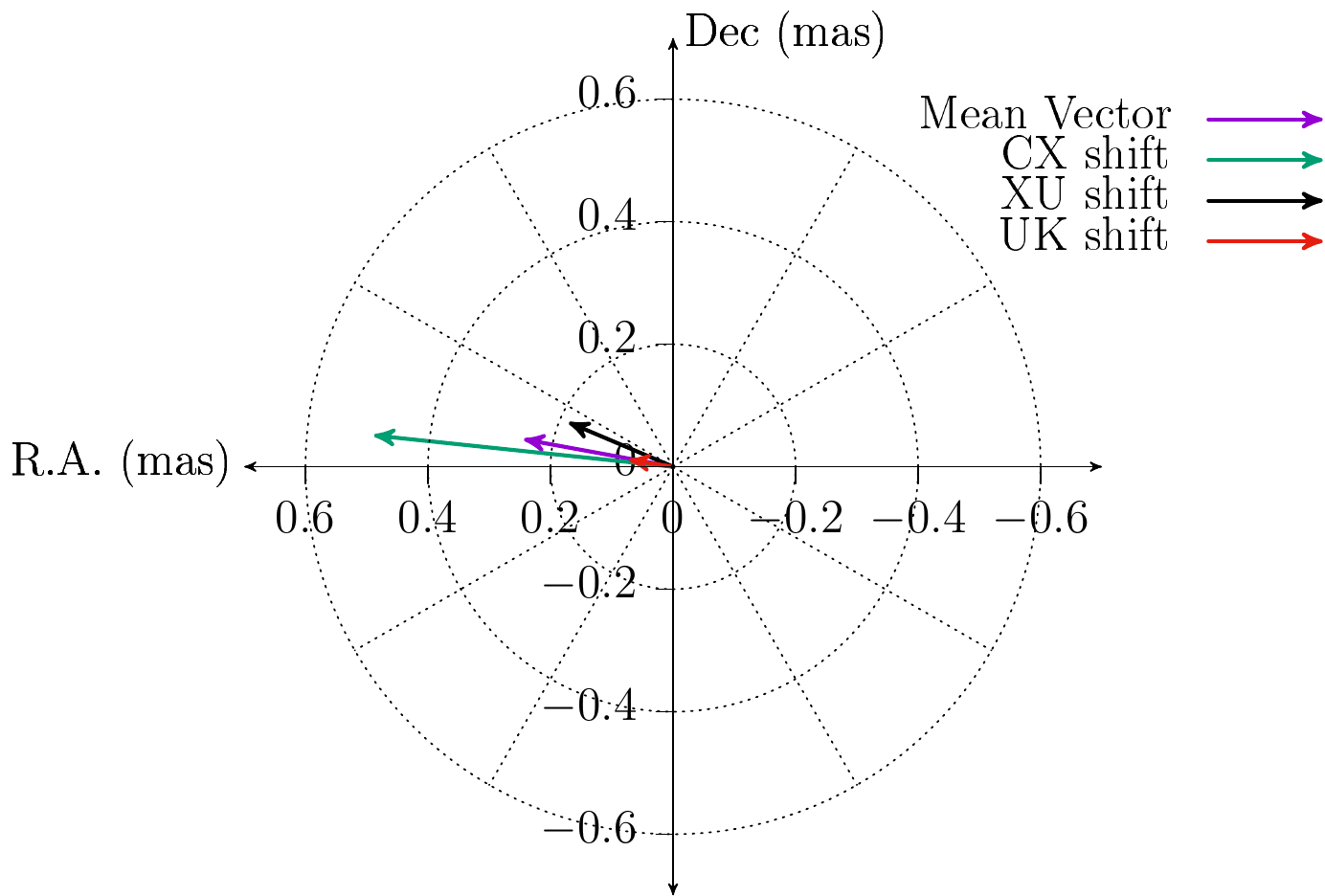}
        \label{}
    }
    \hspace{0.5cm}
    \subfigure[J0006: best fitting curve (red colour) gives $a = 5.8 \pm 1.4$, \hspace{1cm}$k_r = 0.72 \pm 0.08$. The blue curve indicates the fit with $k_r = 1$ (and $b = 3.3 \pm 0.2$).]
    {
        \includegraphics[width=0.47\textwidth]{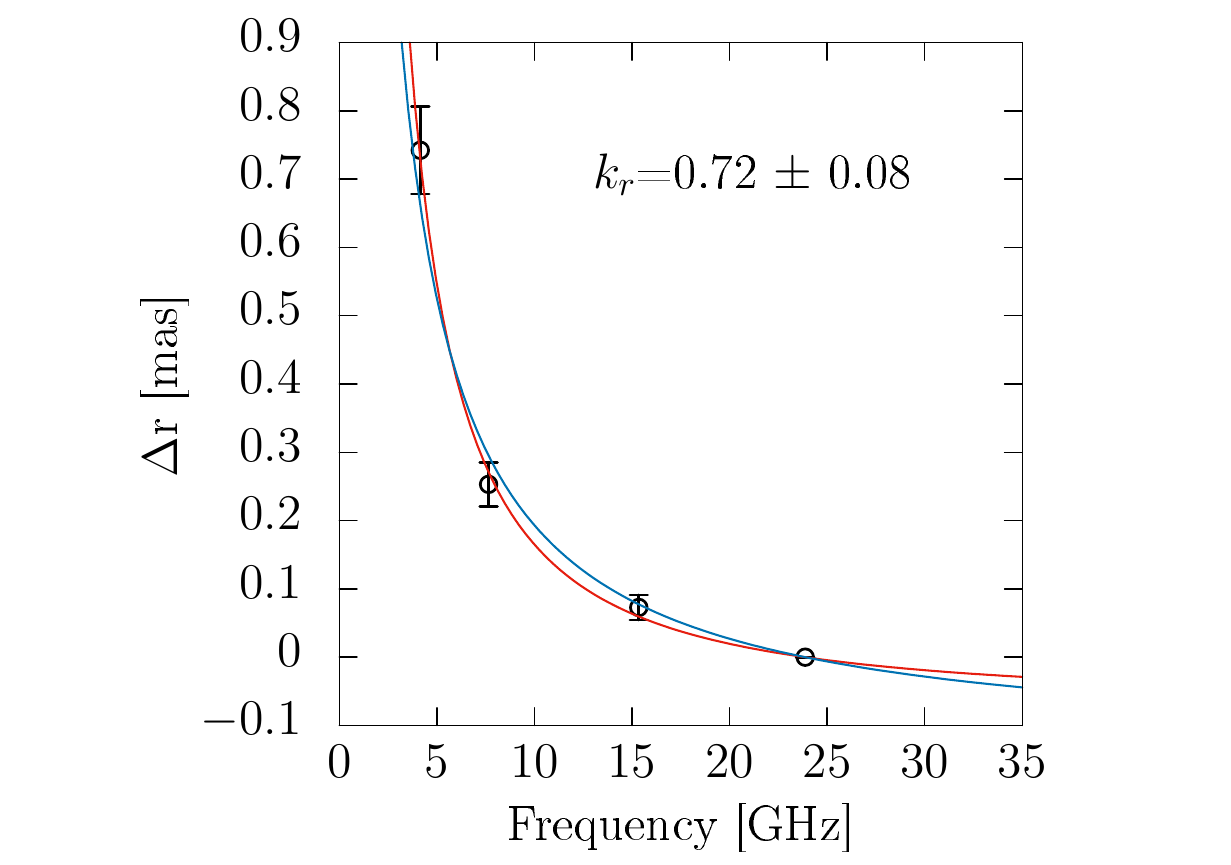}
        \label{}
    }
    \subfigure[J0007: CX, XU and the mean vectors consistently point in the jet direction towards north, while the UK vector points to NW; for comparison see Figure~\ref{fB2}.] 
    {
         \includegraphics[width=0.47\textwidth]{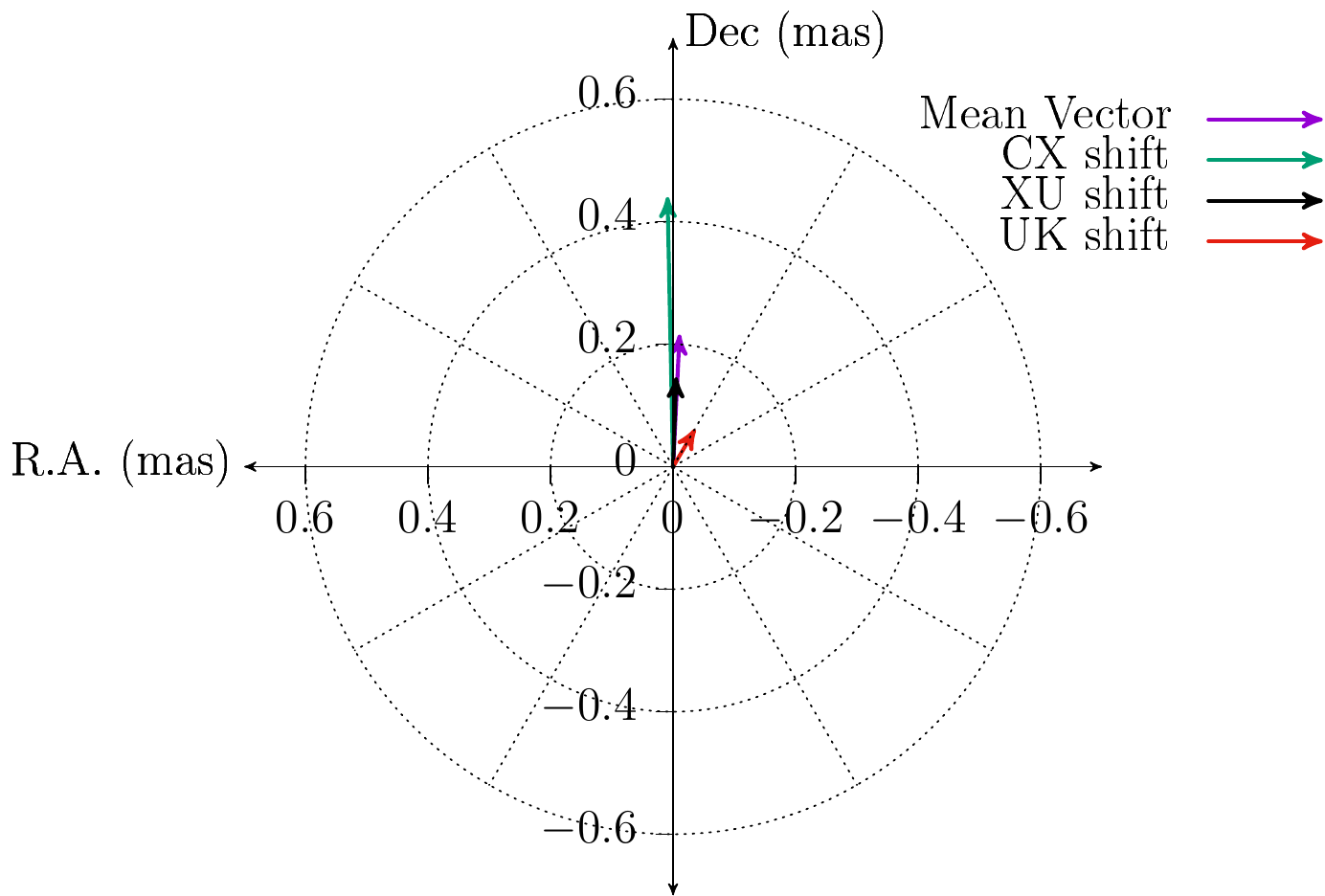}
        \label{}
    }
    \hspace{0.5cm}
    \subfigure[J0007: best fitting curve (red colour) gives $a = 5.04 \pm 1.61$, \hspace{1cm} $k_r = 0.71 \pm 0.11$. The blue curve indicates the fit with $k_r = 1$ (and $b = 2.8 \pm 0.2$). ]  
    {
      \includegraphics[width=0.47\textwidth]{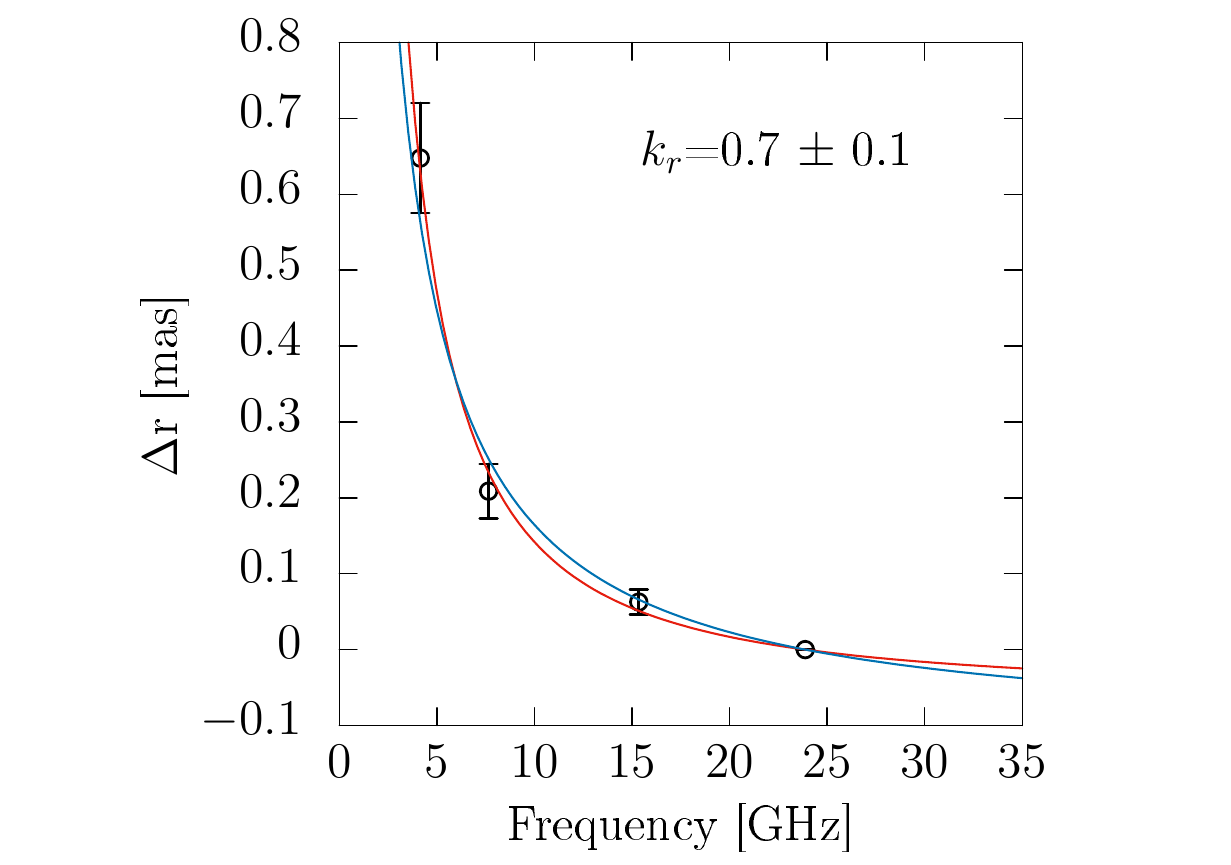}
      \label{}
    }
     \subfigure[J0008: only the CX and the mean vector consistently point in the jet direction towards South-East, for comparison see Figure~\ref{fB3}.] 
    {
        \includegraphics[width=0.47\textwidth]{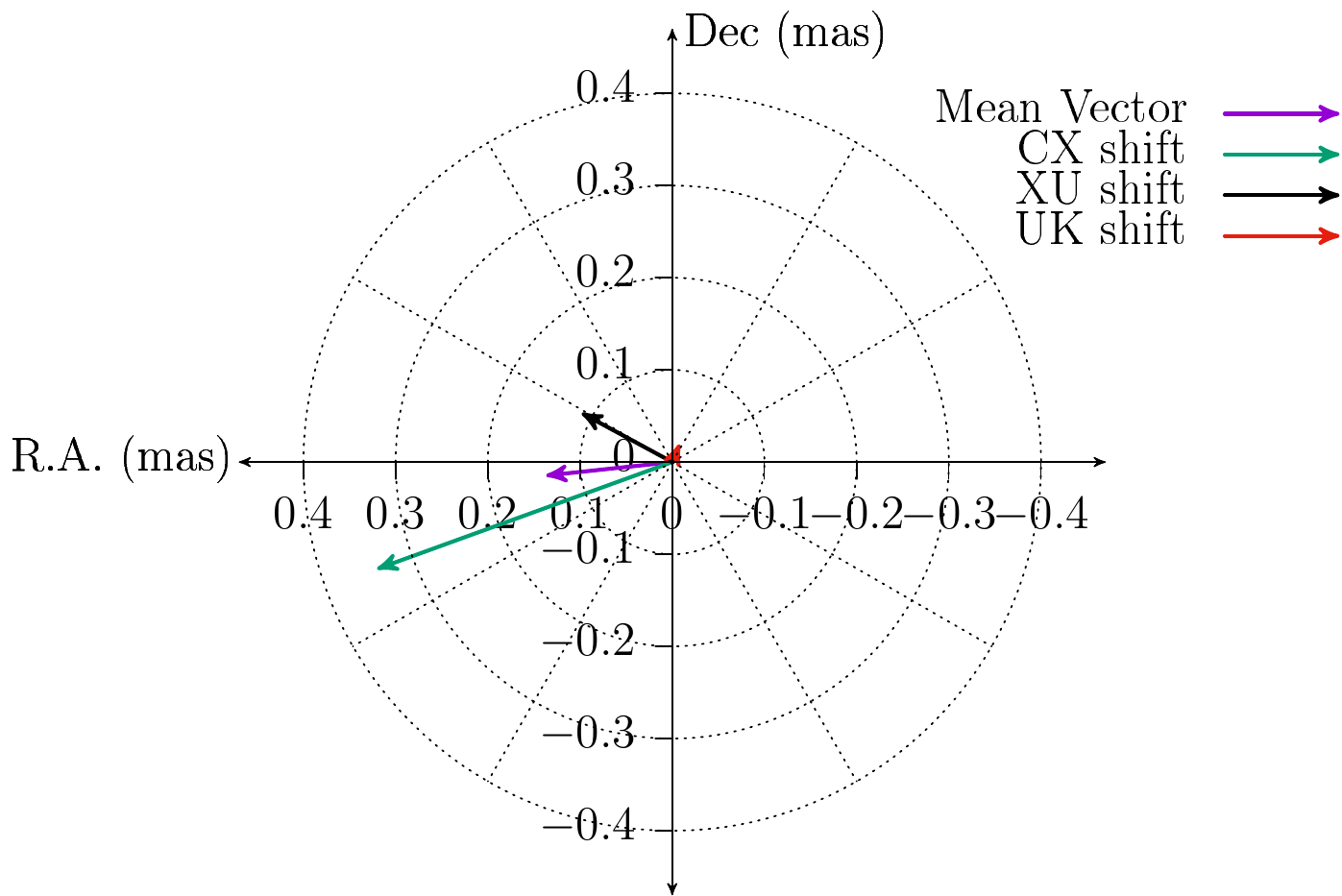}
        \label{}
    }
    \hspace{0.5cm}
    \subfigure[J0008: best fitting curve (red colour) gives $a = 31.7 \pm 38.8$, \hspace{1cm}$k_r = 0.33 \pm 0.09$.  By fitting the data without the U-band, the green curve indicates the fit with $k_r = 1$ (and $b = 1.2 \pm 0.4$).]
    {
        \includegraphics[width=0.47\textwidth]{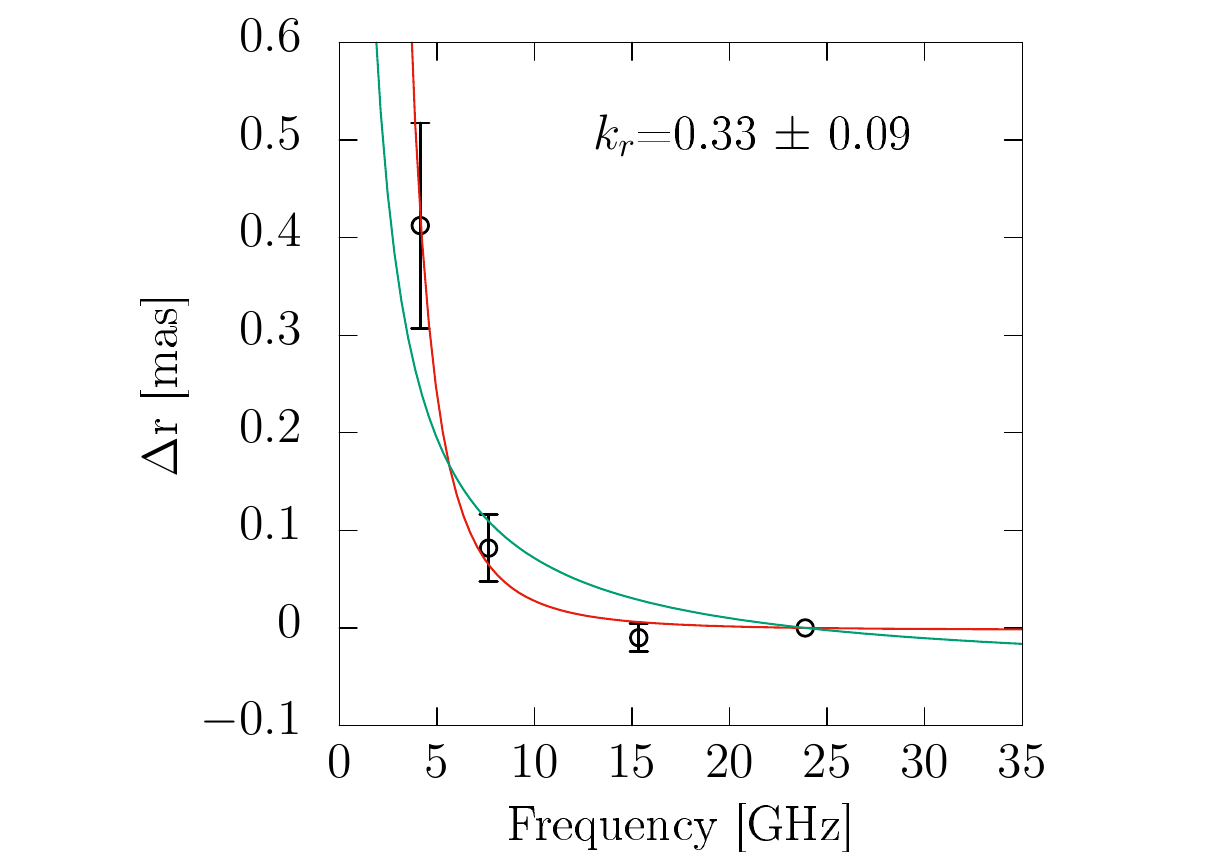}
        \label{}
    }
    \caption{Left: self-referencing core-shift vectors in polar grids. The dotted lines are given at intervals of 30 degrees. Right: core-shift ($\Delta r$) as a function of frequency ($\nu_\mathrm{GHz}$). The red curve indicates the best fit of the form: $\Delta r = a(\nu_\mathrm{GHz}^{-1/k_r} - 23.88^{-1/k_r}$) with the K-band as the reference frequency. The blue and green curves indicate the fits when the power law index $k_r$ is set to 1 with the fitting function: $\Delta r = b(\nu_\mathrm{GHz}^{-1} - 23.88^{-1}$). $a$ and $b$ are fitting parameters.}
    \label{f4}
\end{figure*}

\begin{table*}[]
\centering
\begin{threeparttable}
\caption{Self-referencing core-shifts vectors of each calibrator for different frequency pairs.}
 \label{t2}
\begin{tabular}{cccccc}
\hline \noalign{\smallskip}
\multicolumn{1}{c}{Source} & \begin{tabular}[c]{@{}c@{}}Frequency\\ pair\end{tabular} & \multicolumn{1}{c}{\begin{tabular}[c]{@{}c@{}}R.A. \\ (mas)\end{tabular}} & \multicolumn{1}{c}{\begin{tabular}[c]{@{}c@{}}Dec\\ (mas)\end{tabular}} &
\multicolumn{1}{c}{\begin{tabular}[c]{@{}c@{}}Absolute values\tnote{1} \\ (mas)\end{tabular}} & \multicolumn{1}{c}{\begin{tabular}[c]{@{}c@{}}Projected absolute\tnote{2} \\ values (mas)\end{tabular}}  \\\hline 
 \noalign{\smallskip}
\multirow{3}{*}{J0006} & CX & \multicolumn{1}{c}{$0.49 \pm 0.06$} & $0.05 \pm  0.05$  & $0.49 \pm 0.06$ & $0.49 \pm 0.06$   \\
 & XU & \multicolumn{1}{c}{$0.17 \pm 0.03$} & $0.07 \pm 0.03$  & $0.18 \pm  0.03$ & $0.18 \pm 0.03$\\
 & UK & \multicolumn{1}{c}{$0.07 \pm 0.02$} & $0.01 \pm 0.03$  & $0.07 \pm  0.02$ & $0.07 \pm 0.02$\\ \hline  \noalign{\smallskip}
\multirow{3}{*}{J0007} & CX & \multicolumn{1}{c}{$0.01 \pm 0.04$} & $0.44 \pm  0.06$  & $0.44 \pm 0.06$ & $0.44 \pm 0.06$\\
 & XU & \multicolumn{1}{c}{$-0.01 \pm 0.02$}  & $0.15 \pm 0.03$  & $0.15 \pm  0.03$ & $0.15 \pm 0.03$ \\
 & UK & \multicolumn{1}{c}{$-0.04 \pm 0.01$}  & $0.06 \pm 0.02$  &  $0.07 \pm 0.02$ & $0.06 \pm 0.02$ \\ \hline  \noalign{\smallskip}
\multirow{3}{*}{J0008} & CX & \multicolumn{1}{c}{$0.32 \pm 0.10$} & $-0.12 \pm 0.06$ & $0.34 \pm 0.10$ & $0.33 \pm 0.10$  \\
 & XU & \multicolumn{1}{c}{$0.10 \pm 0.03$}  & $0.05 \pm 0.03$  & $0.11 \pm  0.03$ & $0.09 \pm 0.03$   \\
 & UK & \multicolumn{1}{c}{$-0.01 \pm 0.02$}  & $0.02 \pm 0.02$  & $0.02 \pm 0.02$ & $-0.01 \pm 0.02$ \\ \hline  \noalign{\smallskip}
\end{tabular}
\begin{tablenotes}
\item[1] The magnitude of the vector. \item[2] The scalar product of the core-shift vector with the mean direction vector.
\end{tablenotes}
\end{threeparttable}
\end{table*}

\begin{table*}[ht]
\centering
\begin{threeparttable}
\caption{Phase referencing (combined) core-shift vectors for different frequency pairs.}
 \label{t3}
\begin{tabular}{cccccc}
\hline  \noalign{\smallskip}
Source & \begin{tabular}[c]{@{}c@{}}Frequency \\ pair\end{tabular} & \begin{tabular}[c]{@{}c@{}}R.A.\\ (mas)\end{tabular} & \begin{tabular}[c]{@{}c@{}}Dec\\ (mas)\end{tabular} & \begin{tabular}[c]{@{}c@{}}Absolute values\tnote{1}\\ (mas)\end{tabular} & \begin{tabular}[c]{@{}c@{}}Projected absolute\tnote{2} \\values (mas)\end{tabular} \\ \hline  \noalign{\smallskip}
\multirow{3}{*}{J0006} & CX & 0.38 $\pm$ 0.04 & 0.03 $\pm$ 0.09 & 0.38 $\pm$ 0.04  & 0.37 $\pm$ 0.05 \\
 & XU & 0.12 $\pm$ 0.02 & 0.08 $\pm$ 0.04 & 0.14 $\pm$ 0.03 & 0.13 $\pm$ 0.02 \\
 & UK & 0.10 $\pm$ 0.02 & 0.00 $\pm$ 0.03 & 0.10 $\pm$ 0.02 & 0.10 $\pm$ 0.02 \\\hline  \noalign{\smallskip}
\multirow{3}{*}{J0007} & CX & -0.02 $\pm$ 0.03 & 0.29 $\pm$ 0.12 & 0.29 $\pm$ 0.12 & 0.29 $\pm$ 0.12  \\
 & XU & -0.04 $\pm$ 0.03 & 0.27 $\pm$ 0.05 & 0.27 $\pm$ 0.04  & 0.28 $\pm$ 0.05  \\
 & UK & 0.02 $\pm$ 0.02 & 0.09 $\pm$ 0.02 & 0.09 $\pm$ 0.02 & 0.09 $\pm$ 0.02 \\ \hline  \noalign{\smallskip}
\multirow{3}{*}{J0008} & CX & 0.27 $\pm$ 0.07 & -0.27 $\pm$ 0.08 &  0.38 $\pm$ 0.08 & 0.37 $\pm$ 0.09 \\
 & XU & 0.16 $\pm$ 0.02 & -0.07 $\pm$ 0.05 & 0.17 $\pm$ 0.03 & 0.17 $\pm$ 0.04 \\
 & UK & 0.03 $\pm$ 0.02 & 0.03 $\pm$ 0.03 & 0.04 $\pm$ 0.03 & 0.01 $\pm$ 0.02 \\ \hline  \noalign{\smallskip}
\end{tabular}
\begin{tablenotes}
\item[1] The magnitude of the vector. \item[2] The scalar product of the core-shift vector with the mean direction vector.
\end{tablenotes}
\end{threeparttable}
\end{table*}

\section{Results}
In this section, we describe separately the measurements of the self-referencing and phase-referencing core-shifts.

\subsection{Self-referencing core-shifts}

A well-known method to measure the core-shifts in VLBI images is the so-called self-referencing. The method consists of aligning optically thin regions of the jet at the different frequencies \citep{Lobanov1998, Kovalev2008, Sokolovsky2011, Fromm2013}. To achieve this, we used the 2D cross-correlation technique \citep[e.g.,][]{Walker2000, Croke2008, Pushkarev2012, Fromm2013, Plavin2019}  to align images at adjacent frequency pairs: CX, XU and UK for each calibrator. The cross-correlation analysis was performed with the software described in \citet{Pushkarev2012}. For such an analysis, we first produced images at different frequencies with the same pixel size --- the minor axis of the beam divided by 20 --- and both convolved with the restoring beam of the lower frequency image. The image alignment procedure was run ten times per image pair by selecting slightly different optically thin features every time and excluding the core. The resulting spectral index maps and further details are given in Figures~\ref{fE1}, \ref{fE2} and \ref{fE3} in Appendix~E. We point out that mean image shift values have been used for the core-shift estimations. 

The core was modelled at each frequency (in {\sc Difmap}) with a single circular Gaussian component. The estimation of the core-shift vector for a given frequency pair was performed similarly to the method described by \cite{Pushkarev2012}. In their method, the core-shift coordinates result from the differences between the image shift and core position differences from the map centre of a pair of frequencies ($\nu_1$,$\nu_2$). We used two different approaches to estimate the core-shift's absolute value: a) simply calculating the magnitude of the core-shift vector, b) projecting the core-shift vectors onto the mean direction vector (i.e., the average of the core-shift vectors) to reduce the effect of random errors on core-shift vector directions. In the latter, the absolute values are obtained from the core-shift vector's scalar product with the mean vector.

All measured self-referencing core-shift vectors and their absolute values for all the calibrators are listed in Table~\ref{t2}. The absolute core-shift values measured by both approaches give very similar results. Figure~\ref{f4}ace displays the self-referencing core-shift vectors and mean vectors for each calibrator. 
All absolute core-shifts are significant, except for the UK core-shift in J0008.

The absolute values of the core-shifts and the power-law fits are illustrated in Figure~\ref{f4}bdf. The fit results show $k_r$ indices below one in J0006 and J0007 and well below one in J0008. These results suggest that the jets in these sources may not be conical. Alternatively, the deviation of $k_r$ from unity may also indicate a deviation from the equipartition or that the compact jet in these sources is otherwise not strictly of Blandford \& K\"onigl type. An example of the latter option is a model, where the core at the two highest frequencies does not correspond to a synchrotron photosphere, but instead to a recollimation shock \citep[see e.g.,][and references therein]{Dodson2017}. The position of such a shock is achromatic and that could explain the absence of a core-shift between 15 and 24\,GHz in J0008. 

Furthermore, the total positional uncertainties of the self-referencing core-shift measurement for each coordinate are calculated as
\begin{equation}
\Delta \theta_\mathrm{total,core-shift} = \sqrt{\Delta \theta_\mathrm{2DCC}^2+ \Delta \theta_\mathrm{core-position}^2+\Delta \theta_\mathrm{core-ident}^2},
\end{equation}
where
\begin{itemize}[label=$\bullet$]
    \item $\Delta \theta_\mathrm{2DCC}$ is the error from 2D cross-correlation of optically thin jet regions at the two different frequencies. We aligned the images ten times --- slightly varying the jet features to be compared --- and obtained a statistical error associated to the mean image-shift from the scatter in the ten alignment attempts. Since the 2D cross-correlation software works in full pixel steps, the lower limit to the error is   $\mathrm{pixelsize}/2 = (\theta_\mathrm{beam,min}/40)$ where $\theta_\mathrm{beam,min}$ is the minor axis of beam. The errors are listed in Appendix~\ref{app:errors}, Table \ref{table:2D_err}.
   
    \item $\Delta \theta_\mathrm{core-position}$ is the uncertainty in the core position derived from Gaussian model-fitting directly to the interferometric visibilities (i.e., fitting in the $(u,v)$ plane). Following \citet{Lampton1976} we estimate the positional errors by moving the core component about its best-fit position on a grid with a small step size, fixing its position and letting all the other parameters of the model to vary freely when minimizing $\chi^2$. We then find an area on the grid around the best-fit position for which $\Delta \chi^2 = \chi^2 - \chi^2_\mathrm{min} < C_{p}^{\alpha}$, where $\chi^2_\mathrm{min}$ is the minimum $\chi^2$ corresponding to the best-fit position and $C_{p}^{\alpha}$ is the critical value of the $\chi^2$-distribution with $p$ degrees of freedom and $\alpha$ is the desired significance level. Following \citet{Lampton1976} $p$ is the number of free parameters in the Gaussian model after fixing the position of the core ($p=2$ for a single circular Gaussian). For the significance level we used $\alpha=0.32$, corresponding to 68\% (1-$\sigma$) confidence level. The errors are listed in Table \ref{table:core_pos_err1}.

    \item $\Delta\theta_\mathrm{core-ident}$ is the core identification error. This corresponds to the uncertainty in correctly locating the core in a source that has a jet  which  can  blend  with  the  core  at our  resolution. We estimate $\Delta\theta_\mathrm{core-ident}$ as the difference between the core position from the $(u,v)$ plane Gaussian model-fitting and the brightness peak position in a super-resolved image. The core identification errors are given in Table \ref{table:core_id_err1}.
  \end{itemize}

\subsection{Phase-referencing core-shifts}

We used phase-referenced CLEAN images (no self-calibration applied) to measure the radio core positions of the individual calibrators. The measured core positions are represented by the distances '$d_1$' and '$d_2$' (see Section~3) between the calibrator and the target at two different frequencies. Additionally, a small correction in the target's core position was applied due to a slight offset of the target's peak flux from the map centre (0,0) at each frequency. The model-fitting of the core was performed by identifying the brightest peak in the compact region. We fitted the peak with a single circular Gaussian component. We have also tried fitting an elliptical Gaussian component, which, however, appears to be more affected by the emission from the optically thin jet. This is indicated by the systematically larger fitted flux densities for the elliptical Gaussians than for the circular ones. 

Phase-referencing core-shift vectors, together with the absolute values, are listed in Table \ref{t3}. We found again that the different approaches pointed out in section~4.1 produce almost identical absolute core-shifts values. Note that to obtain the differences $d{_1}-\,d{_2}$ from Equation~5, the notation of the core-shift vectors have to be changed to XC, UX, KU with the coordinate sign changed.

\subsubsection{Phase-referencing error budget}

The astrometric accuracy of phase-referencing depends on random errors due to receiver noise and fast atmospheric phase fluctuations, as well as on various systematic effects that typically dominate. Since it is difficult to empirically estimate the systematic term, we build an error budget by theoretically evaluating each contributing factor. The sources of systematic errors include uncorrected residual delays due to (slowly varying) troposphere, ionosphere, source structure, instrumental effects, and errors in antenna and source positions. First-order corrections are made to most of these during the data calibration stage (see Section~2) and remaining systematics are further reduced by taking phase differences between the target and the calibrator and by taking position differences between frequency bands.

Switching between the target and a nearby calibrator effectively removes the residual instrumental delays (any that remain after clock corrections from geodetic blocks and manual phase calibration) and suppresses the atmospheric propagation delay errors proportionally to their separation on the sky \citep{ReidHonma2014,2017israbook}, 
\begin{equation} \label{eq:pr_err}
    \Delta \theta_\mathrm{pos} \simeq \theta_\mathrm{sep} \frac{|c\Delta \tau|}{B \sqrt{N}},
\end{equation}
where $\theta_\mathrm{sep}$ is the angular distance between sources in radians (see Table~\ref{t1}), $c$ is the speed of light, $\Delta\tau$ is the residual atmospheric delay error, $B$ is the (longest) projected baseline length, and $N$ is the number of stations contributing to the long baselines of the array --- in our case $N=5$ \citep{Orosz2017}. Furthermore, we analyze positional differences between frequency bands, which cancels all the non-dispersive delays that do not change on the time scale in which the frequency bands are switched (the switching times are 0\,min between C and X, $\sim13$\,min between X and U, and $\sim30$\,min between U and K). This removes systematic uncertainties due to antenna position errors and source position errors, and it further suppresses errors due to the slowly varying component of the troposphere. Finally, since III\,Zw\,2 is very compact at all the observed frequencies (see Figure~\ref{f2}), we ignore the errors due to structural delays of the phase calibrator.

\begin{table*}[ht]
\centering
\begin{threeparttable}
\caption{Core-shift vectors of III\,Zw\,2 with each calibrator.}
 \label{t4}
\begin{tabular}{c|cc|cc|cc|c}
\hline 
\multirow{2}{*}{\begin{tabular}[c]{@{}c@{}}Reference\\ source\end{tabular}} & \multicolumn{2}{c|}{CX} & \multicolumn{2}{c|}{XU} & \multicolumn{2}{c|}{UK} & \begin{tabular}[c]{@{}c@{}}CK \\ (upper limit)\end{tabular} \\ \cline{2-8} 
 & \begin{tabular}[c]{@{}c@{}}R.A.\\ (mas)\end{tabular} & \begin{tabular}[c]{@{}c@{}}Dec\\ (mas)\end{tabular} & \begin{tabular}[c]{@{}c@{}}R.A.\\ (mas)\end{tabular} & \begin{tabular}[c]{@{}c@{}}Dec\\ (mas)\end{tabular} & \begin{tabular}[c]{@{}c@{}}R.A.\\ (mas)\end{tabular} & \begin{tabular}[c]{@{}c@{}}Dec\\ (mas)\end{tabular} & \begin{tabular}[c]{@{}c@{}}Absolute value\\ (mas)\end{tabular} \\ \hline
 J0006 & 0.11 $\pm$ 0.07 & 0.02 $\pm$ 0.10 & 0.05 $\pm$ 0.03 & 0.00 $\pm$  0.05  & -0.03 $\pm$ 0.03  & 0.01 $\pm$ 0.04 & 0.14 $\pm$ 0.08 (< 0.22)\\
J0007 & 0.03 $\pm$ 0.05  & 0.15 $\pm$ 0.14  & 0.03 $\pm$ 0.04  & -0.13 $\pm$  0.06  & -0.05 $\pm$ 0.02  & -0.03 $\pm$ 0.03 & 0.00 $\pm$ 0.13 (< 0.13)   \\
J0008 & 0.05 $\pm$ 0.12  & 0.15 $\pm$ 0.10 & -0.06 $\pm$ 0.04 & 0.12 $\pm$ 0.06  & -0.04 $\pm$0.02  & -0.01 $\pm$ 0.04 & 0.27 $\pm$ 0.12 (< 0.39)  \\ \hline
\begin{tabular}[c]{@{}c@{}}Weighted \\ mean\end{tabular} &  0.06 $\pm$ 0.04 &  0.10 $\pm$ 0.07 & 0.01 $\pm$ 0.02  & 0.00 $\pm$ 0.03  & -0.04 $\pm$  0.02  & -0.01 $\pm$ 0.02 & 0.09 $\pm$ 0.07 (< 0.16) \\ \hline
\begin{tabular}[c]{@{}c@{}}Absolute \\ value of the  \\ weighted mean\end{tabular} &\multicolumn{2}{c|}{0.12 $\pm$ 0.06}  & \multicolumn{2}{c|}{ 0.01 $\pm$ 0.02} & \multicolumn{2}{c|}{0.04 $\pm$  0.02}  & 0.17 $\pm$ 0.07 (< 0.24)  \\ \hline
\begin{tabular}[c]{@{}c@{}} Projected \\ value of the  \\ weighted mean\end{tabular} &\multicolumn{2}{c|}{ 0.11 $\pm$ 0.07}  & \multicolumn{2}{c|}{0.00 $\pm$ 0.03} & \multicolumn{2}{c|}{-0.03 $\pm$ 0.03}  & 0.09 $\pm$ 0.08 (< 0.17)\tnote{$\dagger$}  \\ \hline
\end{tabular}
\begin{tablenotes}
\item Note: the absolute values for the CK core-shifts are calculated as $r = \sqrt{(CX+XU+UK)^2_{RA} + (CX+XU+UK)^2_{DEC}}$. The Table contains rounded values.
\item[$\dagger$] Value from Figure \ref{f5}b.
\end{tablenotes}
\end{threeparttable}
\end{table*}

\begin{figure*}[h]
\centering
    \subfigure[]
    {
\includegraphics[width=0.47\textwidth]{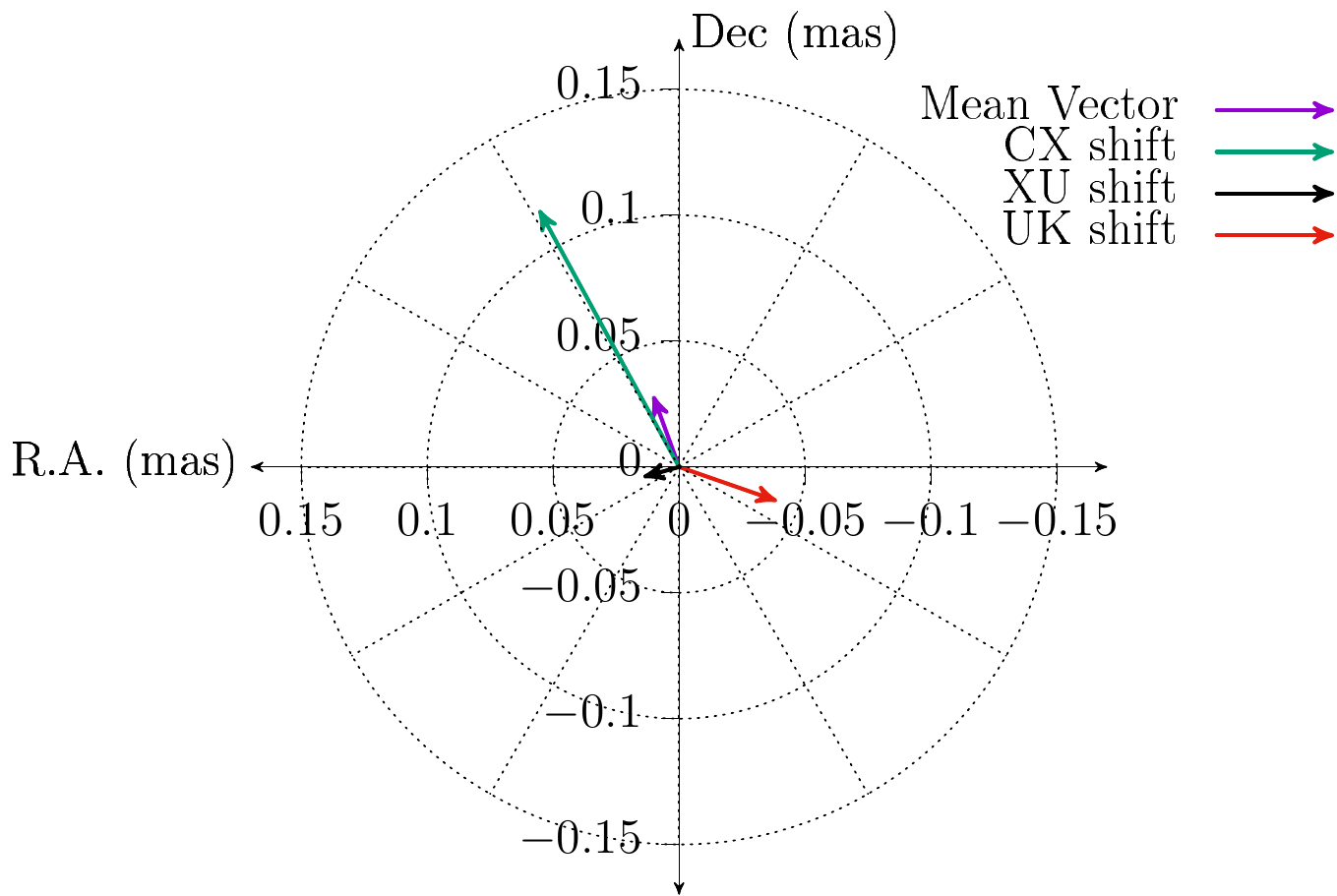}
        \label{}
    }
    \subfigure[]
    {
        \includegraphics[width=0.47\textwidth]{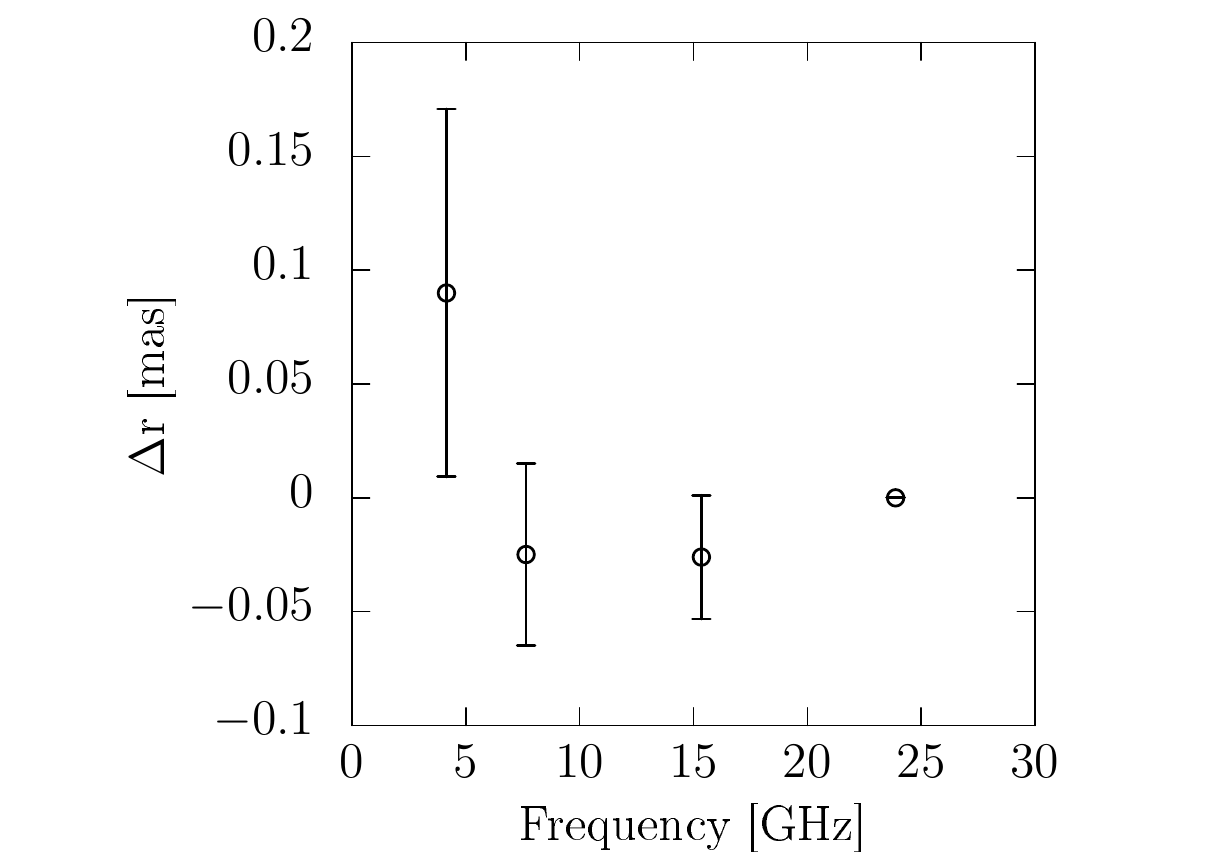}
        \label{}
    }
    \caption{(a) Weighted mean core-shift vectors and mean direction vector of III\,Zw\,2 combining all calibrators. (b) Projected core-shift values with K-band as the reference frequency using the weighted mean core-shift vectors of Table~\ref{t4}. The values relative to K-band are $0.089 \pm 0.081$\,mas for CK, $-0.025 \pm 0.040$\,mas for XK, and $-0.026 \pm 0.027$\,mas for UK.}
    \label{f5}
\end{figure*}

\begin{table*}[ht]
\centering
\begin{threeparttable}
\caption{Physical properties of III\,Zw\,2.}
 \label{t5}
\begin{tabular}{cccccccccc}
\hline\hline \noalign{\smallskip}\noalign{\smallskip}
\multicolumn{1}{c}{$z$} &
\multicolumn{1}{c}{\begin{tabular}[c]{@{}c@{}}$D_\mathrm{L}$\\ (Mpc)\end{tabular}}  &
\multicolumn{1}{c}{$\beta_\mathrm{app}$} 
& \multicolumn{1}{c}{$\delta$} & \multicolumn{1}{c}{$\Gamma$} &  \multicolumn{1}{c}{\begin{tabular}[c]{@{}c@{}} $\theta$  \\ (deg)\end{tabular}} &  \multicolumn{1}{c}{\begin{tabular}[c]{@{}c@{}}$\theta_\mathrm{j}$\\ (deg)\end{tabular}}  & 
\multicolumn{1}{c}{\begin{tabular}[c]{@{}c@{}} $M_\mathrm{BH}$\\ (M$_{\odot}$)\end{tabular}} & 
$a$ &
\multicolumn{1}{c}{\begin{tabular}[c]{@{}c@{}} $L_\mathrm{acc}$\\ (erg s$^{-1}$)\end{tabular}}
\\ 
(1) & (2) & (3) & (4) & (5) & (6) & (7) & (8) & (9) & (10) \\\noalign{\smallskip} \hline \noalign{\smallskip}
0.0898  & 405.5  & 
1.58 & 1.7 & 1.9 & 35 < $\theta$ < 55 & 6.7 < $\theta_j$ < 9.5 & 1.84 $\pm$ 0.27 $\times$ 10$^{8}$ & $\geq$ 0.98 & 8.1 $\times$ 10$^{44}$, 2.5 $\times$ 10$^{45}$   \\ \noalign{\smallskip}\hline
\end{tabular}
\begin{tablenotes}
\item Notes. Column definitions: (1)~redshift, \citet{Hernan2012},
(2)~luminosity distance from \citet{Wright2006},
(3)~apparent velocity from \cite{Lister2019},
(4)~Doppler factor from \cite{Hovatta2009},
(5)~Lorentz factor; $\Gamma=\frac{\beta^2_{app}+\delta^2+1}{2\delta}$,
(6)~viewing angle range from \cite{Brunthaler2000}, \cite{Hovatta2009} and \cite{Chamani2020},
(7)~intrinsic half-opening angle range inferred from \cite{Pushkarev2017},
(8)~black hole mass from \cite{Grier2012},
(9)~black hole spin from \cite{Chamani2020}, 
(10)~accretion disk luminosity values from \cite{Berton2015} and \cite{Falcke1995}, respectively.
\end{tablenotes}
\end{threeparttable}
\end{table*}

The remaining non-negligible terms in the positional error budget of the phase-referencing core-shift are:
\begin{equation}
    \Delta \theta_\mathrm{total,pos} = \sqrt{\Delta \theta_\mathrm{thermal}^2+ \Delta \theta_\mathrm{core-ident}^2 + \Delta \theta_\mathrm{trop}^2 + \Delta \theta_\mathrm{ion}^2 }, 
\end{equation}
where
\begin{itemize} [label=$\bullet$]
    \item $\Delta \theta_\mathrm{thermal}$ is the positional uncertainty due to random (thermal) errors evaluated as $\theta_\mathrm{beam}/(2 \cdot \mathrm{SNR})$, where $\theta_\mathrm{beam}$ is the beam size and $\mathrm{SNR}$ is the peak flux of the phase-referenced image divided by the rms image noise. Since the beam is nearly North-South-oriented, we use minor axis FHWM of the beam, $\theta_\mathrm{beam,min}$, as the beam size for R.A. and the major axis FWHM of the beam, $\theta_\mathrm{beam,max}$, for Dec. The thermal errors are listed in Appendix~\ref{app:errors}, Table~\ref{table:thermal_err}.
    
    \item $\Delta \theta_\mathrm{core-ident}$ is the core identification error. The evaluation of this error was done similarly as for the self-referencing, see previous subsection. The core identification errors are given in Table~\ref{table:core_id_err2}.

    \item $\Delta \theta_\mathrm{trop}$ is the positional uncertainty due to the tropospheric residual delays. These are discussed in detail below.
    
    \item $\Delta \theta_\mathrm{ion}$ is the positional uncertainty due to ionospheric residual delays. These are discussed in detail below.
\end{itemize}

Phase referencing suppresses the fast varying (wet) component of the troposphere and remaining errors due to this component are reflected in $\Delta \theta_\mathrm{thermal}$. Slowly varying  spatial tropospheric delay component is solved from the geodetic blocks and the remaining errors in phase referencing are due to uncertainties in the tropospheric delay gradient between the target and the calibrator after the geodetic block calibration. Since we are interested in positional differences between observing bands and since tropospheric delay is non-dispersive, only the time-variable part of the gradient error remains:

\begin{equation}
    \Delta \tau_\mathrm{trop}^{\nu_2-\nu_1}  
    \approx \Delta t  \frac{d\tau_\mathrm{zenith}}{dt} \sec(z)  \tan(z).
\end{equation}
Here $\Delta t$ is the time interval between the change of the observing bands, $\frac{d\tau_\mathrm{zenith}}{dt}$ is the time derivative of the residual zenith delay (after correction by geodetic blocks) and $z$ is the zenith angle. We (very) conservatively estimate that $\frac{d\tau_\mathrm{zenith}}{dt}$ is of the same order as the time-derivative of the residual zenith delay solved from the geodetic blocks. We have calculated the average $\Delta \tau_\mathrm{trop}^{\nu_2-\nu_1}$ for the telescopes participating in the longest baselines in E-W (MK-SC) and N-S (BR-SC) for elevations above 20 degrees. The resulting  errors from combining $\Delta \tau_\mathrm{trop}^{\nu_2-\nu_1}$ for the both telescopes of a baseline and using Equation~\ref{eq:pr_err} are given in Table~\ref{table:trop_err}.

As mentioned in Section~2, we corrected the dispersive ionospheric delays using global maps of ionospheric total electronic content (TEC) determined from GNSS observations. The analysis centers of the International GNSS Service produce these maps and make them publicly available. The global maps from the Jet Propulsion Laboratory (JPL) analysis center were used in our data processing to calibrate the ionospheric effects. The maps from JPL have a latitude/longitude resolution of 2.5/5.0\,degrees and a temporal resolution of 2.0\,hours, and they are treated as a thin spherical layer at the height of 450\,km. Both slant TEC values and their uncertainties can be obtained from the global TEC maps. In order to derive the slant TEC values for the actual VLBA observations from these maps, we calculated the pierce point at the single layer of the TEC maps for each individual observation. We then determined the vertical TEC at the pierce point at the observing time by interpolating from the TEC maps both in space and in time, and mapped the vertical TEC to the line of sight by using the modified single layer mapping function. The uncertainties for slant TEC values were obtained based on the error propagation. $\Delta \theta_\mathrm{ion}$ was derived for R.A. and Dec by considering slant TEC errors on the longest E-W and N-S baselines, MK-SC and BR-SC, respectively. We calculated the average slant TEC errors for scans with source elevation above 20 degrees and used the relation
\begin{equation} \label{eq:delta_ion}
    |c \Delta \tau_\mathrm{ion}| = 40.3 \Delta I_e \nu^{-2}, 
\end{equation}
where $\Delta I_e$ is the slant TEC uncertainty in TEC units (1\,TECU corresponds to an electron column density of $10^{16}$\,m$^{-2}$) and $\nu$ is the observing frequency \citep{ReidHonma2014}. The results combining Equations~\ref{eq:pr_err} and \ref{eq:delta_ion} are in Table~\ref{table:ion_err}.

\subsection{Core-shift of III\,Zw\,2}

To obtain the core-shift vectors of the target III\,Zw\,2, we use Equation~5, which requires both the self-referencing and phase-referencing core-shifts vectors. Combining the results from Table~\ref{t2} and Table~\ref{t3} we obtained the core-shift vectors of the target, listed in Table~\ref{t4}. Given that the errors are equal or even larger than the measured core-shift coordinates, we combined the measurements from all three calibrators and calculated their weighted mean core-shift vectors for each frequency pair. Furthermore, we also calculated the weighted mean core-shift vectors projection onto the mean direction vector, as illustrated in Figure~\ref{f5}a. 

The results show a disparity of the core-shift vector directions, and none of them points in the target's expected jet direction (towards West). A similar result was found for the projected core-shifts with individual calibrators, see Appendix~F, Figures~\ref{fF1}, \ref{fF2}, and \ref{fF3}. These results imply that the core-shift of III\,Zw\,2, is very small compared to our measurement accuracy which is evident in the vectors' random orientations (see Figure~\ref{f5}b).

It is clear that there is a larger uncertainty of the core-shift along the calibrators jet direction, which implies that the self-referencing core-shifts and/or calibrators core locations along the jet have systematic errors. Combining all three calibrators partly helps to mitigate these issues, since they have different jet directions. However, we note that there is an overall eastward bias in the jet directions of the calibrators, which may affect the combined core-shift.

As is evident from Table~\ref{t4} and Figure~\ref{f5}, we can only estimate an upper limit to the core-shift in III\,Zw\,2. We consider two approaches to calculate the upper limit: a) using the vector component perpendicular to the calibrator's jet direction. For that, we use J0007 for the R.A. component and J0006 for the Declination component (see Table~\ref{t4}). This gives a 1-$\sigma$ upper limit of $<0.15$\,mas ($0.03 \pm 0.12$\,mas). Approach b) consists of using the weighted mean or the projected value of the weighted mean core-shift between 4 and 24\,GHz in Table~\ref{t4}. These give 1-$\sigma$ upper limits of $<0.16$\,mas and $<0.17$\,mas, respectively. The projected weighted mean is not a very useful quantity when all the core-shift vectors point in different directions like in III\,Zw\,2. Therefore, we adopt the upper limit of $<0.16$\,mas for the jet magnetic field strength calculations.

\subsection{Estimation of Magnetic field parameters}

\subsubsection{Magnetic field strength from the core-shift}
Assuming that the magnetic and (radiating) particle energy densities are in equipartition in the jet of III\,Zw\,2, we can calculate $B_1$, the magnetic field strength in the jet\footnote{$B_1$ corresponds to $B$ at a distance of 1\,pc from the jet apex, but the power-law dependence of $B$ on the distance along the jet allows one to calculate $B$ at any distance as long as the jet can be described by the Blandford-K\"onigl model.} \citep{Lobanov1998}. The equipartition assumption together with jet's conical shape also implies that the core-shift index $k_r = 1$ and that the mean jet frame magnetic field is dominated by the azimuthal component on parsec scales. Furthermore, we adopt the flux freezing approximation, which states that the poloidal magnetic flux threading the parsec-scale jet is equal to the poloidal flux threading the black hole. Following \citet{Zamaninasab2014} we can then estimate an upper limit to the poloidal magnetic flux threading the horizon of the central black hole in III\,Zw\,2. In the following, we described briefly the observed parameters and relationships needed to estimate the jet magnetic field strength at 1\,pc as well as the jet magnetic flux. 

Table~\ref{t5} presents the physical parameters of III\,Zw\,2 collected from the literature (see the references below the table). The parameters include the redshift $z$ and luminosity distance $D_\mathrm{L}$, the apparent velocity, $\beta_\mathrm{app}$, the Doppler factor $\delta$, the bulk Lorentz factor $\Gamma$, the viewing angle $\theta$, the half-opening angle $\theta_\mathrm{j}$ of the jet, the black hole mass $M_\mathrm{BH}$, the spin $a$ and the accretion luminosity $L_\mathrm{acc}$. To estimate the range of values for $\theta_\mathrm{j}$, we first took the apparent full opening angle of 23.2\,degrees measured by \cite{Pushkarev2017}. The intrinsic half-opening angle was calculated as $\tan(\theta_\mathrm{j}) = \tan(23.2^\circ/2) \sin \theta$ \citep{Pushkarev2017}. This results in a range of values of $6.7^{\circ} < \theta_\mathrm{j} < 9.5^{\circ}$ and $0.22 < \Gamma \theta_\mathrm{j} < 0.32$ since $35^{\circ} < \theta < 55^{\circ}$. $\Gamma \theta_\mathrm{j}$ agrees well with the typical value derived by \cite{ClausenB2013}.

Following \cite{Lobanov1998}, \cite{Hirotani2005}, \cite{Zamaninasab2014} and the corrections pointed out by \citet{Zdziarski2015}, the core's offset $\Omega_{r\nu}$ and the magnetic field strength in equipartition $B^\mathrm{eq}_\mathrm{1pc}$ at 1\,pc from the apex of the jet are given by the following relations:
\begin{equation}
    \Omega_{r\nu}=4.85\cdot 10^{-9} \frac{\Delta r_\mathrm{mas} D_\mathrm{L}}{(1+z)^2} \frac{\nu^{1/k_r}_{1}\nu^{1/k_r}_{2}}{\nu^{1/k_r}_{2}-\nu^{1/k_r}_{1}}\;[\mathrm{pc\,GHz^{1/k_r}}],
\end{equation}
where $\nu_{1}$ and $\nu_{2}$ are the observed frequencies (in GHz) with $\nu_{2}$ > $\nu_{1}$, $D_\mathrm{L}$ is in pc and $\Delta r_\mathrm{mas}$ is the core-shift in mas between two frequencies $\nu_1$ and $\nu_2$. We use the same cosmology as \citet{Pushkarev2012} with $H_0 = 71$\,km\,s$^{-1}$\,Mpc$^{-1}$, $\Omega_m = 0.27$ and  $\Omega_{\Lambda} = 0.73$.
The magnetic field strength is then given by
\begin{equation} \label{eq:Beq}
    B^\mathrm{eq}_\mathrm{1pc} \approx 0.025 \left [\frac{\sigma_\mathrm{rel}\,\Omega_{r\nu}^{3k_r}\,(1+z)^3} {\delta^2\, \theta_\mathrm{j}\, \sin^{3k_{r}-1}\theta} \right ]^{\frac{1}{4}}\;[\mathrm{G}],
\end{equation}
where $\sigma_\mathrm{rel}$ is the ratio of magnetic and particle energy densities, which we assume to be unity. The above formula also assumes implicitly that  the electron energy distribution has a power-law index of $p=2$, which corresponds to an optically thin spectral index of $\alpha=-0.5$ and that  $\gamma_\mathrm{max}/\gamma_\mathrm{min} = 10^{4.34}$. Setting $k_r = 1$ (conical jet and equipartition) and inserting the core-shift upper limit of 0.16\,mas with the range of values given for $\theta$ and $\theta_\mathrm{j}$, we calculated upper limits to $\Omega_{r\nu}$ and $B^\mathrm{eq}_\mathrm{1pc}$, see Table~\ref{t6}. The resulting mean jet frame magnetic field strength is below 60\,mG.

\begin{figure}[t]
\centering
      \includegraphics[width=0.5\textwidth]{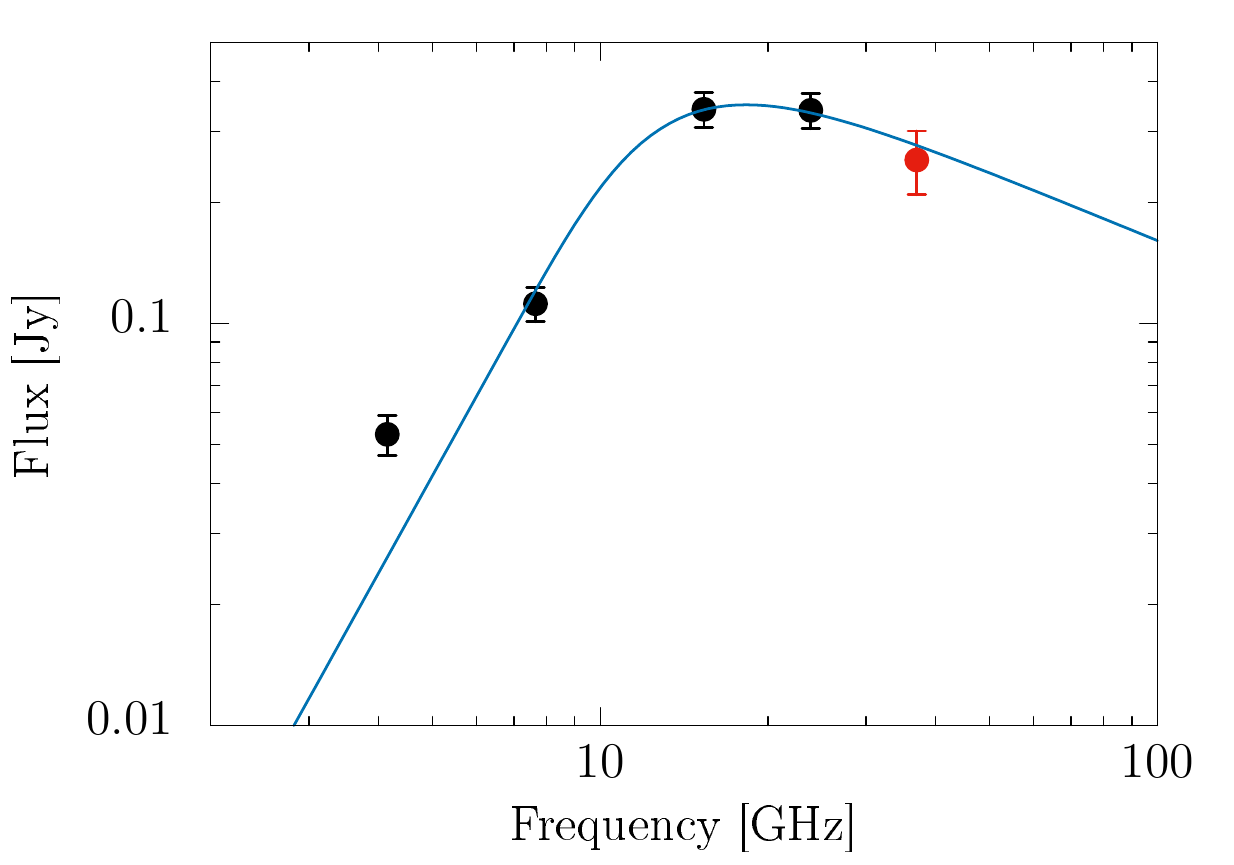}
        \caption{Correlated flux density of III\,Zw\,2 as a function of frequency between 4 and 24\,GHz on November 8, 2017 (black points). The spectrum is clearly inverted between 4 and 15\,GHz. The red point represents the total flux density at 37\,GHz measured by the Mets\"ahovi Radio Observatory quasar monitoring program. The blue curve shows the best-fit self-absorbed synchrotron spectrum with a fixed optically thick spectral index of $+2.5$. See text for more details.}
    \label{f6}
\end{figure}

On the other hand, the magnetic field strength at 1\,pc can be also calculated without assuming equipartition if we have a measurement of the core flux density \citep{Zdziarski2015}:
\begin{equation} \label{eq:Bnoeq}
    B^\mathrm{noeq}_\mathrm{1pc} \simeq \frac{3.35 \times 10^{-11}\, D_\mathrm{L}\, \Delta r^5_\mathrm{mas}\, \delta\, \tan ^2 \theta_\mathrm{j} }{(\nu^{-1}_1-\nu^{-1}_2)^5\, [(1+z) \sin \theta]^3\, F_{\nu}^2} \; [\mathrm{G}],
\end{equation}
where $F_{\nu}$ (in Jy) is the flux density in the flat part of the spectrum, $F_{\nu} \propto \nu^0$. For III\,Zw\,2 the flat part of the spectrum lies between 15 and 24\,GHz (see Figure~\ref{f6}) with a flux density of 0.34\,Jy. Using the upper limit of the core-shift between 4 and 24\,GHz, the upper limit of $B^\mathrm{noeq}_\mathrm{1pc}$ is 4\,mG. This is more than an order-of-magnitude smaller than the upper limit to $B^\mathrm{eq}_\mathrm{1pc}$. Dividing Equation~\ref{eq:Bnoeq} by Equation~\ref{eq:Beq} we see that the ratio $B^\mathrm{noeq}_\mathrm{1pc}/B^\mathrm{eq}_\mathrm{1pc} \propto \Delta r^{17/4}$ for $k_r = 1$. Therefore, the discrepancy between $B^\mathrm{noeq}_\mathrm{1pc}$ and $B^\mathrm{eq}_\mathrm{1pc}$ only increases if the true core-shift value is less than our upper limit. This rises a possibility of a departure from the equipartition conditions in the jet of III\,Zw\,2. However, we would need to have an actual measurement --- instead of an upper limit --- of the core-shift in III\,Zw\,2 to confirm this. 


\subsubsection{Magnetic field strength estimated from the synchrotron spectrum}

Core-shift is not the only method to estimate the jet magnetic field strength. Since the measured radio spectrum of III\,Zw\,2 in Figure~\ref{f6} shows a clear turnover that can be fitted with a self-absorbed synchrotron spectrum, we can also constrain the jet magnetic field strength by measuring the turnover frequency, $\nu_\mathrm{m}$, the maximum flux density at the turnover, $F_\mathrm{m}$, and the size of the emitting region at the turnover frequency, $a(\nu_\mathrm{m})$. We fit the spectrum in Figure~\ref{f6} with a function describing a self-absorbed synchrotron spectrum emitted by electrons with a power-law energy distribution $N(\gamma) = N_0 \gamma^{-p}$ in a homogeneous magnetic field \citep[see e.g.,][]{Pacholczyk1970}:
\begin{equation} \label{eq:synch_spec}
    F_\mathrm{\nu}(\nu) = F_\mathrm{m} \Big( \frac{\nu}{\nu_\mathrm{m}} \Big)^{\alpha_\mathrm{thick}} \frac{1-\exp (-\tau_\mathrm{m}(\nu/\nu_\mathrm{m})^{\alpha-\alpha_\mathrm{thick}})}{1-\exp (-\tau_\mathrm{m})}, 
\end{equation}
where $\tau_\mathrm{m}$ is the optical depth at the turnover, $\alpha = -(p-1)/2$ is the optically thin spectral index, and $\alpha_\mathrm{thick}$ is the spectral index of the optically thick part of the spectrum. Following \citet{Turler1999}, $\tau_\mathrm{m}$ can be approximated as
\begin{equation} \label{eq:tau}
 \tau_\mathrm{m} \approx \frac{3}{2} \bigg( \sqrt{1-\frac{8\alpha}{3\alpha_\mathrm{thick}}} - 1 \bigg)
\end{equation}
We fix the optically thick spectral index to $\alpha_\mathrm{thick}=+2.5$ in order to approximate a homogeneous emission region. We discuss the consequences of this assumption later.

The best-fitting parameters are $F_\mathrm{m}=0.349\pm0.030$\,Jy, $\nu_\mathrm{m}=18.3\pm1.9$\,GHz, and $\alpha=-0.57\pm0.25$. Errors were estimated by fitting 5000 realisations of the spectrum; in each round a set of flux densities were drawn from a set of Gaussian distributions with means corresponding to the measured flux densities and the standard deviations corresponding to the flux density measurement uncertainties at each frequency. We complemented the VLBA flux density measurements with 37\,GHz single-dish data from Mets\"ahovi Radio Observatory monitoring program \citep{Chamani2020,Terasranta1998}. Mets\"ahovi observations were made on November 2 and November 16, 2017 and we interpolated the measured flux densities to the date of the VLBA observations, November 8, 2017. Combining data from different angular scales is in this case justified by the point-like core of III\,Zw\,2 and the steep spectrum of its weak kpc-scale emission \citep[the expected contribution of the extended emission is less than 1\,mJy at 37\,GHz;][]{Brunthaler2005}.

Following \citet{Marscher1983}, the synchrotron self-absorption magnetic field of a homogeneous, spherical source is 
\begin{equation} \label{eq:B_SSA}
    B_\mathrm{SSA} = 10^{-5} b(\alpha) a(\nu_\mathrm{m})^4 \nu_\mathrm{m}^5 F_\mathrm{m}^{'-2} \frac{\delta}{1+z} \hspace{0.2cm} [\mathrm{G}],
\end{equation}
where $a(\nu_\mathrm{m})$ is in mas, $\nu_\mathrm{m}$ is in GHz, $b(\alpha)$ is tabulated in \citet{Marscher1983}, and $F_\mathrm{m}^{'}$ is the flux density at $\nu_\mathrm{m}$ in Jy from a linear extrapolation of the optically thin spectral slope. The latter can be expressed in terms of the fitted $F_\mathrm{m}$ and $\tau_\mathrm{m}$ as 
\begin{equation} \label{eq:Fprime}
    F_\mathrm{m}^{'} = \frac{F_\mathrm{m} \tau_\mathrm{m}}{1-e^{-\tau_\mathrm{m}}}.
\end{equation}
In order to use Equation~\ref{eq:B_SSA} to calculate the magnetic field strength, we need to resolve the emission region at or near the turnover frequency. However, fitting an elliptical Gaussian to the visibility data results in the minor axis of the Gaussian to go to zero both at 15 and 24\,GHz. The major axes of these components are oriented at PA of $-50^\circ$ and $-56^\circ$, respectively. These are relatively close to the orientation of the previous component ejections \citep{Brunthaler2005,Lister2019} and we conclude that at our resolution we can only (barely) resolve the jet in the longitudinal direction --- not in the transverse direction. Therefore, any estimate of $B_\mathrm{SSA}$ is only an upper limit, like our magnetic field estimates from the core-shift analysis.

To estimate an upper limit to $B_\mathrm{SSA}$ we fitted both 15 and 24\,GHz visibility data with a circular Gaussian and obtained FWHM sizes of $0.066 \pm 0.003$\,mas and $0.045 \pm 0.003$\,mas, respectively. The uncertainties are estimated following \citet{Lampton1976} like in Section~4.1. These size measurements follow the expected $a \propto r \propto \nu^{-1}$ dependence and we interpolate the Gaussian FWHM size at $\nu_\mathrm{m}$ using this relation. To approximate the size of a partially opaque spherical source --- as assumed in Equation~\ref{eq:B_SSA} --- from a Gaussian FWHM, we approximate $a(\nu_\mathrm{m}) \approx 1.6 \times \mathrm{FWHM}$. This corresponds to the size of an optically thick disk for which the visibility amplitude drops to 50\% at the same baseline length as for a Gaussian with the given FWHM. 

Finally, we get $B_\mathrm{SSA} = 42 \pm 14$\,mG. Since this is an upper limit, we can say that $B_\mathrm{SSA} \lesssim 60$\,mG (1-$\sigma$ upper limit) at the location of the core at $\sim 18$\,GHz. To compare this with $B_\mathrm{1pc}$, we need to estimate an upper limit to the distance of the core at 18\,GHz from the jet apex. Following \citet{Lobanov2005} we can estimate that the minimum resolvable size of a Gaussian for our 24\,GHz data, which has on-source SNR of 217, is 0.03\,mas. With this upper limit to the transverse width of the jet at 24\,GHz and by assuming a conical jet, we can estimate an upper limit to the distance of 18\,GHz core from the jet apex, $r(18\,\mathrm{GHz}) \lesssim 0.4$\,pc. Therefore, we can put a 1-$\sigma$ upper limit to $B_\mathrm{1pc} \lesssim 20$\,mG from our $B_\mathrm{SSA}$ upper limit. This is about a factor of three smaller than the upper limit from the core-shift analysis. 

If left as a free parameter in the spectral fit, $\alpha_\mathrm{thick} = 1.58 \pm 0.18$, which indicates that the emission region has a non-uniform structure in a sense that the magnetic field and the relativistic particle density have gradients inside it. This is what we also expect from the Blandford-K\"onigl jet model and assume in the core-shift analysis. Therefore, the assumption of a homogeneous self-absorbed synchrotron source does not, strictly speaking, hold. However, as shown in \citet{Marscher1977}, the difference in derived magnetic field strength is modest when $p=2$ and gradients are not overly steep. The ratio $B_\mathrm{non-uniform}/B_\mathrm{uniform} = 0.68$ for $m=1$, $n=3$ and 1.8 for $m=2$, $n=2$ (see Section~3.1. for the definition of the indices $m$ and $n$). This does not change our result --- the upper limit to $B_\mathrm{1pc}$ estimated from the turnover in the synchrotron spectrum is comparable to or smaller than the upper limit estimated from the core-shift limit in equipartition case.

\begin{table}[t!]
\centering
\begin{threeparttable}
\caption{Parameters for the magnetic field of III\,Zw\,2, using the core-shift upper limit of 0.16\,mas; and the synchrotron self-absorption magnetic field strength, $B_{SSA}$.}
\label{t6}
\begin{tabular}{lcc}
\hline \noalign{\smallskip}
\multicolumn{1}{c}{Parameter} & \multicolumn{1}{c}{Value} & \multicolumn{1}{c}{Units} \\ \hline \hline \noalign{\smallskip}
$\Omega_{r\nu}$   & $\leqslant 1.3$     & pc GHz   \\\noalign{\smallskip}
$B^\mathrm{eq}_\mathrm{1pc}$   &  $\leqslant  60$     &  mG   \\\noalign{\smallskip}
$B^\mathrm{noeq}_\mathrm{1pc}$  & $\leqslant  4 $     &  mG   \\ \noalign{\smallskip}
$B_\mathrm{1pc,SSA}$  & $\leqslant  20 $ &  mG   \\ \noalign{\smallskip}
$\Phi^\mathrm{eq}_\mathrm{jet}$ &  $\leqslant 10^{32}$ & G cm$^2$ \\ \noalign{\smallskip}
$\Phi^\mathrm{noeq}_\mathrm{jet}$ &  $\leqslant 7 \times 10^{30}$ & G cm$^2$ \\ \noalign{\smallskip}
$\Phi_\mathrm{jet,SSA}$ &  $\leqslant 3.6 \times 10^{31}$ & G cm$^2$ \\ \noalign{\smallskip}
$\Phi_\mathrm{BH,MAD}$\tnote{(a)}  & $\sim 5 \times 10^{32}$ & G cm$^2$  \\ \noalign{\smallskip}
$\Phi_\mathrm{BH,MAD}$\tnote{(b)}  & $(7-9) \times 10^{32}$ & G cm$^2$  \\  \noalign{\smallskip}
$\Phi_\mathrm{BH,MAD}$\tnote{(c)}  &  $1.1 \times 10^{33}$ & G cm$^2$  \\ \hline \noalign{\smallskip} 
\end{tabular}
\begin{tablenotes}
\item[(a)] With the average value of L$_\mathrm{acc}$ (values taken from Table~\ref{t5}) and adopting $\eta=0.4$.
\item[(b)] With the average value of L$_\mathrm{acc}$ and adopting $\eta=0.13-0.19$ from \cite{DavisLaor2011}. 
\item[(c)] Using the L$_\mathrm{acc}$ value from \citet{Falcke1995} and adopting $\eta=0.13$ from \cite{DavisLaor2011}. 
\end{tablenotes}
\end{threeparttable}
\end{table}

\begin{figure*}[t]
\centering
      \includegraphics[width=0.7\textwidth]{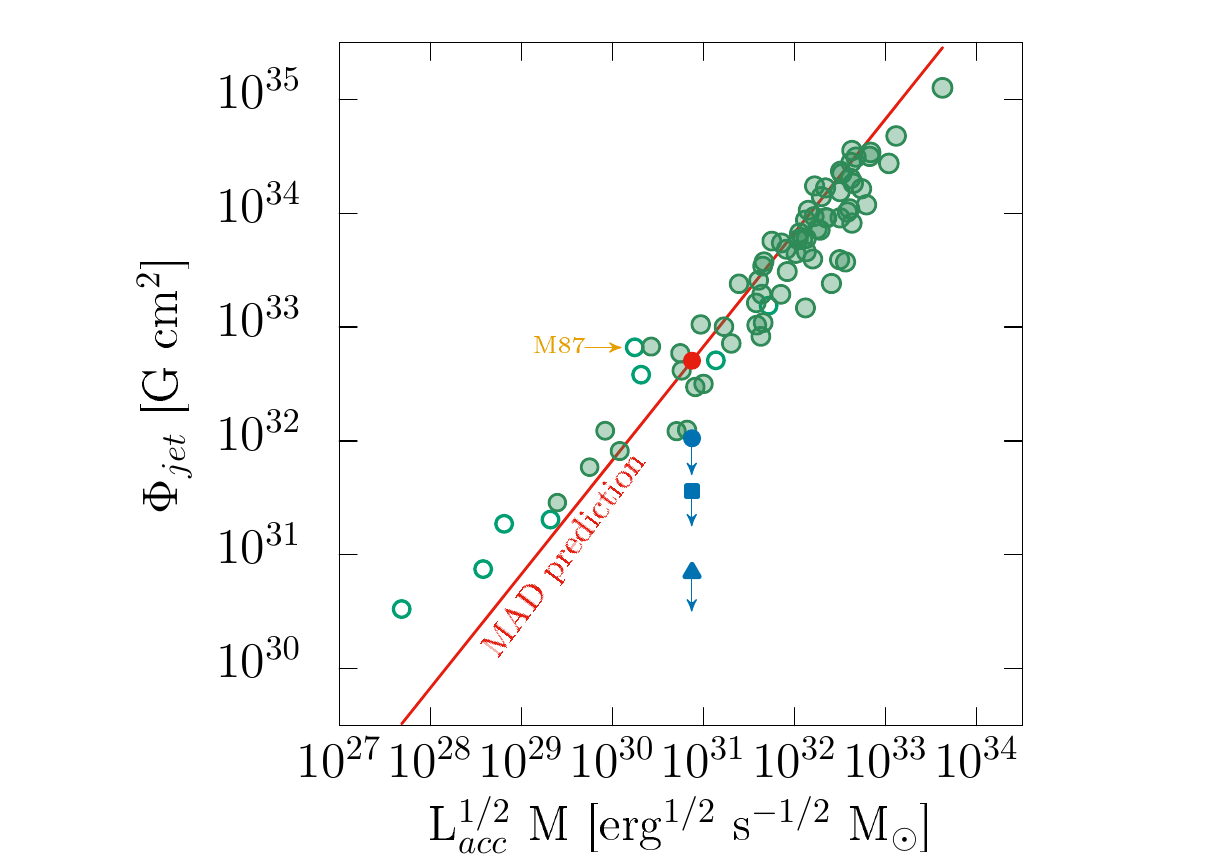}
        \caption{Plot adapted from \cite{Zamaninasab2014}. In their work, they assumed $\Gamma \theta_\mathrm{j} = 1$ for the whole sample. The open and filled green circles represent our corrected values with $\Gamma \theta_\mathrm{j} \neq 1$ for radio galaxies and $\Gamma \theta_\mathrm{j} = 0.13$ for blazars, respectively. The red and blue filled-circles represent the MAD predicted and measured (upper limit in equipartition) values of the jet magnetic flux of III\,Zw\,2. The blue triangle represents the measured upper limit without equipartition assumption. The blue square represents the measured upper limit from the synchrotron self-absorption magnetic field $B_\mathrm{SSA}$.}
    \label{f7}
\end{figure*}
\subsubsection{Limits to the jet magnetic flux}

In order to calculate the jet magnetic flux, $\Phi_\mathrm{jet}$, we follow again \citet{Zdziarski2015}, who modified the \citep{Zamaninasab2014} analysis to self-consistently include the observed condition $\Gamma\theta_j \sim 0.1$ and that the observed synchrotron self-absorption magnetic field corresponds to the transverse-average azimuthal magnetic field. From \citet{Zdziarski2015},
\begin{equation}
    \Phi_\mathrm{jet}=\frac{2^{3/2}\,\pi\, r_\mathrm{H}\, s\, h\, B \, (1+\sigma)^{1/2}}{l\,a},
\end{equation}
where the jet magnetization parameter (ratio of Poynting flux to kinetic energy flux) $\sigma = (\Gamma \theta_\mathrm{j}/s)^2$, $s \lesssim 1$ \citep{Komissarov2009}, $r_\mathrm{H}$ represents the black hole event horizon radius, $r_\mathrm{H} = r_{g} \left(1+(1-a^2)^{1/2} \right)$, $r_{g} = G M_\mathrm{BH} / c^2$ is the black hole gravitational radius, $h$ is the distance along the jet in parsecs, $l$ is the ratio of the angular frequency of the field lines to the BH angular frequency, and $a$ the black hole spin. Setting $l=0.5$, $s=1$, and $h=1$\,pc, the jet magnetic flux is
\begin{equation}
     \Phi_\mathrm{jet} = 8 \times 10^{33} f(a_*)\, [1+\sigma]^{1/2} \left[\frac{M_\mathrm{BH}}{10^{9} \mathrm{M}_{\odot}}  \right ] \left [\frac{B}{\mathrm{G}} \right ]\, \hspace{0.2cm} [\mathrm{G \,cm^{2}}],
\end{equation}
where
\begin{equation}
    f(a)=\frac{1}a\frac{r_{H}}{r_{g}}=\frac{1+(1-a^2)^{1/2}}{a}.
\end{equation}
For $a=1$, $f(a) = 1$ \citep{Zamaninasab2014}, and for  III\,Zw\,2 with $a \ge 0.98$, $1 \le f(a) \le 1.2$.  We calculated the jet magnetic flux using the core-shift magnetic field strength both with and without equipartition assumption as well as using the SSA magnetic field strength. The upper limit to the magnetic flux in the equipartition case is $\Phi^\mathrm{eq}_\mathrm{jet} \lesssim 10^{32}$\,G\,cm$^2$, while the upper limit of the magnetic flux in the non-equipartition case, $\Phi^\mathrm{noeq}_\mathrm{jet}$, is lower by more than an order of magnitude. The upper limit to the magnetic flux estimated from the SSA spectrum is $\Phi_\mathrm{jet,SSA} \lesssim 4 \times 10^{31}$\,G\,cm$^2$. See Table~\ref{t6}.

Recalling the relation for the predicted magnetic flux threading the black hole (which by flux freezing should be equal $\Phi_\mathrm{jet}$) in the MAD state \citep{Narayan2003, Tchekhovskoy2011},
\begin{equation}
    \Phi_\mathrm{BH, MAD} = 50 \left (\dot{M} r_g^2 c  \right )^{1/2},
\end{equation}
which is equivalent to
\begin{align}
    \Phi_\mathrm{BH, MAD} & = 2.4 \times 10^{34} \left [\frac{\eta}{0.4}  \right ]^{-1/2} \left [ \frac{M_\mathrm{BH}}{10^9\,\mathrm{M}_{\odot}} \right ]\nonumber \\& \times \left [\frac{L_\mathrm{acc}}{1.26\times 10^{47}\mathrm{\,erg\,s}^{-1}}  \right ]^{1/2} \,[\mathrm{G\,cm^2}],
\end{align}
where $\eta$ is the radiative efficiency of the accretion disk. We note that the higher $L_\mathrm{acc}$ for III\,Zw\,2 from \citet{Falcke1995} is the luminosity of the UV-bump derived by spectral fitting, while the lower $L_\mathrm{acc}$ from \citet{Berton2015} is derived by a scaling relation from the luminosity of the H$\beta$ line (see Table~\ref{t5}).

By applying Equation~22 and assuming $\eta=0.4$ as in \cite{Zamaninasab2014}, we find that the upper limit of $\Phi^\mathrm{eq}_\mathrm{jet}$ for III\,Zw\,2 is smaller by a factor of five with respect to the predicted $\Phi_\mathrm{BH, MAD}$ value (see $\Phi^\mathrm{(a)}_\mathrm{BH, MAD}$ in Table~\ref{t6}). Since the accretion in the MAD state proceeds via interchange instability, it is very difficult to know the actual radiative efficiency of a MAD. However, based on the discussion in \citet{Zdziarski2015} and on the relatively low Eddington ratio of III\,Zw\,2 \citep[0.04$-$0.1; ][]{Berton2015, Inoue2007}, we argue here that the radiative efficiency is likely smaller than 0.4 for III\,Zw\,2. For instance, \cite{DavisLaor2011} derived the radiative efficiency for a sample of PG quasars (including III\,Zw\,2) by employing optical spectroscopy measurements assuming a standard thin accretion disk that gives $\eta=0.19$, and from BH mass and stellar velocity dispersion relationship, resulting in $\eta=0.13$. By adopting these efficiencies --- although not directly applicable in a MAD case --- the upper limit to the jet magnetic flux is smaller than $\Phi_\mathrm{BH, MAD}$ by a factor of seven and nine respectively; for comparisons see $\Phi^\mathrm{(b)}_\mathrm{BH, MAD}$ in Table~\ref{t6}. Since the $L_\mathrm{acc}$ from \citet{Falcke1995} is likely to be more accurate, we also calculate $\Phi_\mathrm{BH, MAD}$ using that value only (see $\Phi^\mathrm{(c)}_\mathrm{BH, MAD}$ in Table~\ref{t6}). This results in more than an order of magnitude higher value than our upper limit to $\Phi^\mathrm{eq}_\mathrm{jet}$.

\section{Summary and discussion}

Active Galactic Nuclei show a wide range of radio-loudness that can be considered as a proxy for their jet production efficiency. It is still unclear which mechanisms are responsible for this variety even among sources with otherwise similar properties. One explanation is that the spread in the BH spin alone determines the range of observed radio-loudness. Another possible controlling parameter of the radio-loudness is the magnetic flux threading the BH. In this so-called magnetic flux paradigm, it is the accretion history of the AGN that determines how much magnetic flux has accumulated in the inner parts of the accretion disk and only those sources that have developed a magnetically arrested disk can have high jet production efficiency and consequently radio-loudness \citep{Sikora2013, Xie2019}. The latter model was successfully tested with a powerful radio-loud AGN sample that showed jet magnetic flux values consistent with the prediction of the MAD scenario \citep{Zamaninasab2014}. However, the model has not yet been tested with low radio-loudness sources, which would be expected to show lower jet magnetic fluxes --- well below the MAD level --- for similar BH spins. 

In this work, we tested the magnetic flux paradigm with the radio intermediate quasar III\,Zw\,2. First, we analyzed multi-frequency phase-referencing VLBA data and obtained an upper limit for the core-shift. The latter allowed us to put upper limits to the jet magnetic field strength and the magnetic flux both with and without assuming equipartition conditions in the jet. We found that III\,Zw\,2 has not reached the MAD level magnetic flux in either scenario. By assuming equipartition, the magnetic flux is less than 20\% of the MAD level. The failure to reach the MAD state is reinforced under the non-equipartition scenario where the magnetic flux is below the MAD limit by more than an order of magnitude. Another supporting result comes from the fitting of the synchrotron spectrum which leads to an upper limit of the magnetic flux below the MAD limit by one order of magnitude. The values are displayed in Figure~\ref{f7} where the observed and the predicted MAD magnetic fluxes are plotted in the figure adapted from \citet{Zamaninasab2014}\footnote{
We have corrected $B'_\mathrm{1pc}$ in their whole sample by multiplying it by a factor of $(1+z)^{1/4}$ which is missing in their original work \citep[see][]{Zdziarski2015}. The magnetic fluxes are estimated with Equation~19, and they are slightly lower than the  values presented in \citet{Zamaninasab2014} due to the difference in the value of $\Gamma\theta_j$.} These results can naturally explain the relatively low jet production efficiency in this RIQ despite it hosting a fast spinning SMBH. The results are in good agreement with the magnetic flux paradigm.

Another way to look at this is the notion by \citet{Nalewajko2014} that the $\Phi_\mathrm{jet} \sim \Phi_\mathrm{BH,MAD}$ condition expected in the MAD case is equivalent to the condition that the jet magnetic power, $L_B = 7.14 \times 10^{46} (\Gamma \theta_\mathrm{j})^2 B'^2_\mathrm{1pc}$\,[erg s$^{-1}$], is comparable to the accretion disk luminosity, $L_\mathrm{acc}$. In the case of III\,Zw\,2,  $L_B \lesssim 10^{43}$\,erg\,s$^{-1}$ (for $\theta_\mathrm{j} = 6.7^{\circ}$), which is two orders of magnitude lower than the mean disk luminosity of $1.7 \times 10^{45}$\,erg\,s$^{-1}$. This calculation consistently shows that the source has not reached the MAD state, which, however, is not surprising considering that it uses the same $B'_\mathrm{1pc}$, $\Gamma \theta_\mathrm{j}$ and $L_\mathrm{acc}$ as the magnetic flux calculations. 

The order-of-magnitude discrepancy between the upper limits to $B^\mathrm{noeq}_\mathrm{1pc}$ and $B^\mathrm{eq}_\mathrm{1pc}$ requires some attention. Since the ratio $B^\mathrm{eq}_\mathrm{1pc}/B^\mathrm{noeq}_\mathrm{1pc}$ scales as $\Delta r^{-17/4}$, the discrepancy only becomes larger if the true core-shift value is less than our upper limit. Therefore, we have a lower limit to the ratio $B^\mathrm{eq}_\mathrm{1pc}/B^\mathrm{noeq}_\mathrm{1pc}$, which would in principle indicate that the VLBI core in III\,Zw\,2 is particle energy dominated by quite a large factor of $\gtrsim 5 \times 10^4$. However, we note that B$^\mathrm{noeq}_\mathrm{1pc}$ is very sensitive to the measured quantities. 
Confirming the possible deviation from the equipartition will therefore require an accurate measurement of the core-shift instead of the upper limit that we have now.

Magnetic flux paradigm appears to explain the RIQ-level jet production efficiency in III\,Zw\,2, but what is the reason for low accumulated magnetic flux in this source? Below we discuss some possibilities:
\begin{itemize}
    \item \citet{Sikora2013} suggest that RIQ sources could be linked to a stochastic jet production triggered by magnetic field fluctuations due to turbulence in the inner hot or geometrically thick region of the accretion flow. If the fluctuations have large enough spatial coherence length and strength, they could deposit enough magnetic flux to the BH to produce an intermittent jet. There is some evidence that the jet in III\,Zw\,2 could be of intermittent nature. First, the source exhibits highly fluctuating radio flux density at 37\,GHz \citep[more than a factor of 10 peak-to-through ratio;][]{Chamani2020}, which could be due to variable jet power. It undergoes a recurrent activity roughly every five years \citep{Li2010} with a detected variable gamma-ray emission in the short term \citep{Liao2016}. Secondly, the radio morphology is point-like most of the time, but occasionally a blob is ejected \citep{Brunthaler2005,Lister2019}. 
   
    \item  Early morphology studies indicated that III\,Zw\,2 is hosted in a spiral galaxy \citep{Hutchings1983, Taylor1996}. If the host galaxy of III\,Zw\,2 indeed is a disk, this may be connected to the failure of the development of a powerful jet. After all, radio-loud AGN are typically associated with elliptical galaxies --- although there are exceptions, such as jetted Narrow Line Seyfert~1 galaxies \citep[e.g.,][]{Lahteenmaki2018}. \citet{Blandford2019} speculate that the mostly equatorial large-scale gas inflow towards the central region in spiral galaxies could lead to difficulties in accumulating and trapping magnetic flux within the BH radius of influence. In such an equatorial flow, field lines become radial on the surface of the disk and may easily dissipate through reconnection. However, the gas inflow in spiral galaxies may not stay equatorial from the kiloparsec scales all the way down to the BH given the complex nuclear structures \citep[see e.g.,][]{Hopkins2010, Hopkins2011, Combes2014, Chamani2017, Pjanka2017}. It is therefore unclear if a difficulty in transporting large-scale magnetic fields from the galactic scale to the central $\lesssim$\,kpc is enough to hinder the build-up of the magnetic flux in the BH sphere of influence.   
    

    \item Later observations with the Hubble Space Telescope reported by \citet{Veilleux2009} suggest that III\,Zw\,2 resides in an elliptical galaxy instead. In such a case, magnetic flux could have been supplied from the large scales in a short hot accretion phase, which --- for some reason --- has not been long enough to build up enough magnetic flux to create a magnetically arrested disk --- or perhaps the accreted field was too disordered. As long as the host galaxy morphology is unclear, it is difficult to address these questions. Interestingly, it appears that III\,Zw\,2 is in an on-going merger \citep{Surace2001, Veilleux2009}. Since the interaction can efficiently drive gas into the inner region, the merger may have triggered the on-going cold mode accretion phase of the AGN in III\,Zw\,2. As a sidenote, \citet{Sikora2013} propose that a merger of two disk galaxies where there has never been an extended hot accretion phase --- and therefore little accumulation of magnetic flux --- represents one possible evolutionary track to a radio-quiet quasar. 
\end{itemize}

The core-shift in III\,Zw\,2 is below our measurement accuracy and we note that this is likely partly due to our calibrators. The multi-frequency images of III\,Zw\,2 show that the source is point-like with no strong or extended jet structure.  Unfortunately, this is not true for the calibrators and our measurement accuracy of the core-shift of III\,Zw\,2 was degraded by all calibrators being extended sources that exhibit significant core-shifts (up to 0.77\,mas between 4 and 24\,GHz) themselves. Removing these calibrator core-shifts introduces additional uncertainties as demonstrated in Section~4.1. In the case of the calibrator J0007, the jet exhibits a significant bend which introduces further uncertainty on the resulting core-shift values with our method. Another source of uncertainty has to do with the proper core-identification in phase-referenced images, especially in the C-band. 

We conclude by noting that we have here and in \citet{Chamani2020} demonstrated a new method for testing the magnetic flux paradigm by combining BH spin measurements from X-ray reflection spectroscopy with jet magnetic field measurements from sub-milliarcsecond multi-frequency VLBI astrometry. We have shown for one particular RIQ source that its measured black hole spin and jet magnetic flux are in a good agreement with the \citet{Sikora2013} model, in which the amount of magnetic flux threading the BH is the parameter controlling the AGN jet production efficiency. Obviously, we have tested our method only for one source and these studies should be extended to a larger sample of radio-intermediate and radio-quiet sources in order to cover a range of radio-loudness and Eddington ratio. 

\begin{acknowledgements}
     The authors thank Karri Koljonen for his help in the preparation of the observing proposals, Maria Rioja, Richard Dodson, Andreas Brunthaler and Andrzej Zdziarski for helpful discussions, and Eduardo Ros for careful reading of the manuscript. This work was supported partly by the Academy of Finland projects 274477 and 315721. W.~Chamani is thankful for the visiting joint research grant supported by the National Astronomical Observatory of Japan (NAOJ). The Very Long Baseline Array (VLBA) is an instrument of the National Radio Astronomy Observatory. The National Radio Astronomy Observatory is a facility of the National Science Foundation operated under cooperative agreement by Associated Universities, Inc. This work made use of the Swinburne University of Technology software correlator, developed as part of the Australian Major National Research Facilities Programme and operated under licence. This publication makes use of data obtained at the Mets\"ahovi Radio Observatory, operated by the Aalto University in Finland. 
\end{acknowledgements}

%
%

\bibliographystyle{aa}
\bibliography{citations}

\begin{thebibliography}{93}
\expandafter\ifx\csname natexlab\endcsname\relax\def\natexlab#1{#1}\fi

\bibitem[{{Berton} {et~al.}(2015){Berton}, {Foschini}, {Ciroi}, {Cracco}, {La
  Mura}, {Lister}, {Mathur}, {Peterson}, {Richards}, \&
  {Rafanelli}}]{Berton2015}
{Berton}, M., {Foschini}, L., {Ciroi}, S., {et~al.} 2015, \aap, 578, A28

\bibitem[{{Blandford} {et~al.}(2019){Blandford}, {Meier}, \&
  {Readhead}}]{Blandford2019}
{Blandford}, R., {Meier}, D., \& {Readhead}, A. 2019, \araa, 57, 467

\bibitem[{{Blandford} \& {K{\"o}nigl}(1979)}]{Blandford1979}
{Blandford}, R.~D. \& {K{\"o}nigl}, A. 1979, \apj, 232, 34

\bibitem[{{Blandford} \& {Znajek}(1977)}]{Blandford1977}
{Blandford}, R.~D. \& {Znajek}, R.~L. 1977, \mnras, 179, 433

\bibitem[{{Blundell} \& {Beasley}(1998)}]{Blundell1998}
{Blundell}, K.~M. \& {Beasley}, A.~J. 1998, \mnras, 299, 165

\bibitem[{{Brunthaler} {et~al.}(2005){Brunthaler}, {Falcke}, {Bower}, {Aller},
  {Aller}, \& {Ter{\"a}sranta}}]{Brunthaler2005}
{Brunthaler}, A., {Falcke}, H., {Bower}, G.~C., {et~al.} 2005, \aap, 435, 497

\bibitem[{{Brunthaler} {et~al.}(2000){Brunthaler}, {Falcke}, {Bower}, {Aller},
  {Aller}, {Ter{\"a}sranta}, {Lobanov}, {Krichbaum}, \&
  {Patnaik}}]{Brunthaler2000}
{Brunthaler}, A., {Falcke}, H., {Bower}, G.~C., {et~al.} 2000, \aap, 357, L45

\bibitem[{{Chamani} {et~al.}(2017){Chamani}, {D{\"o}rschner}, \&
  {Schleicher}}]{Chamani2017}
{Chamani}, W., {D{\"o}rschner}, S., \& {Schleicher}, D. R.~G. 2017, \aap, 602,
  A84

\bibitem[{{Chamani} {et~al.}(2020){Chamani}, {Koljonen}, \&
  {Savolainen}}]{Chamani2020}
{Chamani}, W., {Koljonen}, K., \& {Savolainen}, T. 2020, \aap, 635, A172

\bibitem[{{Clausen-Brown} {et~al.}(2013){Clausen-Brown}, {Savolainen},
  {Pushkarev}, {Kovalev}, \& {Zensus}}]{ClausenB2013}
{Clausen-Brown}, E., {Savolainen}, T., {Pushkarev}, A.~B., {Kovalev}, Y.~Y., \&
  {Zensus}, J.~A. 2013, \aap, 558, A144

\bibitem[{{Combes} {et~al.}(2014){Combes}, {Garc{\'\i}a-Burillo}, {Casasola},
  {Hunt}, {Krips}, {Baker}, {Boone}, {Eckart}, {Marquez}, {Neri}, {Schinnerer},
  \& {Tacconi}}]{Combes2014}
{Combes}, F., {Garc{\'\i}a-Burillo}, S., {Casasola}, V., {et~al.} 2014, \aap,
  565, A97

\bibitem[{{Cooper} {et~al.}(2007){Cooper}, {Lister}, \&
  {Kochanczyk}}]{Cooper2007}
{Cooper}, N.~J., {Lister}, M.~L., \& {Kochanczyk}, M.~D. 2007, \apjs, 171, 376

\bibitem[{{Croke} \& {Gabuzda}(2008)}]{Croke2008}
{Croke}, S.~M. \& {Gabuzda}, D.~C. 2008, \mnras, 386, 619

\bibitem[{{Davis} \& {Laor}(2011)}]{DavisLaor2011}
{Davis}, S.~W. \& {Laor}, A. 2011, \apj, 728, 98

\bibitem[{{Dodson} {et~al.}(2017){Dodson}, {Rioja}, {Molina}, \&
  {G{\'o}mez}}]{Dodson2017}
{Dodson}, R., {Rioja}, M.~J., {Molina}, S.~N., \& {G{\'o}mez}, J.~L. 2017,
  \apj, 834, 177

\bibitem[{{Falcke} {et~al.}(1995){Falcke}, {Malkan}, \&
  {Biermann}}]{Falcke1995}
{Falcke}, H., {Malkan}, M.~A., \& {Biermann}, P.~L. 1995, \aap, 298, 375

\bibitem[{{Falcke} {et~al.}(1996{\natexlab{a}}){Falcke}, {Patnaik}, \&
  {Sherwood}}]{Falcke1996b}
{Falcke}, H., {Patnaik}, A.~R., \& {Sherwood}, W. 1996{\natexlab{a}}, \apjl,
  473, L13

\bibitem[{{Falcke} {et~al.}(1996{\natexlab{b}}){Falcke}, {Sherwood}, \&
  {Patnaik}}]{Falcke1996a}
{Falcke}, H., {Sherwood}, W., \& {Patnaik}, A.~R. 1996{\natexlab{b}}, \apj,
  471, 106

\bibitem[{{Fromm} {et~al.}(2013){Fromm}, {Ros}, {Perucho}, {Savolainen},
  {Mimica}, {Kadler}, {Lobanov}, \& {Zensus}}]{Fromm2013}
{Fromm}, C.~M., {Ros}, E., {Perucho}, M., {et~al.} 2013, \aap, 557, A105

\bibitem[{{Ghisellini} {et~al.}(2004){Ghisellini}, {Haardt}, \&
  {Matt}}]{Ghisellini2004}
{Ghisellini}, G., {Haardt}, F., \& {Matt}, G. 2004, \aap, 413, 535

\bibitem[{{Ghisellini} {et~al.}(2014){Ghisellini}, {Tavecchio}, {Maraschi},
  {Celotti}, \& {Sbarrato}}]{Ghisellini2014}
{Ghisellini}, G., {Tavecchio}, F., {Maraschi}, L., {Celotti}, A., \&
  {Sbarrato}, T. 2014, \nat, 515, 376

\bibitem[{{Greisen}(2003)}]{Greisen2003}
{Greisen}, E.~W. 2003, {AIPS, the VLA, and the VLBA}, Vol. 285 (Kluwer Academic
  Publishers), 109

\bibitem[{{Grier} {et~al.}(2012){Grier}, {Peterson}, {Pogge}, {Denney},
  {Bentz}, {Martini}, {Sergeev}, {Kaspi}, {Minezaki}, {Zu}, {Kochanek},
  {Siverd}, {Shappee}, {Stanek}, {Araya Salvo}, {Beatty}, {Bird}, {Bord},
  {Borman}, {Che}, {Chen}, {Cohen}, {Dietrich}, {Doroshenko}, {Drake},
  {Efimov}, {Free}, {Ginsburg}, {Henderson}, {King}, {Koshida}, {Mogren},
  {Molina}, {Mosquera}, {Nazarov}, {Okhmat}, {Pejcha}, {Rafter}, {Shields},
  {Skowron}, {Szczygiel}, {Valluri}, \& {van Saders}}]{Grier2012}
{Grier}, C.~J., {Peterson}, B.~M., {Pogge}, R.~W., {et~al.} 2012, \apj, 755, 60

\bibitem[{{Hada} {et~al.}(2011){Hada}, {Doi}, {Kino}, {Nagai}, {Hagiwara}, \&
  {Kawaguchi}}]{Hada2011}
{Hada}, K., {Doi}, A., {Kino}, M., {et~al.} 2011, \nat, 477, 185

\bibitem[{{Haga} {et~al.}(2015){Haga}, {Doi}, {Murata}, {Sudou}, {Kameno}, \&
  {Hada}}]{Haga2015}
{Haga}, T., {Doi}, A., {Murata}, Y., {et~al.} 2015, \apj, 807, 15

\bibitem[{{Hern{\'a}n Caballero}(2012)}]{Hernan2012}
{Hern{\'a}n Caballero}, A. 2012, \mnras, 427, 816

\bibitem[{{Hirotani}(2005)}]{Hirotani2005}
{Hirotani}, K. 2005, \apj, 619, 73

\bibitem[{{Hodgson} {et~al.}(2017){Hodgson}, {Krichbaum}, {Marscher},
  {Jorstad}, {Rani}, {Marti-Vidal}, {Bach}, {Sanchez}, {Bremer}, {Lindqvist},
  {Uunila}, {Kallunki}, {Vicente}, {Fuhrmann}, {Angelakis}, {Karamanavis},
  {Myserlis}, {Nestoras}, {Chidiac}, {Sievers}, {Gurwell}, \&
  {Zensus}}]{Hodgson2017}
{Hodgson}, J.~A., {Krichbaum}, T.~P., {Marscher}, A.~P., {et~al.} 2017, \aap,
  597, A80

\bibitem[{{Hopkins} \& {Quataert}(2010)}]{Hopkins2010}
{Hopkins}, P.~F. \& {Quataert}, E. 2010, \mnras, 407, 1529

\bibitem[{{Hopkins} \& {Quataert}(2011)}]{Hopkins2011}
{Hopkins}, P.~F. \& {Quataert}, E. 2011, \mnras, 415, 1027

\bibitem[{{Hovatta} \& {Lindfors}(2019)}]{Hovatta2019}
{Hovatta}, T. \& {Lindfors}, E. 2019, \nar, 87, 101541

\bibitem[{{Hovatta} {et~al.}(2009){Hovatta}, {Valtaoja}, {Tornikoski}, \&
  {L{\"a}hteenm{\"a}ki}}]{Hovatta2009}
{Hovatta}, T., {Valtaoja}, E., {Tornikoski}, M., \& {L{\"a}hteenm{\"a}ki}, A.
  2009, \aap, 494, 527

\bibitem[{{Hutchings} \& {Campbell}(1983)}]{Hutchings1983}
{Hutchings}, J.~B. \& {Campbell}, B. 1983, \nat, 303, 584

\bibitem[{{Inoue} {et~al.}(2007){Inoue}, {Terashima}, \& {Ho}}]{Inoue2007}
{Inoue}, H., {Terashima}, Y., \& {Ho}, L.~C. 2007, \apj, 662, 860

\bibitem[{{Kellermann} {et~al.}(1989){Kellermann}, {Sramek}, {Schmidt},
  {Shaffer}, \& {Green}}]{Kellermann1989}
{Kellermann}, K.~I., {Sramek}, R., {Schmidt}, M., {Shaffer}, D.~B., \& {Green},
  R. 1989, \aj, 98, 1195

\bibitem[{{Komissarov} {et~al.}(2009){Komissarov}, {Vlahakis}, {K{\"o}nigl}, \&
  {Barkov}}]{Komissarov2009}
{Komissarov}, S.~S., {Vlahakis}, N., {K{\"o}nigl}, A., \& {Barkov}, M.~V. 2009,
  \mnras, 394, 1182

\bibitem[{{Kovalev} {et~al.}(2008){Kovalev}, {Lobanov}, {Pushkarev}, \&
  {Zensus}}]{Kovalev2008}
{Kovalev}, Y.~Y., {Lobanov}, A.~P., {Pushkarev}, A.~B., \& {Zensus}, J.~A.
  2008, \aap, 483, 759

\bibitem[{{Kutkin} {et~al.}(2014){Kutkin}, {Sokolovsky}, {Lisakov}, {Kovalev},
  {Savolainen}, {Voytsik}, {Lobanov}, {Aller}, {Aller}, {Lahteenmaki},
  {Tornikoski}, {Volvach}, \& {Volvach}}]{Kutkin2014}
{Kutkin}, A.~M., {Sokolovsky}, K.~V., {Lisakov}, M.~M., {et~al.} 2014, \mnras,
  437, 3396

\bibitem[{{L{\"a}hteenm{\"a}ki} {et~al.}(2018){L{\"a}hteenm{\"a}ki},
  {J{\"a}rvel{\"a}}, {Ramakrishnan}, {Tornikoski}, {Tammi}, {Vera}, \&
  {Chamani}}]{Lahteenmaki2018}
{L{\"a}hteenm{\"a}ki}, A., {J{\"a}rvel{\"a}}, E., {Ramakrishnan}, V., {et~al.}
  2018, \aap, 614, L1

\bibitem[{{Lampton} {et~al.}(1976){Lampton}, {Margon}, \&
  {Bowyer}}]{Lampton1976}
{Lampton}, M., {Margon}, B., \& {Bowyer}, S. 1976, \apj, 208, 177

\bibitem[{Li {et~al.}(2010)Li, Xie, Dai, Chen, Yi, Tang, Bao, Lü, Na, \&
  Ren}]{Li2010}
Li, H., Xie, G., Dai, H., {et~al.} 2010, New Astronomy, 15, 254

\bibitem[{{Liao} {et~al.}(2016){Liao}, {Xin}, {Fan}, {Weng}, {Li}, {Chen}, \&
  {Fan}}]{Liao2016}
{Liao}, N.-H., {Xin}, Y.-L., {Fan}, X.-L., {et~al.} 2016, \apjs, 226, 17

\bibitem[{Lister {et~al.}(2019)Lister, Homan, Hovatta, Kellermann, Kiehlmann,
  Kovalev, Max-Moerbeck, Pushkarev, Readhead, Ros, \& Savolainen}]{Lister2019}
Lister, M.~L., Homan, D.~C., Hovatta, T., {et~al.} 2019, The Astrophysical
  Journal, 874, 43

\bibitem[{{Lobanov}(1998)}]{Lobanov1998}
{Lobanov}, A.~P. 1998, \aap, 330, 79

\bibitem[{{Lobanov}(2005)}]{Lobanov2005}
{Lobanov}, A.~P. 2005, arXiv e-prints, arXiv:0503225

\bibitem[{{Marscher}(1977)}]{Marscher1977}
{Marscher}, A.~P. 1977, \apj, 216, 244

\bibitem[{{Marscher}(1983)}]{Marscher1983}
{Marscher}, A.~P. 1983, \apj, 264, 296

\bibitem[{McKinney {et~al.}(2012)McKinney, Tchekhovskoy, \&
  Blandford}]{McKinney2012}
McKinney, J.~C., Tchekhovskoy, A., \& Blandford, R.~D. 2012, Monthly Notices of
  the Royal Astronomical Society, 423, 3083

\bibitem[{{McNamara} {et~al.}(2011){McNamara}, {Rohanizadegan}, \&
  {Nulsen}}]{McNamara2011}
{McNamara}, B.~R., {Rohanizadegan}, M., \& {Nulsen}, P.~E.~J. 2011, \apj, 727,
  39

\bibitem[{{Mioduszewski} \& {Kogan}(2009)}]{Mioduszewski2009}
{Mioduszewski}, A.~J. \& {Kogan}, L. 2009, {Strategy for removing tropospheric
  and clock errors using \textsc{delzn}}, AIPS Memo 110, NRAO

\bibitem[{Moderski {et~al.}(1998)Moderski, Sikora, \& Lasota}]{Moderski1998}
Moderski, R., Sikora, M., \& Lasota, J.-P. 1998, Monthly Notices of the Royal
  Astronomical Society, 301, 142

\bibitem[{{Nalewajko} {et~al.}(2014){Nalewajko}, {Sikora}, \&
  {Begelman}}]{Nalewajko2014}
{Nalewajko}, K., {Sikora}, M., \& {Begelman}, M.~C. 2014, \apjl, 796, L5

\bibitem[{{Narayan} {et~al.}(2003){Narayan}, {Igumenshchev}, \&
  {Abramowicz}}]{Narayan2003}
{Narayan}, R., {Igumenshchev}, I.~V., \& {Abramowicz}, M.~A. 2003, \pasj, 55,
  L69

\bibitem[{{Nyland} {et~al.}(2020){Nyland}, {Dong}, {Patil}, {Lacy}, {van
  Velzen}, {Kimball}, {Sarbadhicary}, {Hallinan}, {Baldassare}, {Clarke},
  {Goulding}, {Greene}, {Hughes}, {Kassim}, {Kunert-Bajraszewska}, {Maccarone},
  {Mooley}, {Mukherjee}, {Peters}, {Petrov}, {Polisensky}, {Rujopakarn},
  {Whittle}, \& {Vaccari}}]{Nyland2020}
{Nyland}, K., {Dong}, D.~Z., {Patil}, P., {et~al.} 2020, arXiv e-prints,
  arXiv:2011.08872

\bibitem[{{Orosz} {et~al.}(2017){Orosz}, {Imai}, {Dodson}, {Rioja}, {Frey},
  {Burns}, {Etoka}, {Nakagawa}, {Nakanishi}, {Asaki}, {Goldman}, \&
  {Tafoya}}]{Orosz2017}
{Orosz}, G., {Imai}, H., {Dodson}, R., {et~al.} 2017, \aj, 153, 119

\bibitem[{{O'Sullivan} \& {Gabuzda}(2009)}]{OSullivan2009}
{O'Sullivan}, S.~P. \& {Gabuzda}, D.~C. 2009, \mnras, 400, 26

\bibitem[{{Pacholczyk}(1970)}]{Pacholczyk1970}
{Pacholczyk}, A.~G. 1970, {Radio astrophysics. Nonthermal processes in galactic
  and extragalactic sources}

\bibitem[{{Pjanka} {et~al.}(2017){Pjanka}, {Greene}, {Seth}, {Braatz},
  {Henkel}, {Lo}, \& {L{\"a}sker}}]{Pjanka2017}
{Pjanka}, P., {Greene}, J.~E., {Seth}, A.~C., {et~al.} 2017, \apj, 844, 165

\bibitem[{{Plavin} {et~al.}(2019){Plavin}, {Kovalev}, {Pushkarev}, \&
  {Lobanov}}]{Plavin2019}
{Plavin}, A.~V., {Kovalev}, Y.~Y., {Pushkarev}, A.~B., \& {Lobanov}, A.~P.
  2019, \mnras, 485, 1822

\bibitem[{Pushkarev {et~al.}(2018)Pushkarev, Butuzova, Kovalev, \&
  Hovatta}]{Pushkarev2018}
Pushkarev, A.~B., Butuzova, M.~S., Kovalev, Y.~Y., \& Hovatta, T. 2018, Monthly
  Notices of the Royal Astronomical Society, 482, 2336

\bibitem[{{Pushkarev} {et~al.}(2012){Pushkarev}, {Hovatta}, {Kovalev},
  {Lister}, {Lobanov}, {Savolainen}, \& {Zensus}}]{Pushkarev2012}
{Pushkarev}, A.~B., {Hovatta}, T., {Kovalev}, Y.~Y., {et~al.} 2012, \aap, 545,
  A113

\bibitem[{{Pushkarev} {et~al.}(2017){Pushkarev}, {Kovalev}, {Lister}, \&
  {Savolainen}}]{Pushkarev2017}
{Pushkarev}, A.~B., {Kovalev}, Y.~Y., {Lister}, M.~L., \& {Savolainen}, T.
  2017, \mnras, 468, 4992

\bibitem[{{Raginski} \& {Laor}(2016)}]{Raginski2016}
{Raginski}, I. \& {Laor}, A. 2016, \mnras, 459, 2082

\bibitem[{{Rawlings} \& {Saunders}(1991)}]{RawlingsSa1991}
{Rawlings}, S. \& {Saunders}, R. 1991, \nat, 349, 138

\bibitem[{{Reid} \& {Honma}(2014)}]{ReidHonma2014}
{Reid}, M.~J. \& {Honma}, M. 2014, \araa, 52, 339

\bibitem[{{Reid} {et~al.}(2009){Reid}, {Menten}, {Brunthaler}, {Zheng},
  {Moscadelli}, \& {Xu}}]{Reid2009}
{Reid}, M.~J., {Menten}, K.~M., {Brunthaler}, A., {et~al.} 2009, \apj, 693, 397

\bibitem[{Reynolds(2013)}]{Reynolds2013}
Reynolds, C.~S. 2013, Space Science Reviews, 183, 277–294

\bibitem[{{Reynolds}(2020)}]{Reynolds2020}
{Reynolds}, C.~S. 2020, arXiv e-prints, arXiv:2011.08948

\bibitem[{{Savolainen} {et~al.}(2008){Savolainen}, {Wiik}, {Valtaoja}, \&
  {Tornikoski}}]{Savolainen2008}
{Savolainen}, T., {Wiik}, K., {Valtaoja}, E., \& {Tornikoski}, M. 2008, in
  Astronomical Society of the Pacific Conference Series, Vol. 386,
  Extragalactic Jets: Theory and Observation from Radio to Gamma Ray, ed. T.~A.
  {Rector} \& D.~S. {De Young}, 451

\bibitem[{{Shepherd}(1997)}]{Shepherd1997}
{Shepherd}, M.~C. 1997, in Astronomical Society of the Pacific Conference
  Series, Vol. 125, Astronomical Data Analysis Software and Systems VI, ed.
  G.~{Hunt} \& H.~{Payne}, 77

\bibitem[{{Sikora} \& {Begelman}(2013)}]{Sikora2013}
{Sikora}, M. \& {Begelman}, M.~C. 2013, \apjl, 764, L24

\bibitem[{{Sikora} {et~al.}(2007){Sikora}, {Stawarz}, \& {Lasota}}]{Sikora2007}
{Sikora}, M., {Stawarz}, {\L}., \& {Lasota}, J.-P. 2007, \apj, 658, 815

\bibitem[{{Snellen} {et~al.}(2002){Snellen}, {McMahon}, {Hook}, \&
  {Browne}}]{Snellen2002}
{Snellen}, I.~A.~G., {McMahon}, R.~G., {Hook}, I.~M., \& {Browne}, I.~W.~A.
  2002, \mnras, 329, 700

\bibitem[{{Sokolovsky} {et~al.}(2011){Sokolovsky}, {Kovalev}, {Pushkarev}, \&
  {Lobanov}}]{Sokolovsky2011}
{Sokolovsky}, K.~V., {Kovalev}, Y.~Y., {Pushkarev}, A.~B., \& {Lobanov}, A.~P.
  2011, \aap, 532, A38

\bibitem[{{Strittmatter} {et~al.}(1980){Strittmatter}, {Hill}, {Pauliny-Toth},
  {Steppe}, \& {Witzel}}]{Strittmatter1980}
{Strittmatter}, P.~A., {Hill}, P., {Pauliny-Toth}, I.~I.~K., {Steppe}, H., \&
  {Witzel}, A. 1980, \aap, 88, L12

\bibitem[{{Surace} {et~al.}(2001){Surace}, {Sanders}, \& {Evans}}]{Surace2001}
{Surace}, J.~A., {Sanders}, D.~B., \& {Evans}, A.~S. 2001, \aj, 122, 2791

\bibitem[{Taylor {et~al.}(1996)Taylor, Dunlop, Hughes, \& Robson}]{Taylor1996}
Taylor, G.~L., Dunlop, J.~S., Hughes, D.~H., \& Robson, E.~I. 1996, Monthly
  Notices of the Royal Astronomical Society, 283, 930

\bibitem[{{Tchekhovskoy} {et~al.}(2010){Tchekhovskoy}, {Narayan}, \&
  {McKinney}}]{Tchekhovskoy2010}
{Tchekhovskoy}, A., {Narayan}, R., \& {McKinney}, J.~C. 2010, \apj, 711, 50

\bibitem[{{Tchekhovskoy} {et~al.}(2011){Tchekhovskoy}, {Narayan}, \&
  {McKinney}}]{Tchekhovskoy2011}
{Tchekhovskoy}, A., {Narayan}, R., \& {McKinney}, J.~C. 2011, \mnras, 418, L79

\bibitem[{{Ter\"asranta} {et~al.}(1998){Ter\"asranta}, {Tornikoski}, {Mujunen},
  {Karlamaa}, {Valtonen}, {Henelius}, {Urpo}, {Lainela}, {Pursimo}, {Nilsson},
  {Wiren}, {Laehteenmaeki}, {Korpi}, {Rekola}, {Heinaemaeki}, {Hanski},
  {Nurmi}, {Kokkonen}, {Keinaenen}, {Joutsamo}, {Oksanen}, {Pietilae},
  {Valtaoja}, {Valtonen}, \& {Koenoenen}}]{Terasranta1998}
{Ter\"asranta}, H., {Tornikoski}, M., {Mujunen}, A., {et~al.} 1998, \aaps, 132,
  305

\bibitem[{{Thompson} {et~al.}(2017){Thompson}, {Moran}, \&
  {Swenson}}]{2017israbook}
{Thompson}, A.~R., {Moran}, J.~M., \& {Swenson}, George~W., J. 2017,
  {Interferometry and Synthesis in Radio Astronomy, 3rd Edition} (Springer)

\bibitem[{{T{\"u}rler} {et~al.}(1999){T{\"u}rler}, {Courvoisier}, \&
  {Paltani}}]{Turler1999}
{T{\"u}rler}, M., {Courvoisier}, T.~J.~L., \& {Paltani}, S. 1999, \aap, 349, 45

\bibitem[{{Unger} {et~al.}(1987){Unger}, {Lawrence}, {Wilson}, {Elvis}, \&
  {Wright}}]{Unger1987}
{Unger}, S.~W., {Lawrence}, A., {Wilson}, A.~S., {Elvis}, M., \& {Wright},
  A.~E. 1987, \mnras, 228, 521

\bibitem[{{Veilleux} {et~al.}(2009){Veilleux}, {Kim}, {Rupke}, {Peng},
  {Tacconi}, {Genzel}, {Lutz}, {Sturm}, {Contursi}, {Schweitzer}, {Dasyra},
  {Ho}, {Sanders}, \& {Burkert}}]{Veilleux2009}
{Veilleux}, S., {Kim}, D.~C., {Rupke}, D.~S.~N., {et~al.} 2009, \apj, 701, 587

\bibitem[{{Voitsik} {et~al.}(2018){Voitsik}, {Pushkarev}, {Kovalev}, {Plavin},
  {Lobanov}, \& {Ipatov}}]{Voitsik2018}
{Voitsik}, P.~A., {Pushkarev}, A.~B., {Kovalev}, Y.~Y., {et~al.} 2018,
  Astronomy Reports, 62, 787

\bibitem[{{Volonteri} {et~al.}(2013){Volonteri}, {Sikora}, {Lasota}, \&
  {Merloni}}]{Volonteri2013}
{Volonteri}, M., {Sikora}, M., {Lasota}, J.~P., \& {Merloni}, A. 2013, \apj,
  775, 94

\bibitem[{{Walker} {et~al.}(2000){Walker}, {Dhawan}, {Romney}, {Kellermann}, \&
  {Vermeulen}}]{Walker2000}
{Walker}, R.~C., {Dhawan}, V., {Romney}, J.~D., {Kellermann}, K.~I., \&
  {Vermeulen}, R.~C. 2000, \apj, 530, 233

\bibitem[{{White} {et~al.}(2017){White}, {Jarvis}, {Kalfountzou}, {Hardcastle},
  {Verma}, {Cao Orjales}, \& {Stevens}}]{White2017}
{White}, S.~V., {Jarvis}, M.~J., {Kalfountzou}, E., {et~al.} 2017, \mnras, 468,
  217

\bibitem[{{Wright}(2006)}]{Wright2006}
{Wright}, E.~L. 2006, \pasp, 118, 1711

\bibitem[{{Xie} \& {Zdziarski}(2019)}]{Xie2019}
{Xie}, F.-G. \& {Zdziarski}, A.~A. 2019, \apj, 887, 167

\bibitem[{{Zakamska} \& {Greene}(2014)}]{Zakamska2014}
{Zakamska}, N.~L. \& {Greene}, J.~E. 2014, \mnras, 442, 784

\bibitem[{{Zamaninasab} {et~al.}(2014){Zamaninasab}, {Clausen-Brown},
  {Savolainen}, \& {Tchekhovskoy}}]{Zamaninasab2014}
{Zamaninasab}, M., {Clausen-Brown}, E., {Savolainen}, T., \& {Tchekhovskoy}, A.
  2014, \nat, 510, 126

\bibitem[{{Zdziarski} {et~al.}(2015){Zdziarski}, {Sikora}, {Pjanka}, \&
  {Tchekhovskoy}}]{Zdziarski2015}
{Zdziarski}, A.~A., {Sikora}, M., {Pjanka}, P., \& {Tchekhovskoy}, A. 2015,
  \mnras, 451, 927

\end{thebibliography}

\begin{appendix}

\onecolumn    
\section{Phase-referenced images}

Phase-referenced images of the calibrator sources at the four frequencies displayed below in Figures~A.1, A.2, and A.3.
\begin{figure*}[!h]
\centering
   \subfigure[]
   {
       \includegraphics[width=0.4\linewidth]{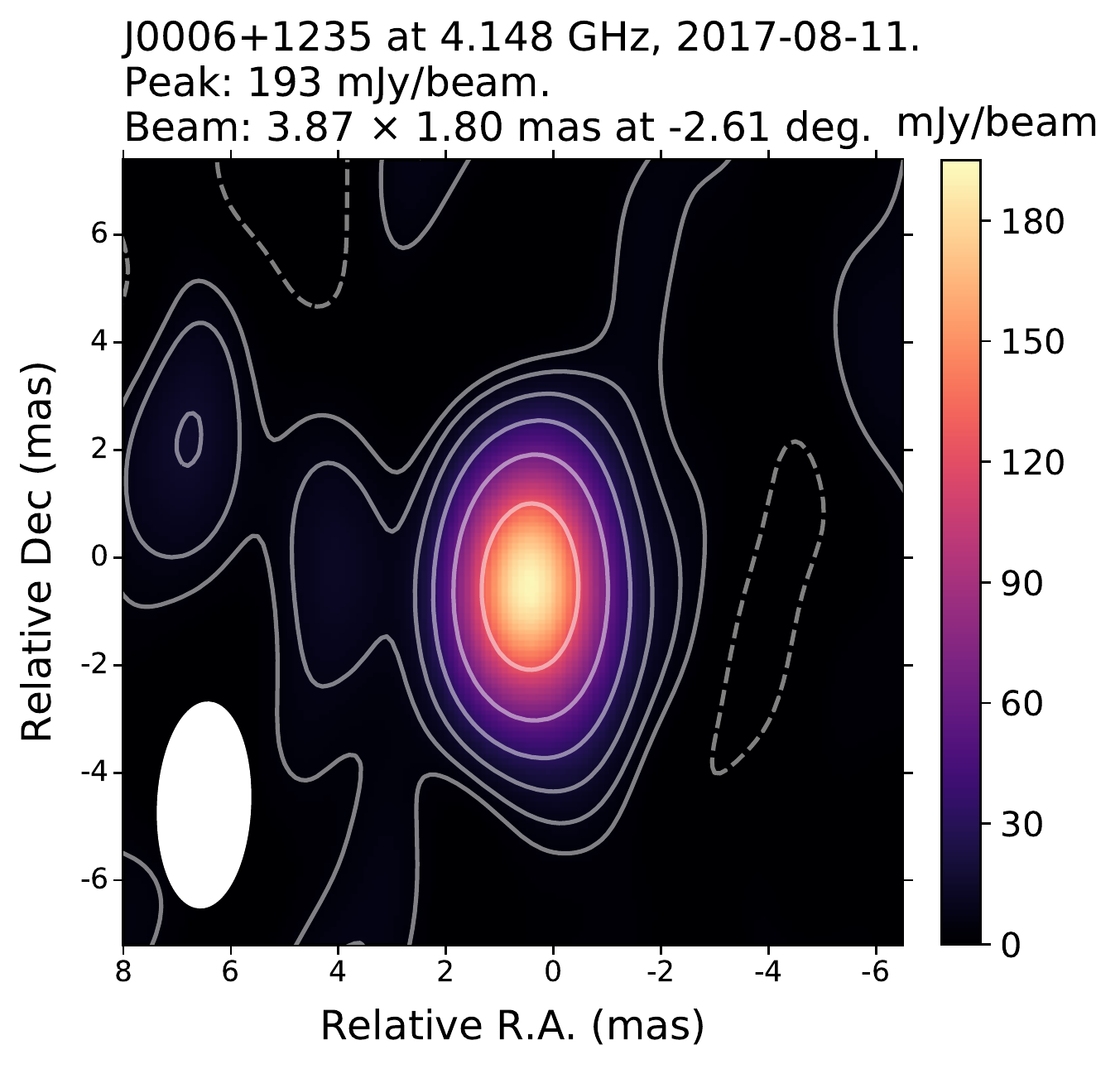}
        \label{}
    }
     \subfigure[]
    {
        \includegraphics[width=0.4\linewidth]{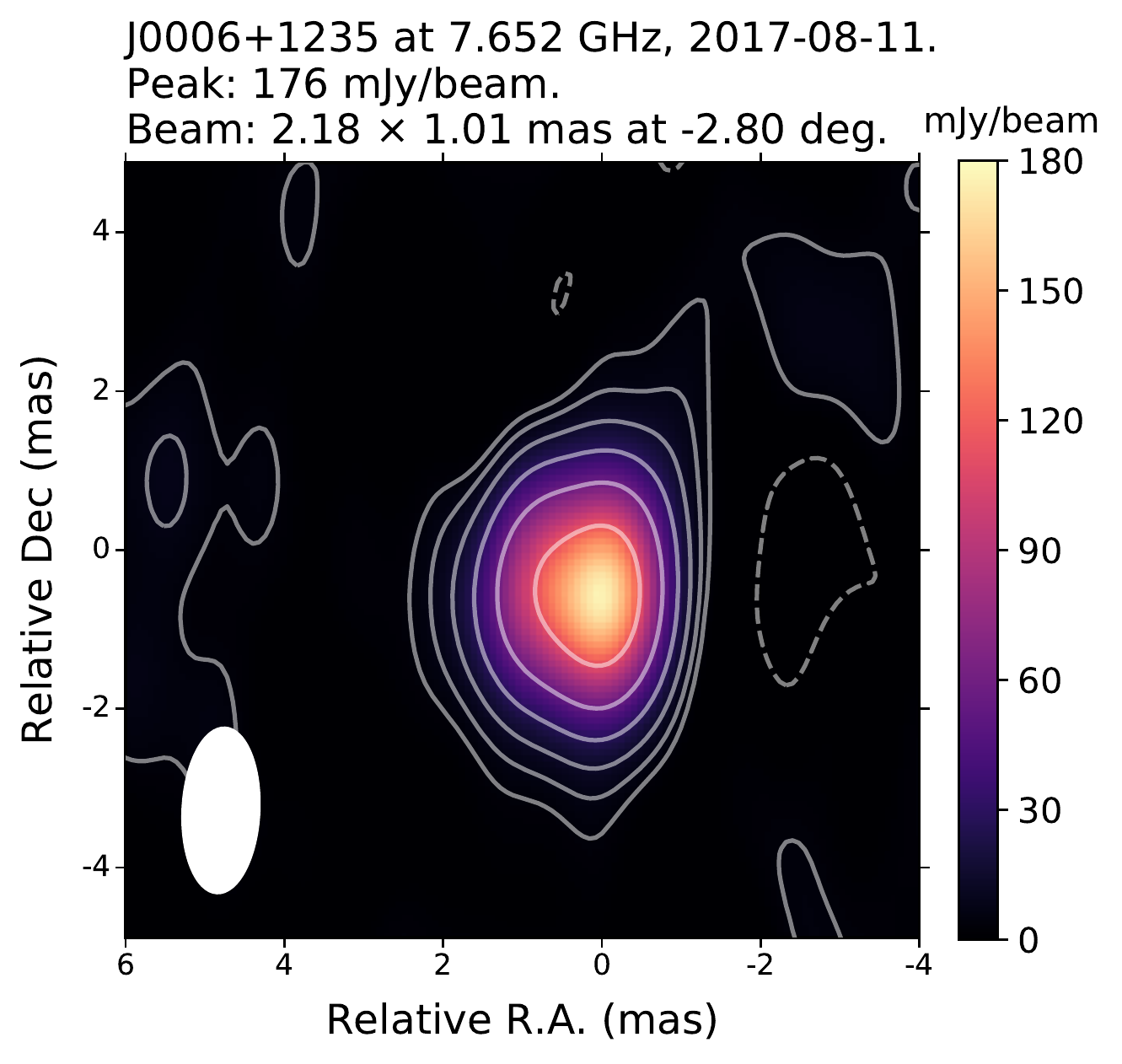}
        \label{}
    }
    \subfigure[]
    {
        \includegraphics[width=0.4\linewidth]{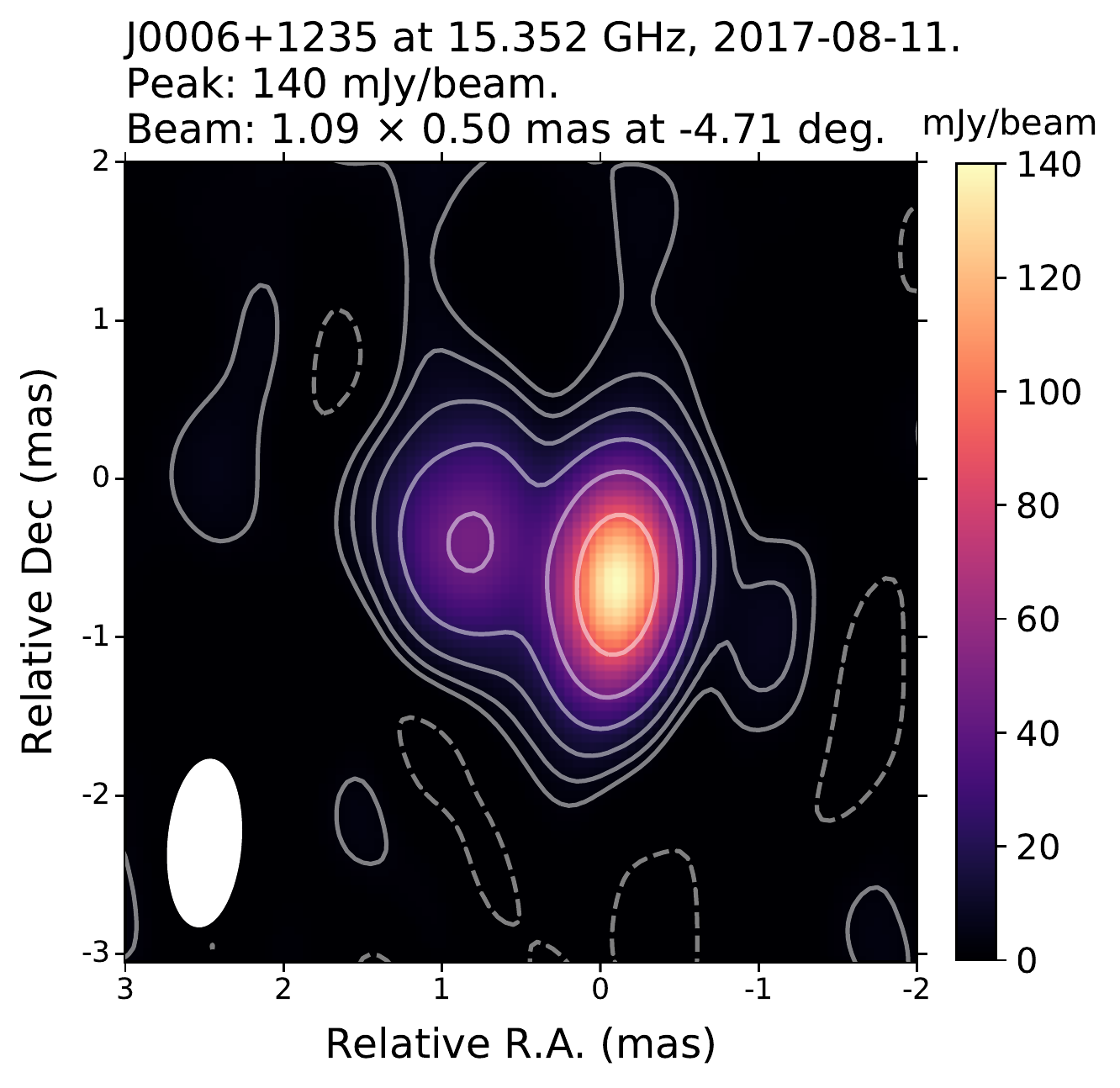}
        \label{}
    }
     \subfigure[]
    {
        \includegraphics[width=0.4\linewidth]{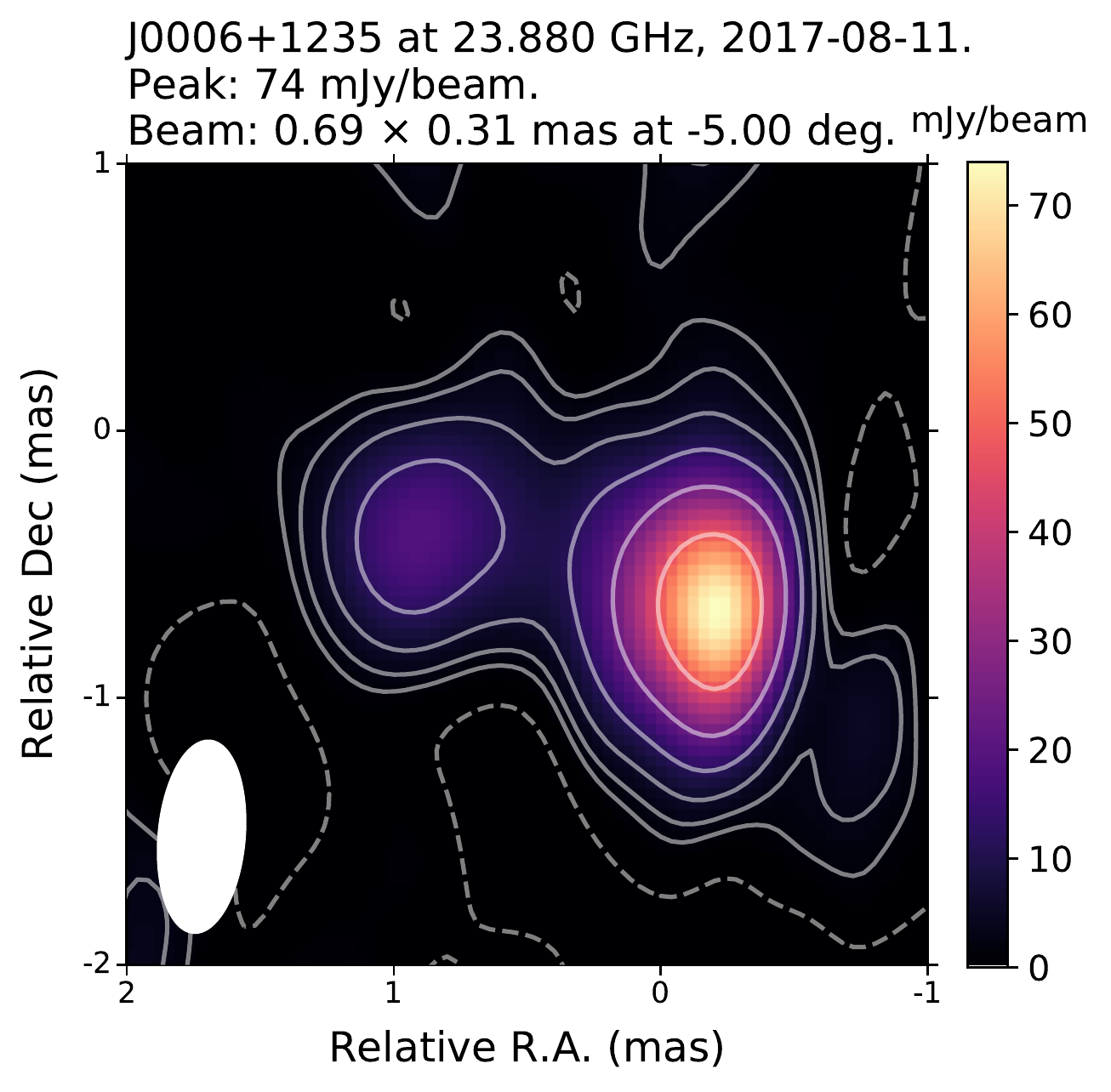}
        \label{}
    } 

    \caption{Phase-referenced clean images of J0006+1235. The images were obtained at a) 4.148\,GHz, b) 7.652\,GHz, c) 15.352\,GHz and d) 23.88\,GHz. The rms noise level from the lower to the higher frequency is 2.6, 2, 1.7, 1.3  mJy/beam. The interferometric beam (ellipse) is displayed on the bottom-left corner of each image. Contours represent -2\%, 2\%, 4\%, 8\%, 16\%, 32\%, and 64\% of the peak intensity at each image.}
    \label{fA1}
\end{figure*}

\begin{figure*}[t]
\centering
   \subfigure[]
   {
        \includegraphics[width=0.4\linewidth]{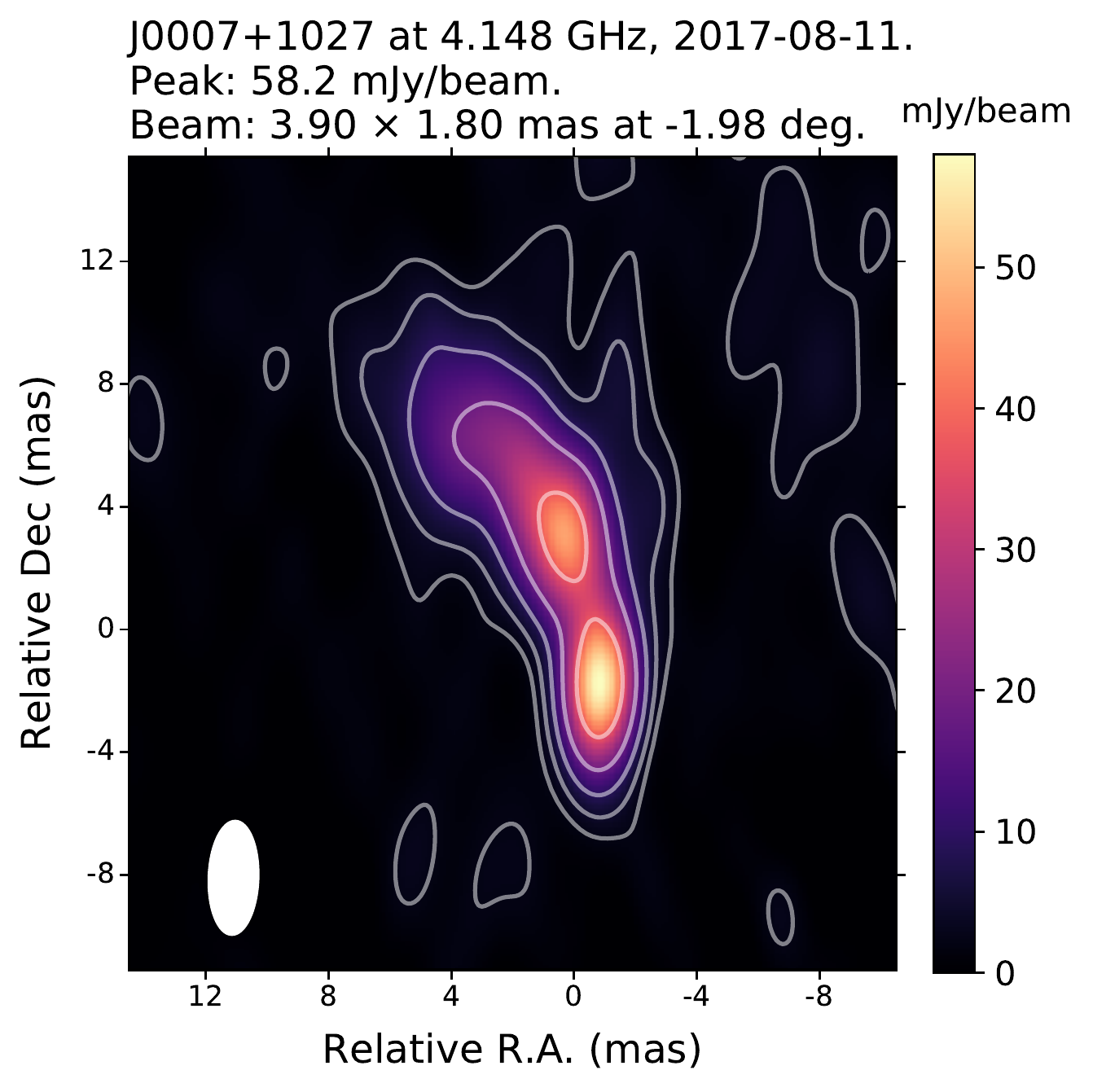}
        \label{}
    }
     \subfigure[]
    {
        \includegraphics[width=0.4\linewidth]{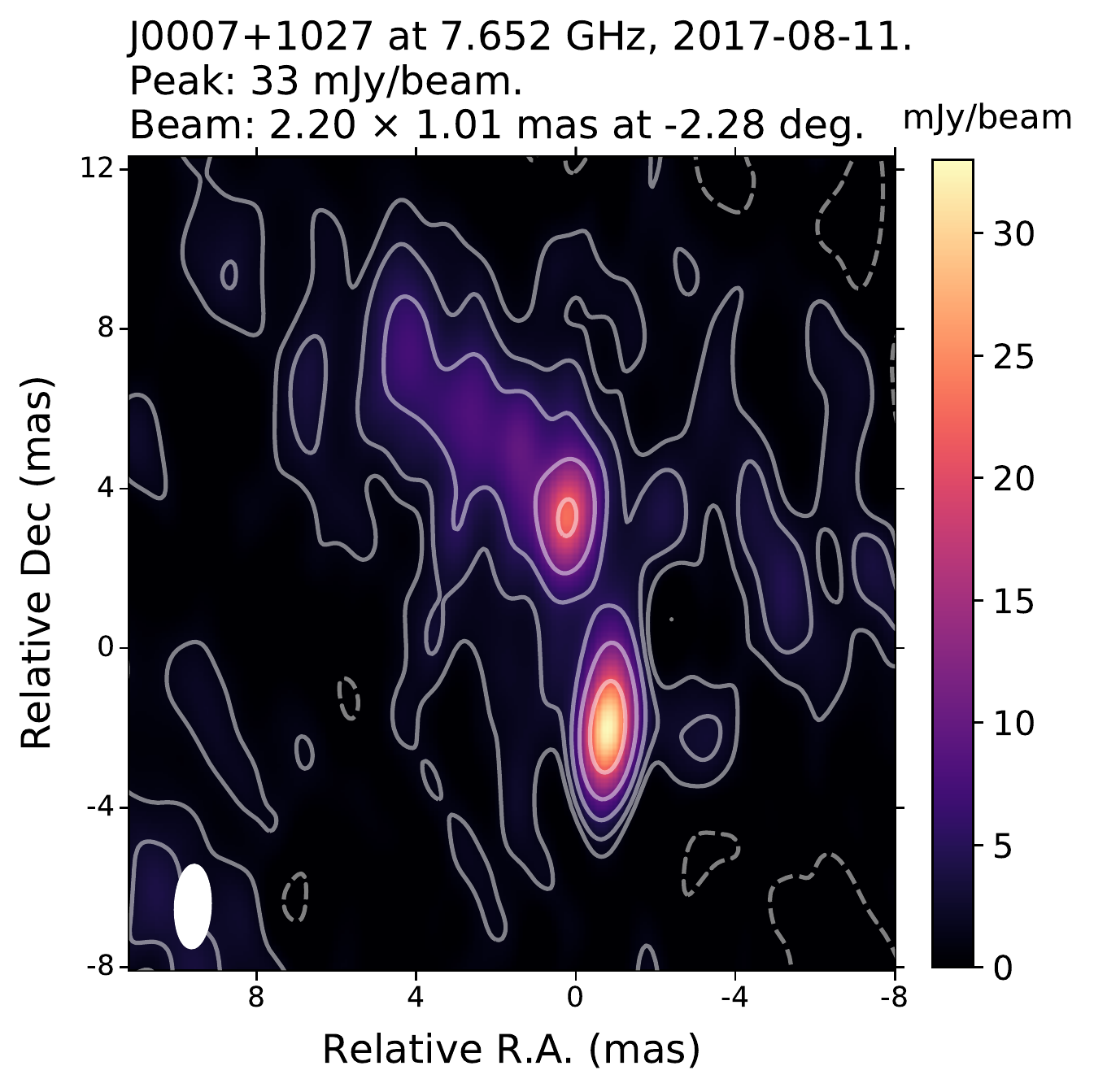}
        \label{}
    }
    \subfigure[]
    {
        \includegraphics[width=0.4\linewidth]{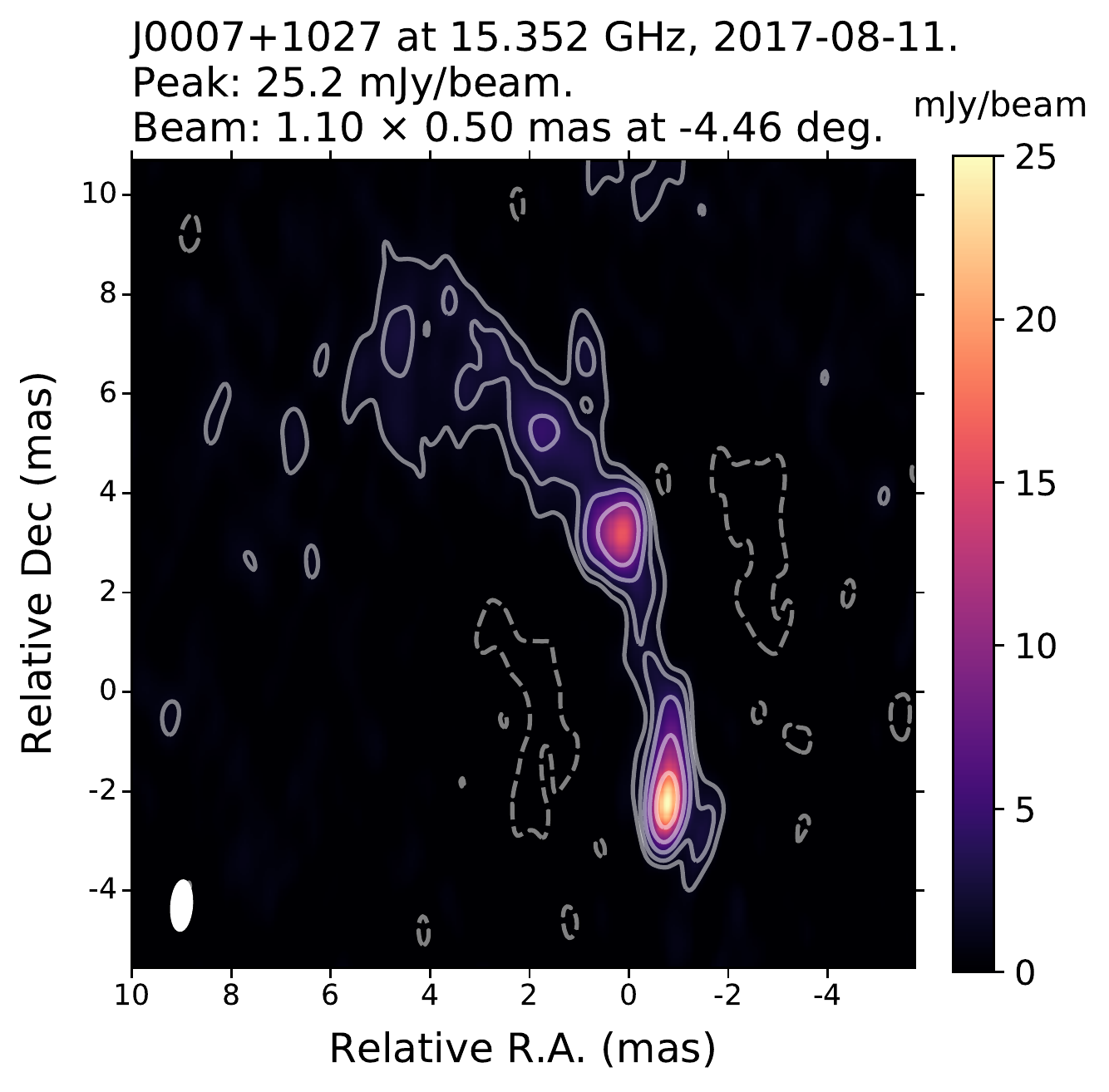}
        \label{}
    }
     \subfigure[]
    {
        \includegraphics[width=0.4\linewidth]{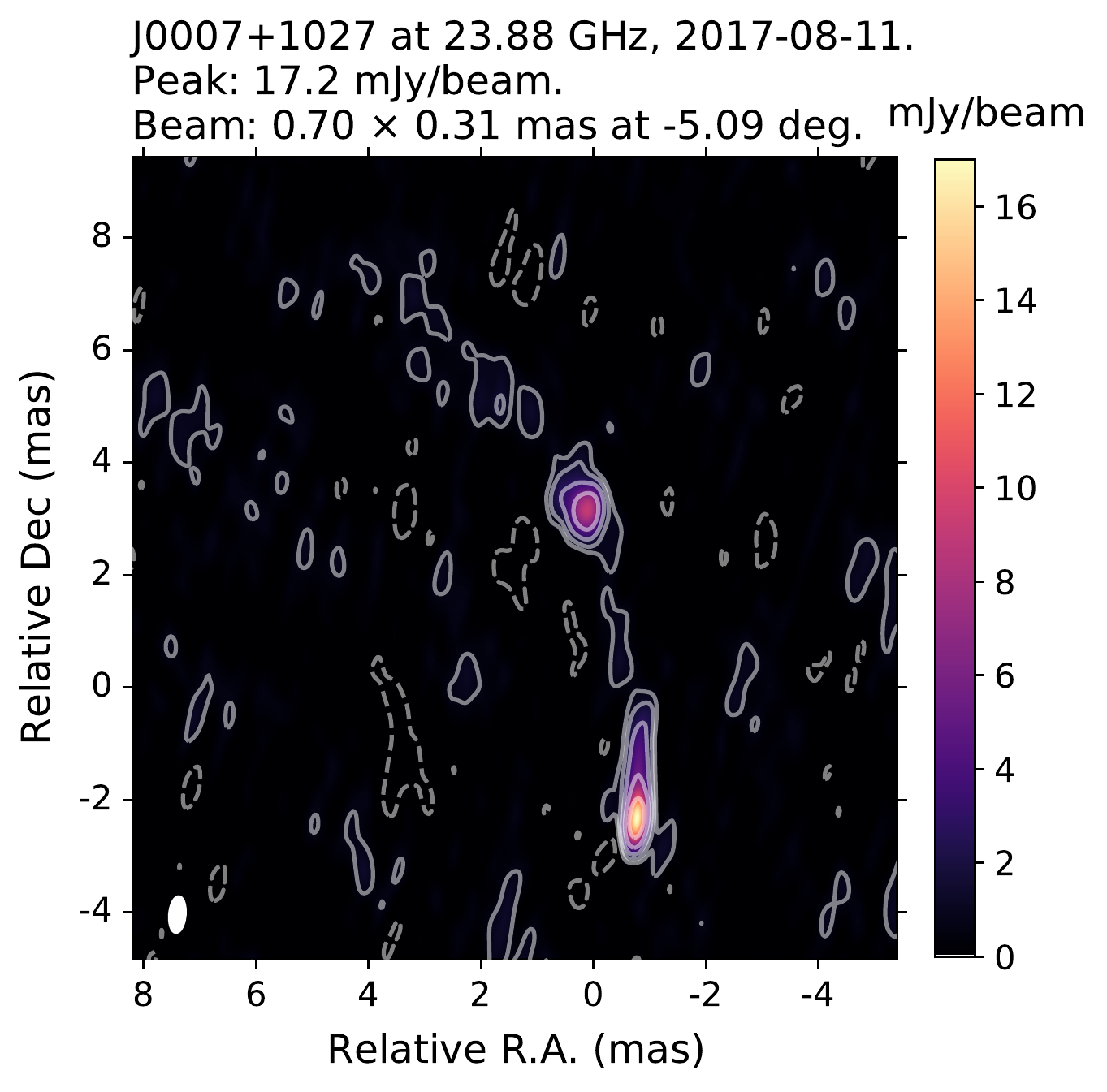}
        \label{}
    } 

    \caption{Phase-referenced clean images of J0007+1027. The images were obtained at a) 4.148\,GHz, b) 7.652\,GHz, c) 15.352\,GHz and d) 23.88\,GHz.  The rms noise level from the lower to the higher frequency is 1, 1.2, 0.5, 0.3  mJy/beam. The interferometric beam (ellipse) is displayed on the bottom-left corner of each image. Contours represent -4\%, 4\%, 8\%, 16\%, 32\%, and 64\% of the peak intensity at each image.}
    \label{fA2}
\end{figure*}

\begin{figure*}[t]
\centering
   \subfigure[]
   {
        \includegraphics[width=0.4\linewidth]{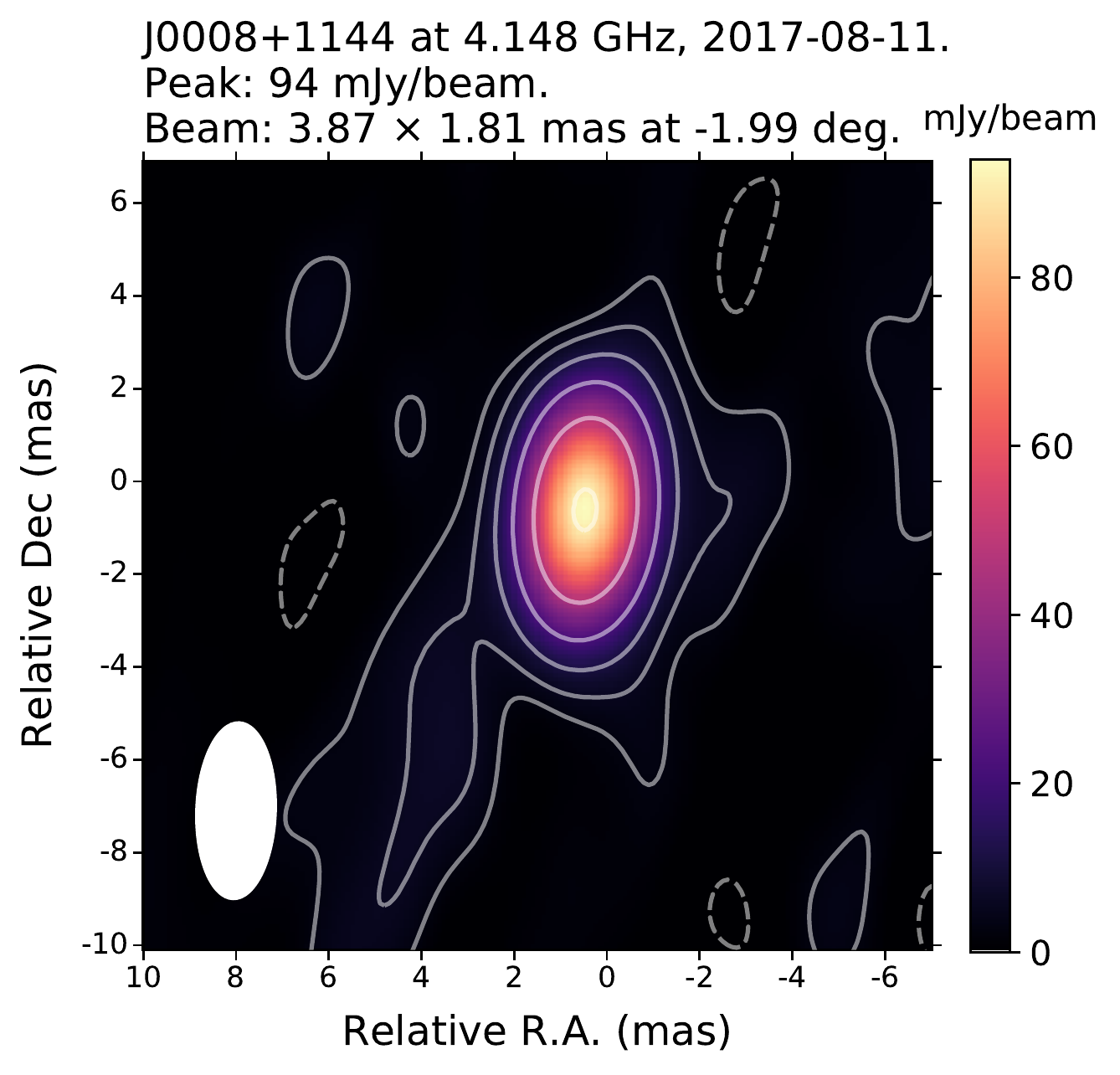}
        \label{}
    }
     \subfigure[]
    {
        \includegraphics[width=0.4\linewidth]{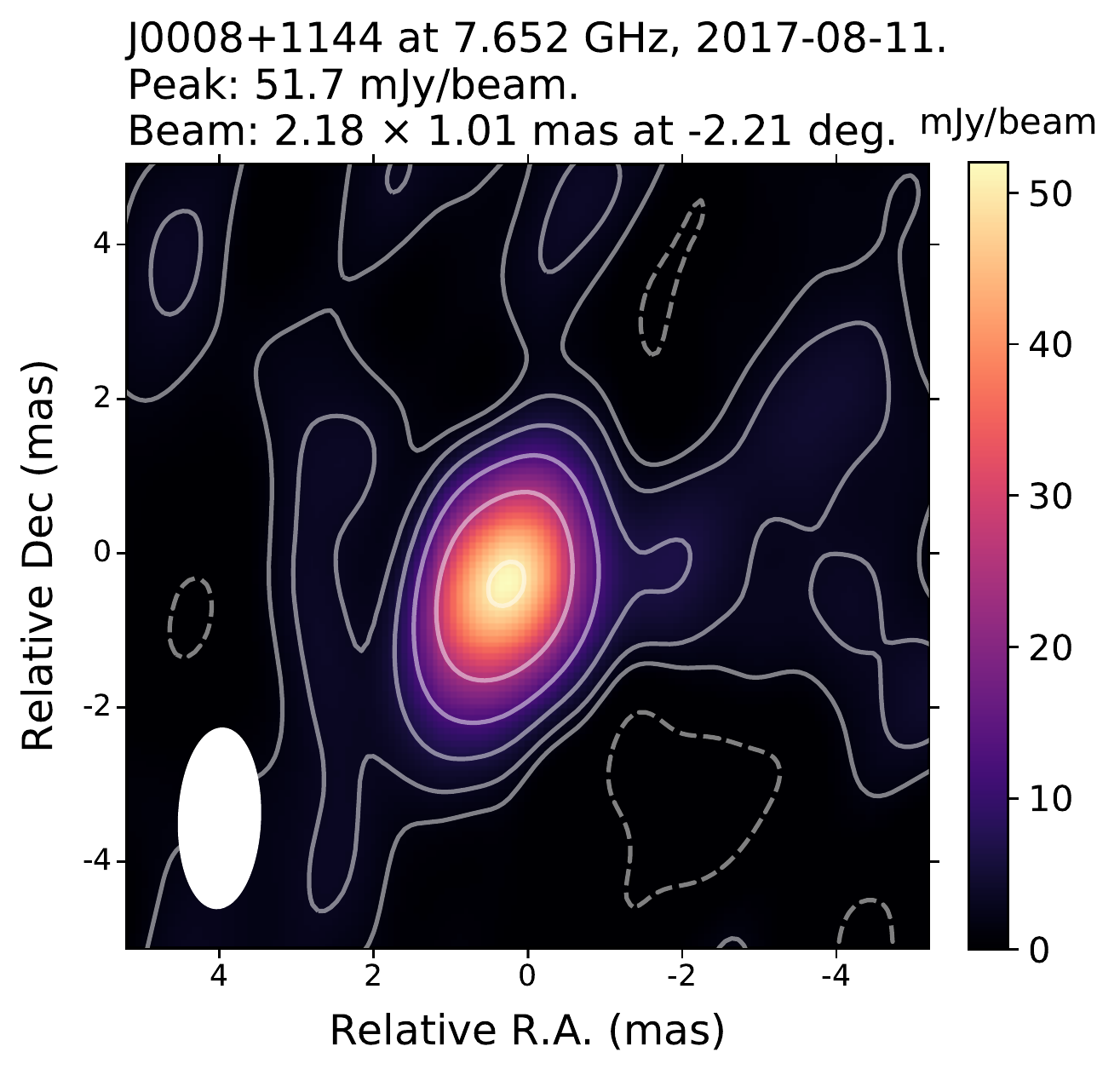}
        \label{}
    }
    \subfigure[]
    {
        \includegraphics[width=0.4\linewidth]{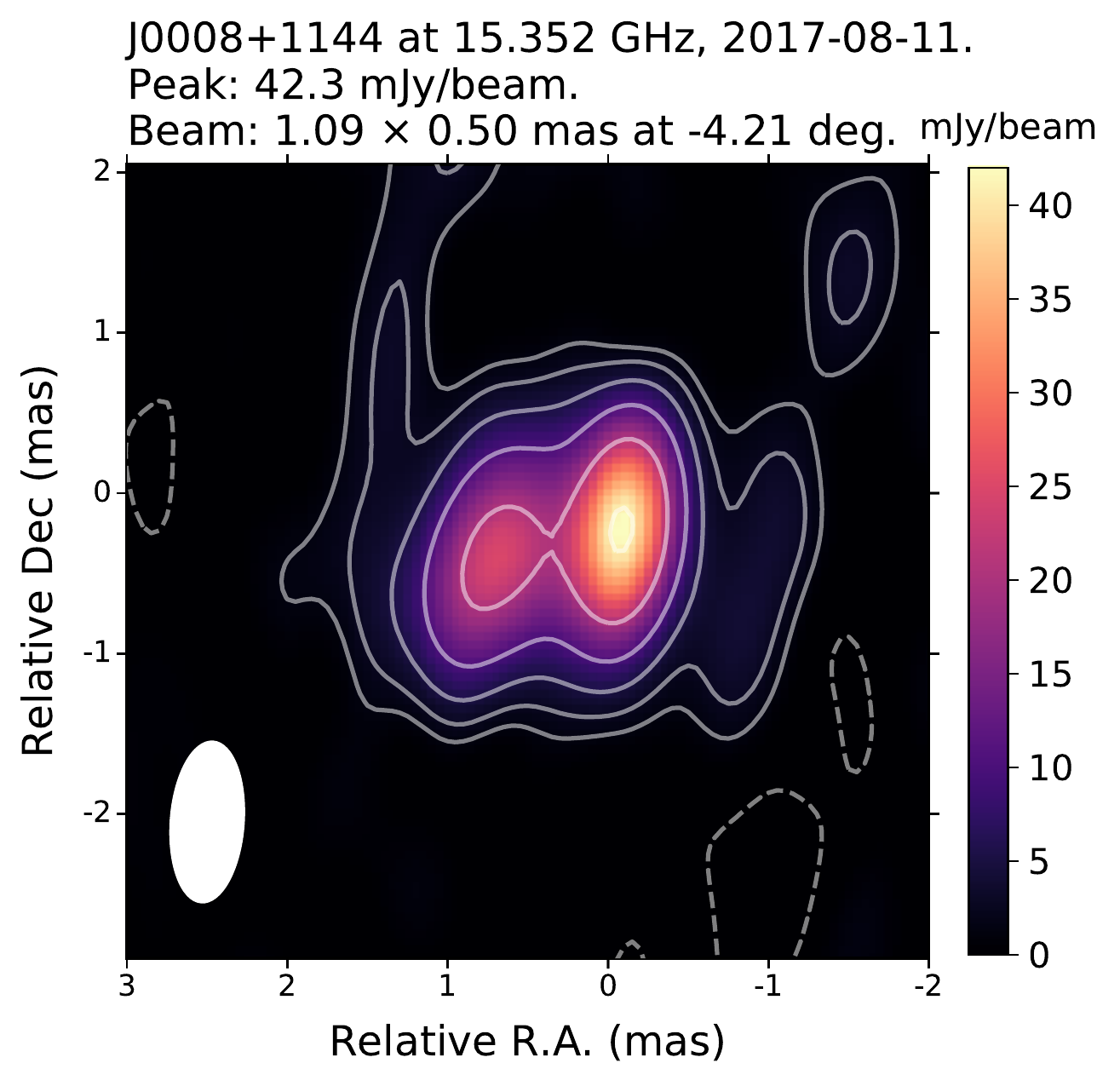}
        \label{}
    }
     \subfigure[]
    {
        \includegraphics[width=0.4\linewidth]{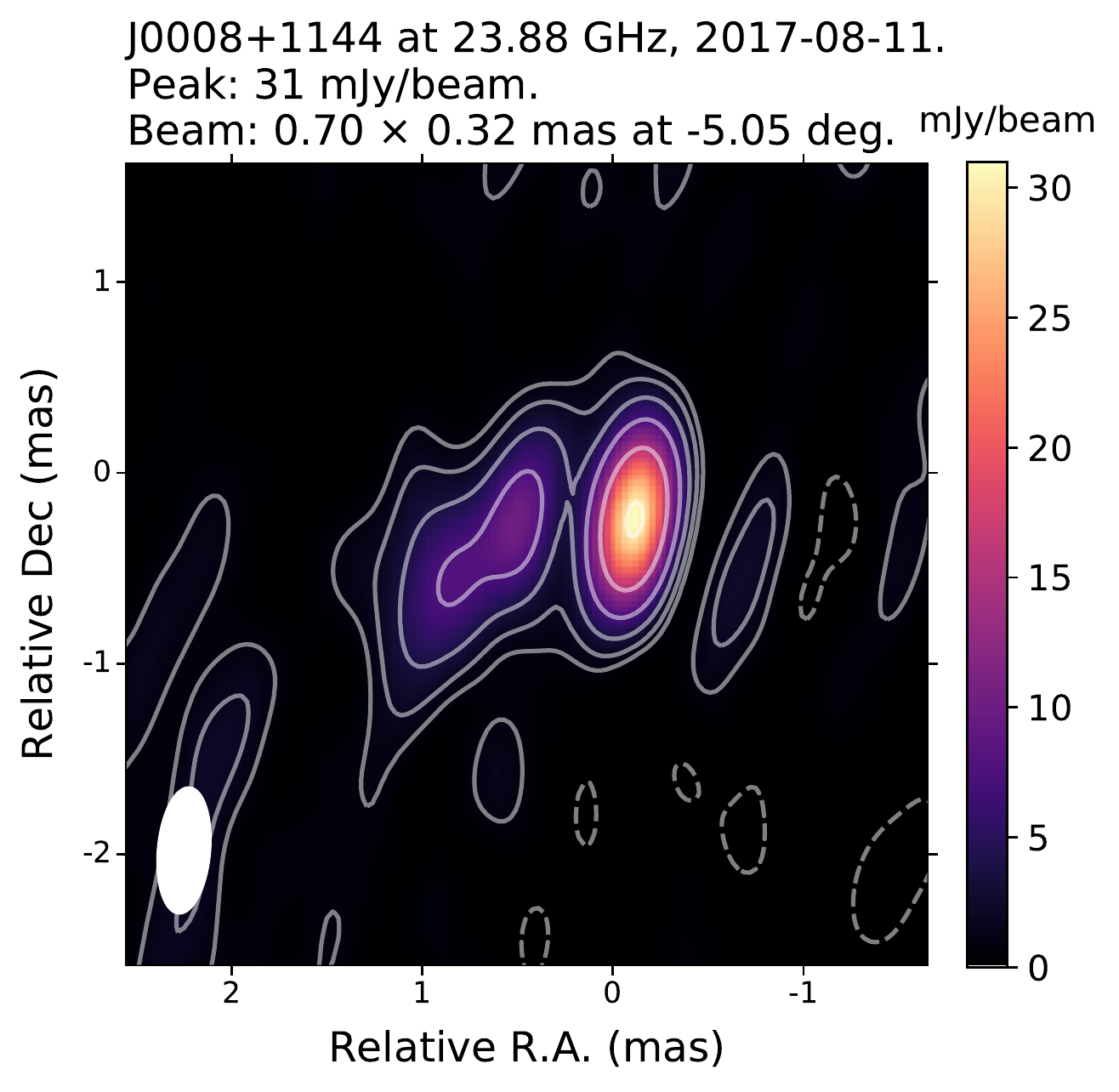}
        \label{}
    } 

    \caption{Phase-referenced clean images of J0008+1144. The images were obtained at a) 4.148\,GHz, b) 7.652\,GHz, c) 15.352\,GHz and d) 23.88\,GHz. The rms noise level from the lower to the higher frequency is 1.1, 1.4, 0.7, 0.5 mJy/beam. The interferometric beam (ellipse) is displayed on the bottom-left corner of each image. Contours represent -2\%, 2\%, 4\%, 8\%, 16\%, 32\%, and 64\% of the peak intensity at each image.}
    \label{fA3}
\end{figure*}

\clearpage

\section{Self-calibrated images}

The CLEAN images of the calibrator sources after self-calibration in amplitudes and in phases are displayed below in Figures~B.1, B.2, and B.3.



\begin{figure*}[!h]
\centering
   \subfigure[]
   {
        \includegraphics[width=0.4\linewidth]{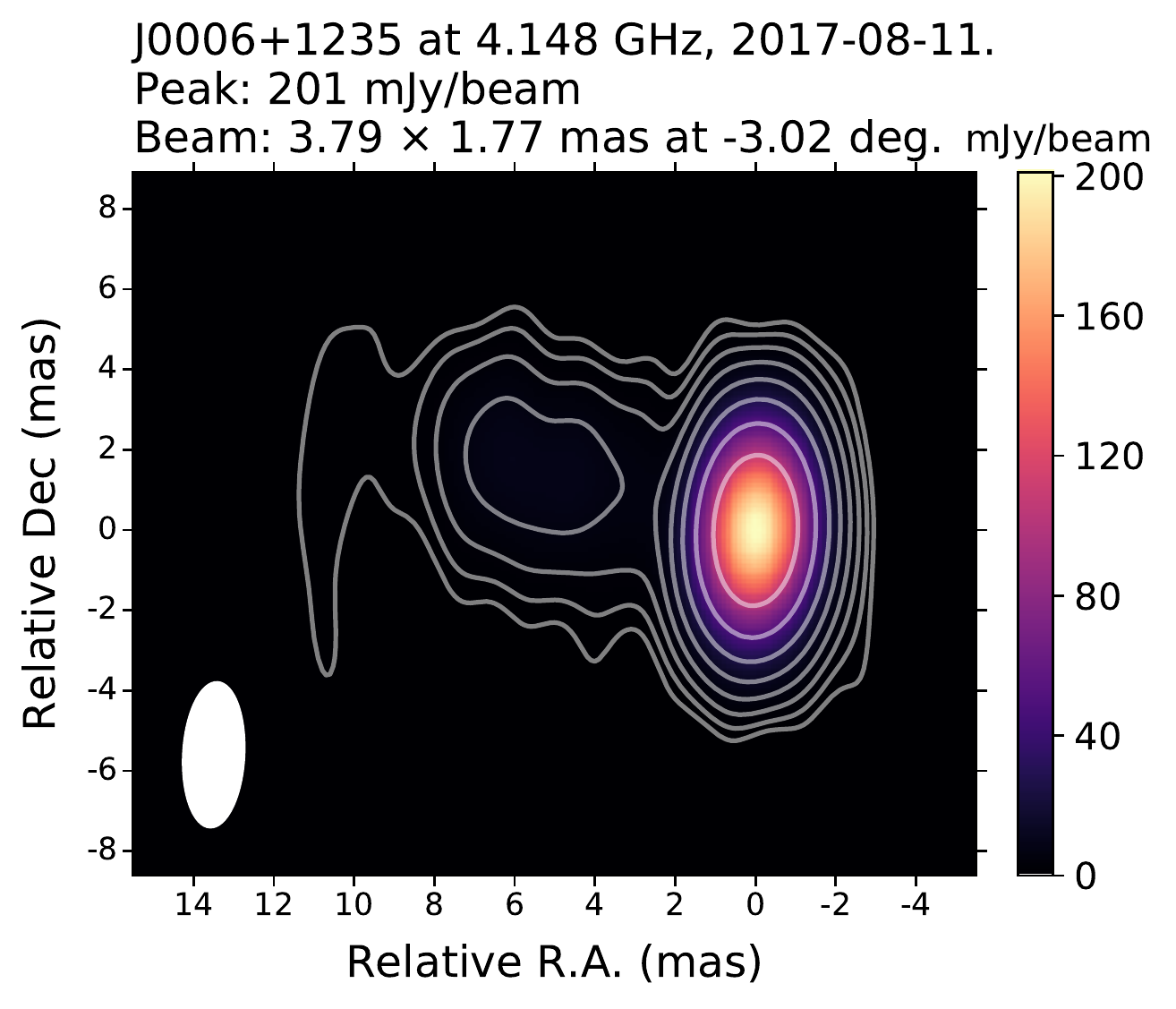}
        \label{}
    }
     \subfigure[]
    {
        \includegraphics[width=0.4\linewidth]{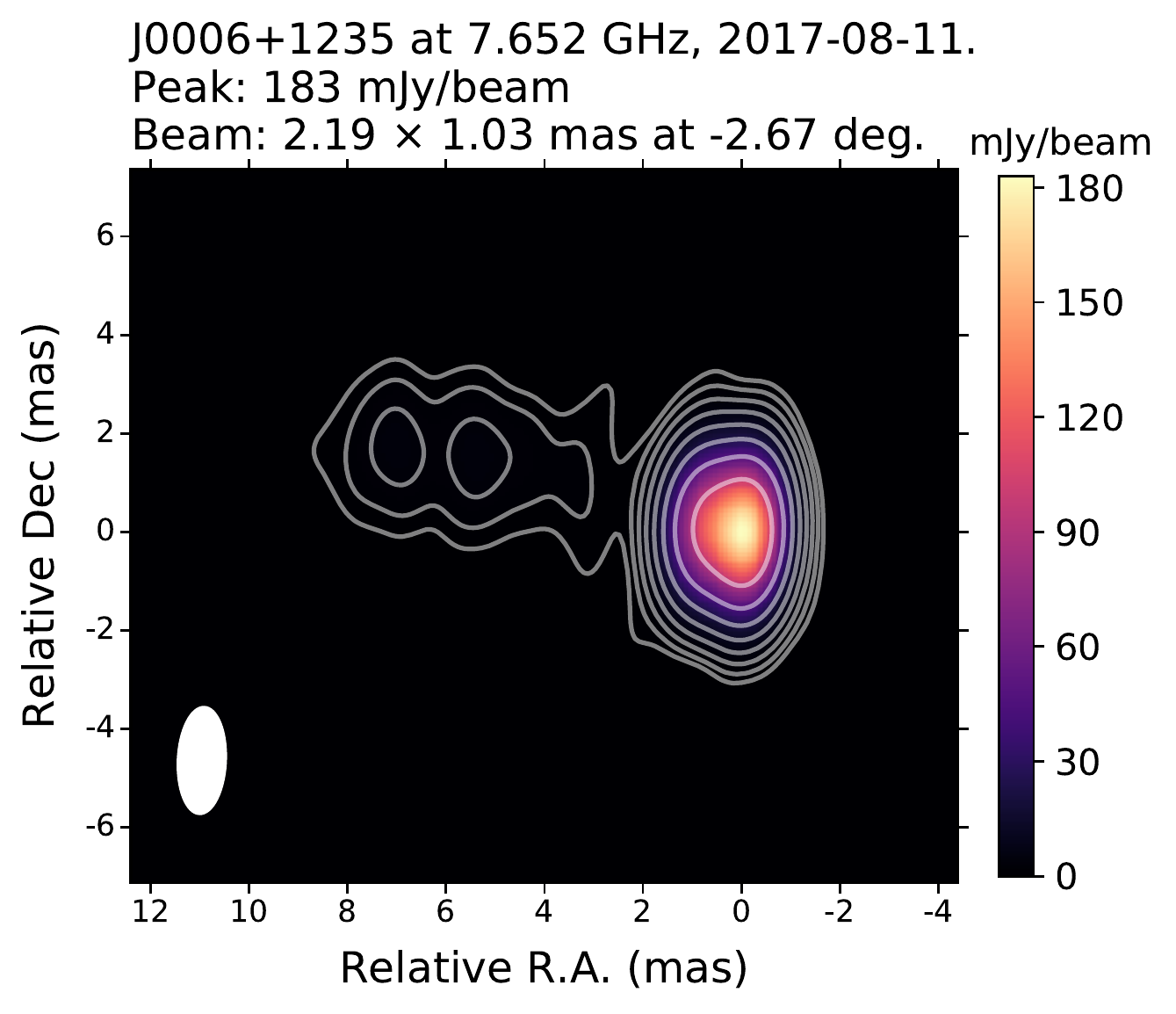}
        \label{}
    }
    \subfigure[]
    {
        \includegraphics[width=0.4\linewidth]{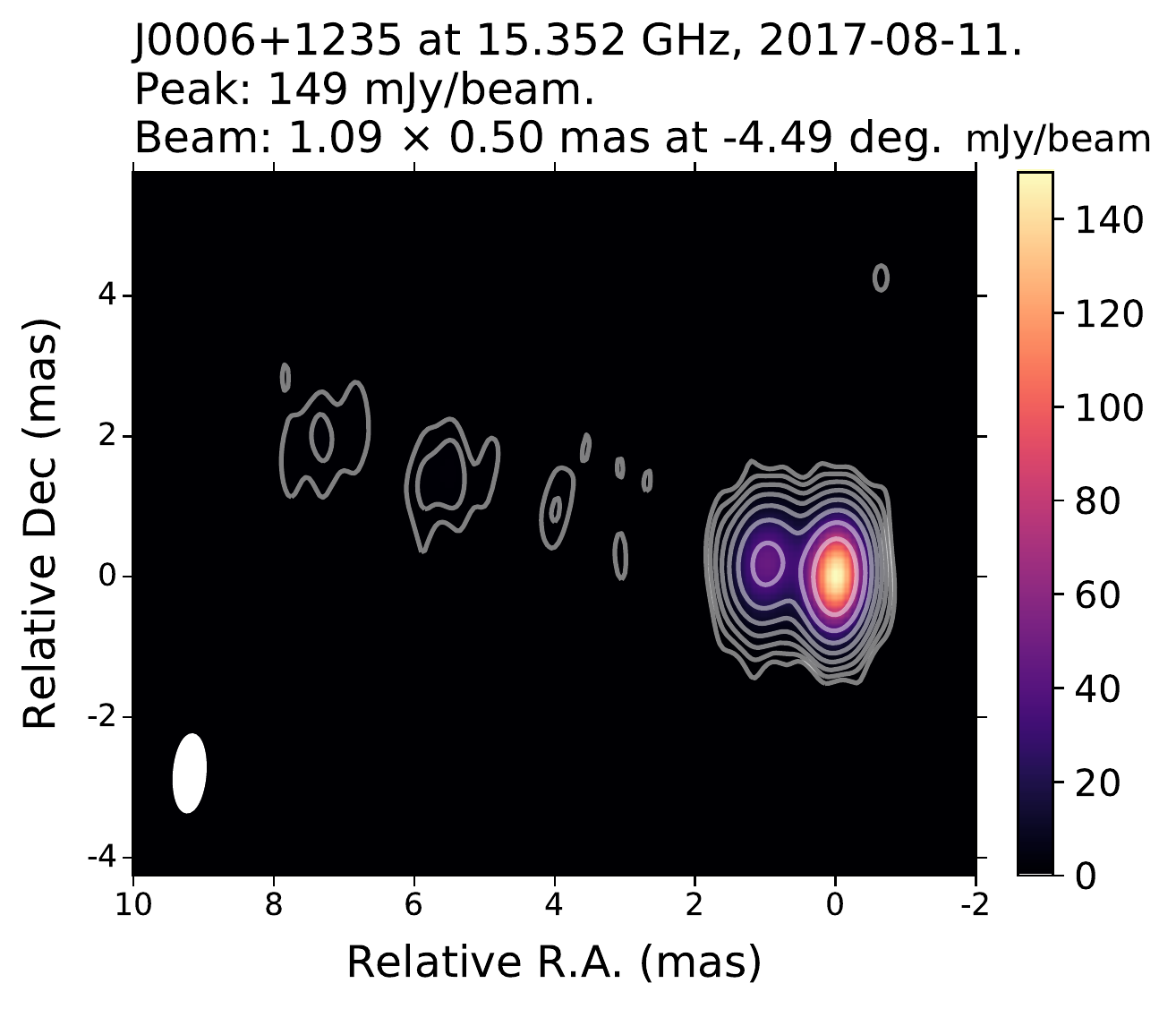}
        \label{}
    }
     \subfigure[]
    {
        \includegraphics[width=0.4\linewidth]{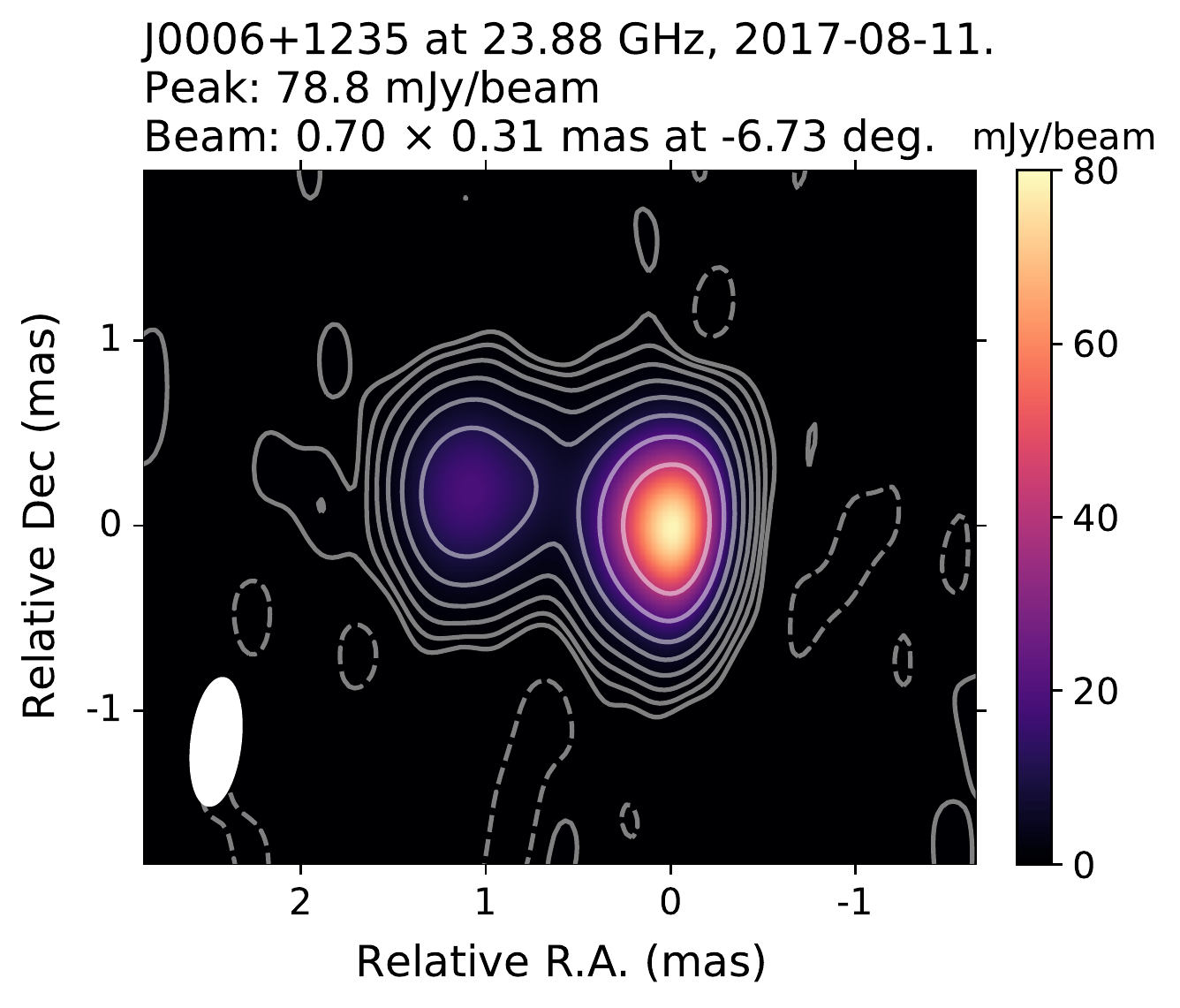}
        \label{}
    } 

    \caption{Self-calibrated images of J0006+1235. The images were obtained at a) 4.148\,GHz, b) 7.652\,GHz, c) 15.352\,GHz and d) 23.88\,GHz. The rms noise level from the lower to the higher frequency is 0.3, 0.1, 0.2, 0.2  mJy/beam. The interferometric beam (ellipse) is displayed on the bottom-left corner of each image. Contours represent -0.4\%, 0.4\%, 0.8\%, 1.6\%, 3.2\%, 6.4\%, 12.8\%, 25.6\%, and 51.2\% of the peak intensity at each image. }
    \label{fB1} 
\end{figure*}

\begin{figure*}[t]
\centering
   \subfigure[]
   {
        \includegraphics[width=0.4\linewidth]{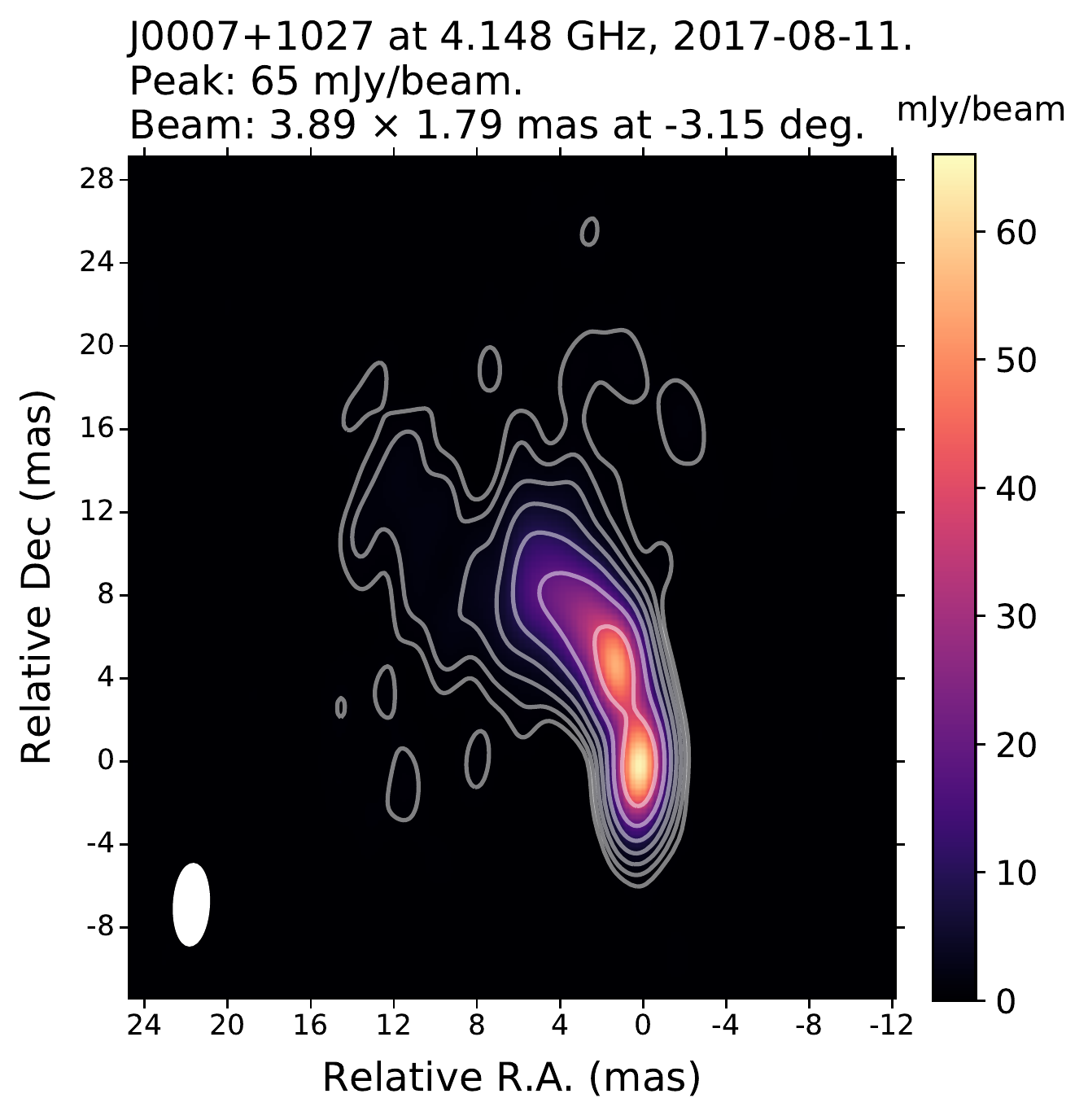}
        \label{}
    }
     \subfigure[]
    {
        \includegraphics[width=0.4\linewidth]{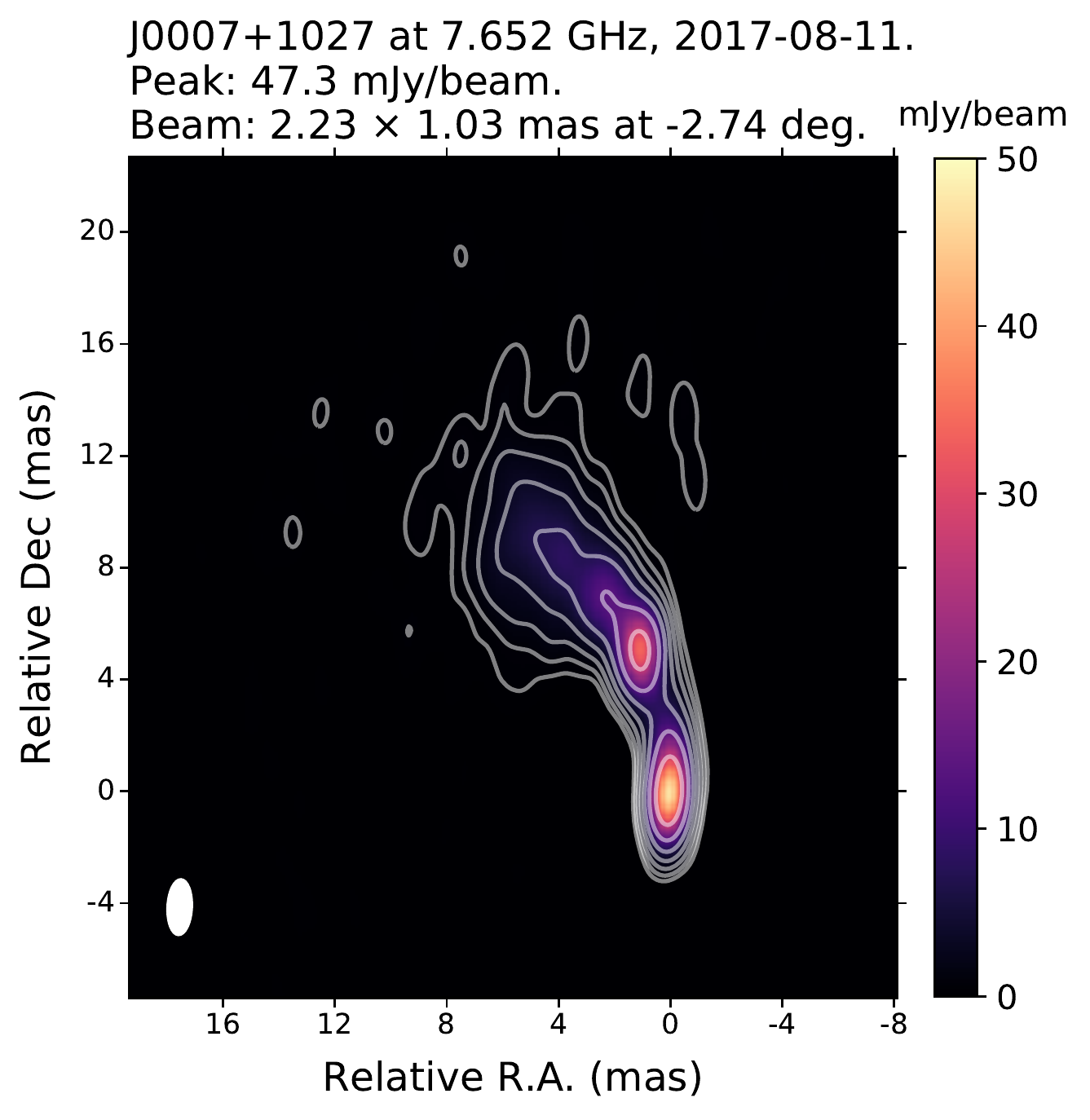}
        \label{}
    }
    \subfigure[]
    {
        \includegraphics[width=0.4\linewidth]{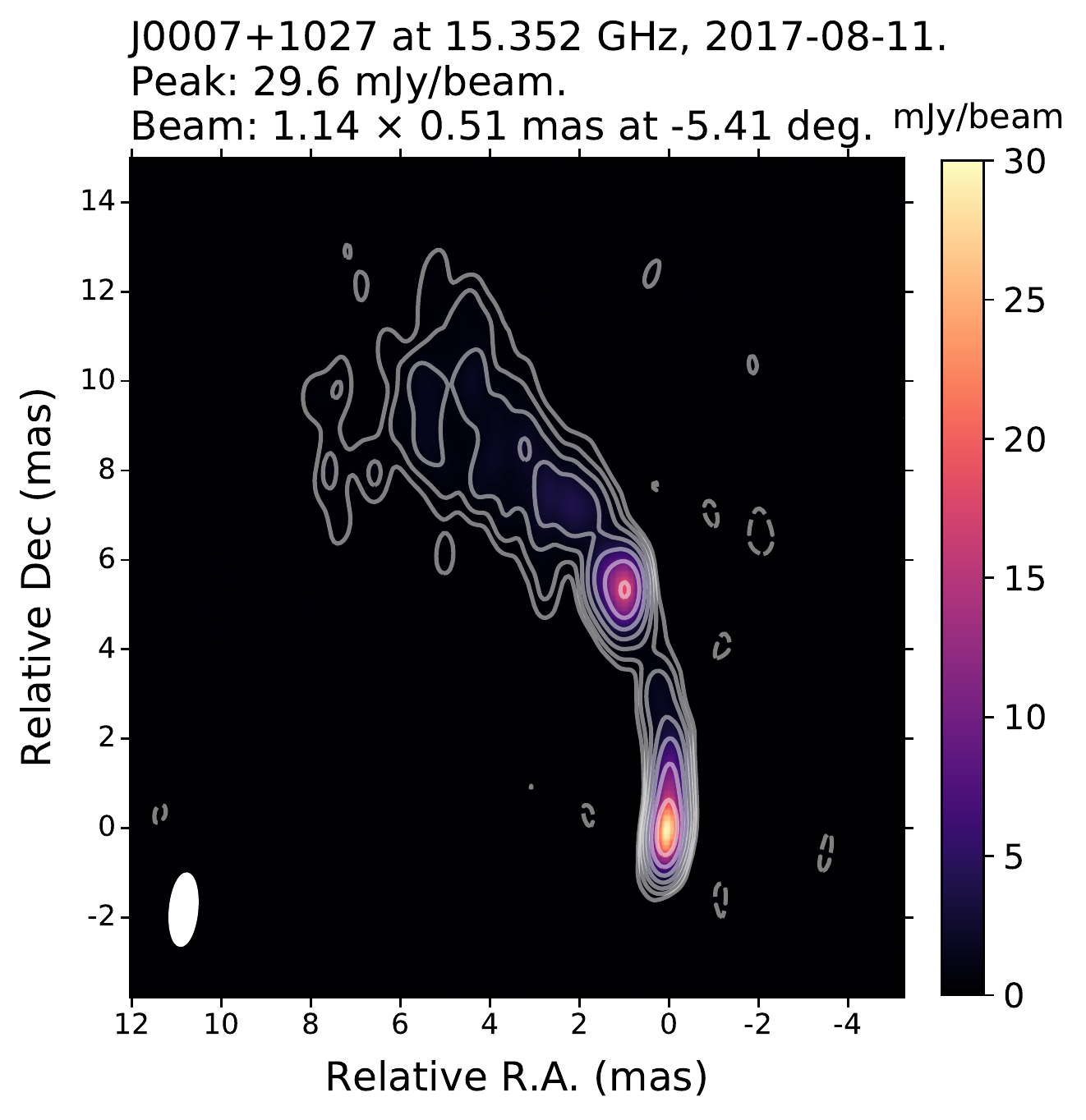}
        \label{}
    }
     \subfigure[]
    {
        \includegraphics[width=0.4\linewidth]{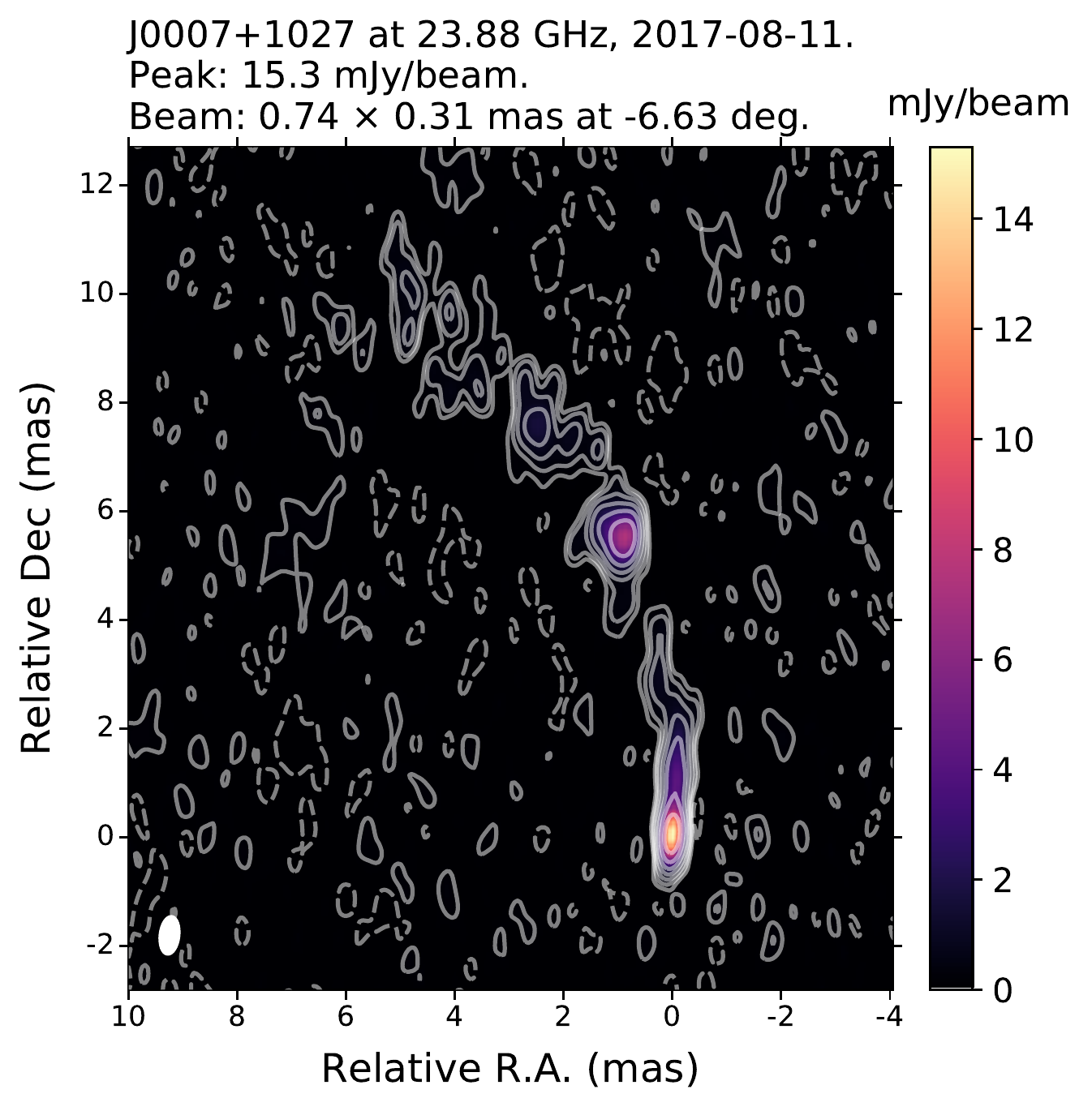}
        \label{}
    } 

    \caption{Self-calibrated images of J0007+1027. The images were obtained at a) 4.148\,GHz, b) 7.652\,GHz, c) 15.352\,GHz and d) 23.88\,GHz. The rms noise level from the lower to the higher frequency is 0.2, 0.1, 0.1, 0.09 mJy/beam. The interferometric beam (ellipse) is displayed on the bottom-left corner of each image. Contours represent -0.9\%, 0.9\%, 1.8\%, 3.6\%, 7.2\%, 14.4\%, 28.8\%, and 57.6\% of the peak intensity at each image.}
    \label{fB2} 
\end{figure*}

\begin{figure*}[t]
\centering
   \subfigure[]
   {
        \includegraphics[width=0.4\linewidth]{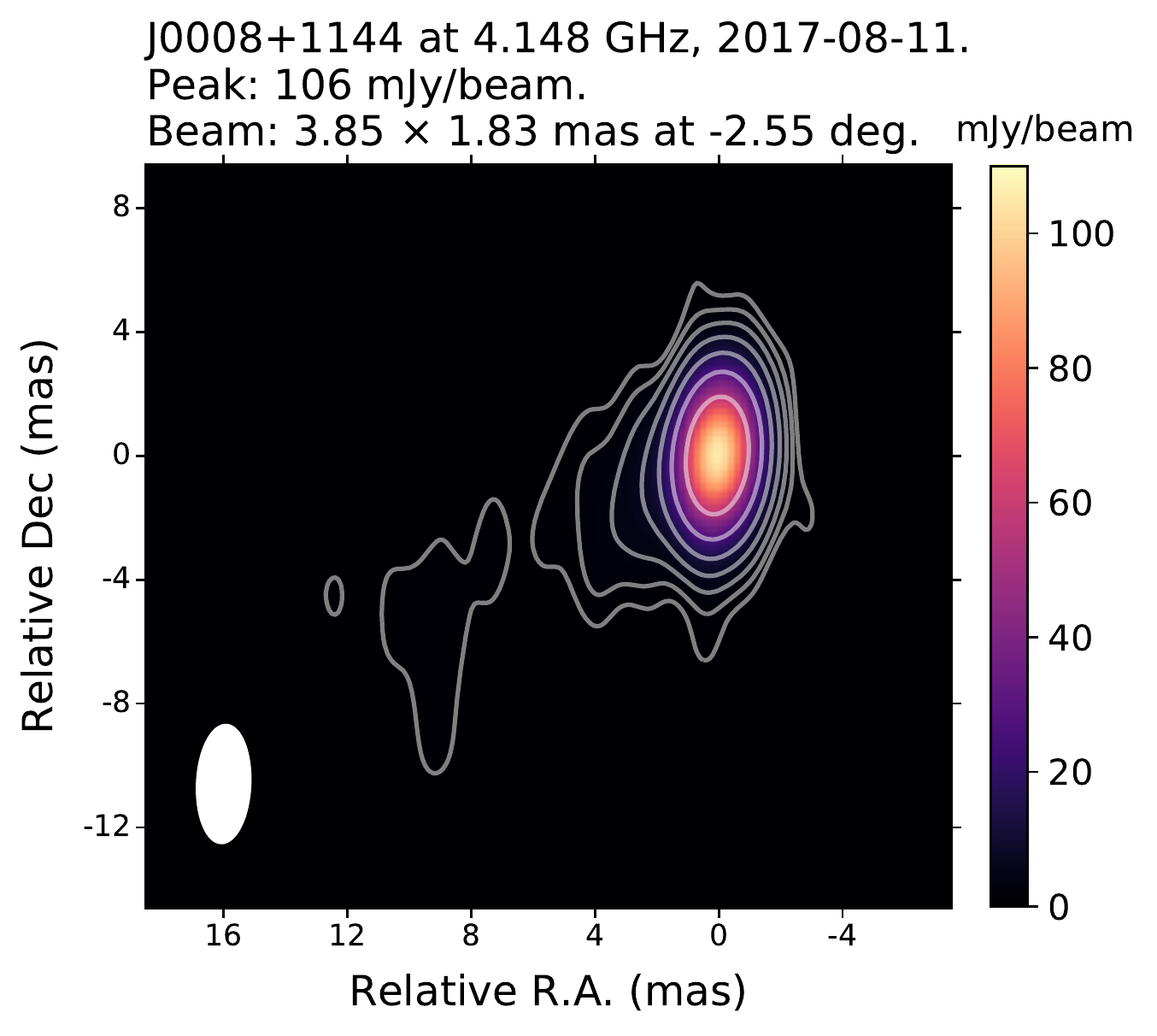}
        \label{}
    }
     \subfigure[]
    {
        \includegraphics[width=0.4\linewidth]{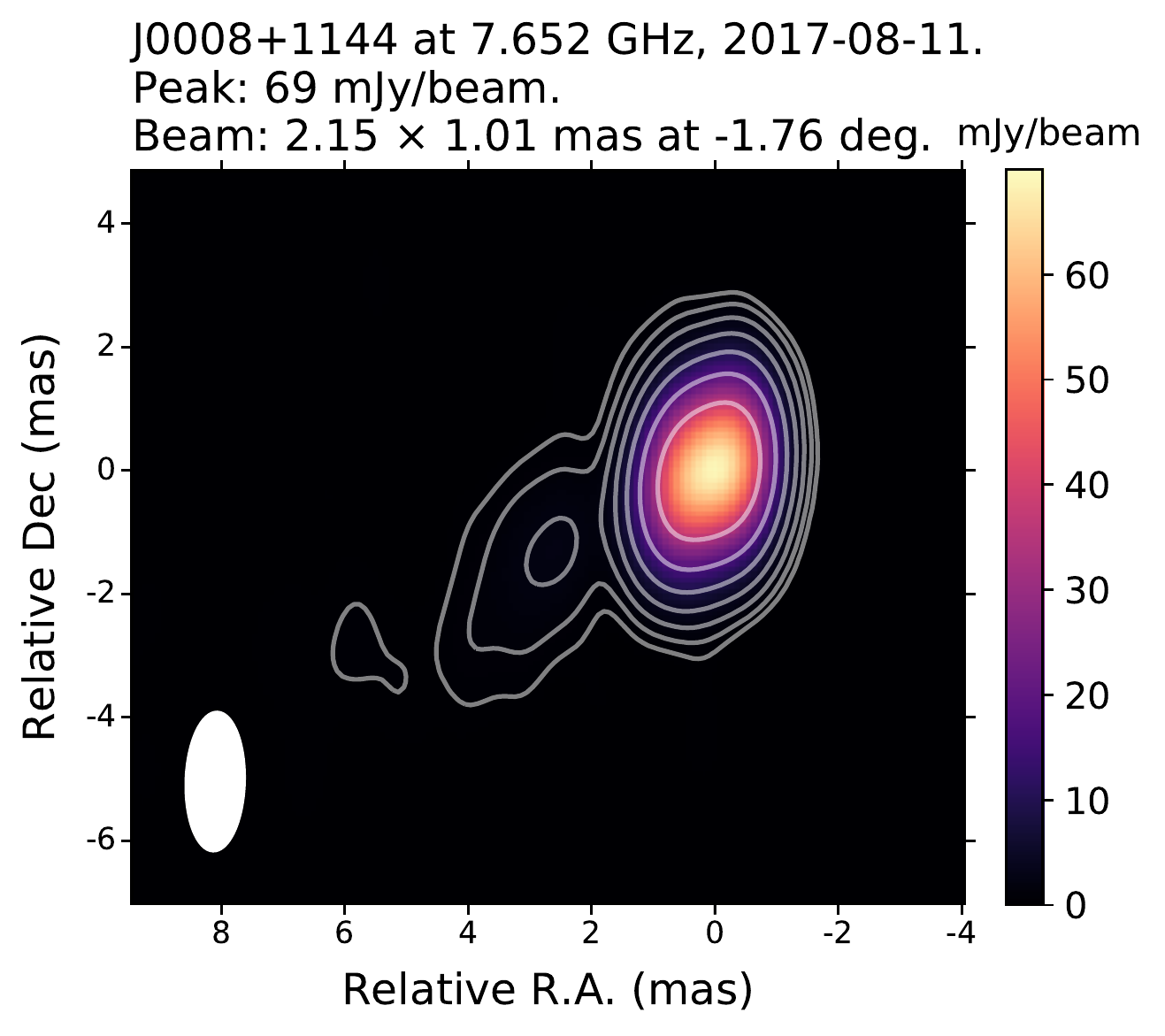}
        \label{}
    }
    \subfigure[]
    {
        \includegraphics[width=0.4\linewidth]{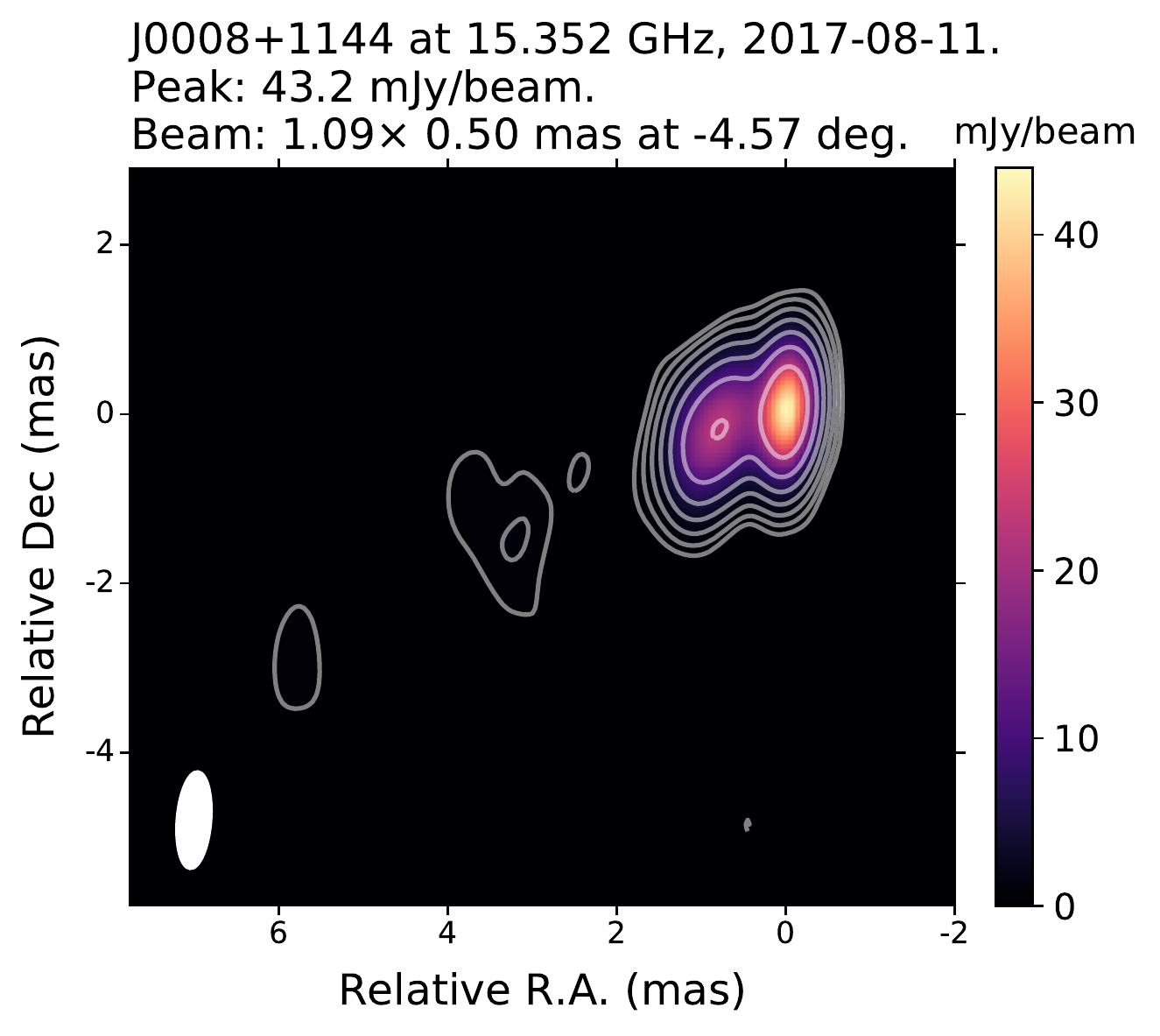}
        \label{}
    }
     \subfigure[]
    {
        \includegraphics[width=0.4\linewidth]{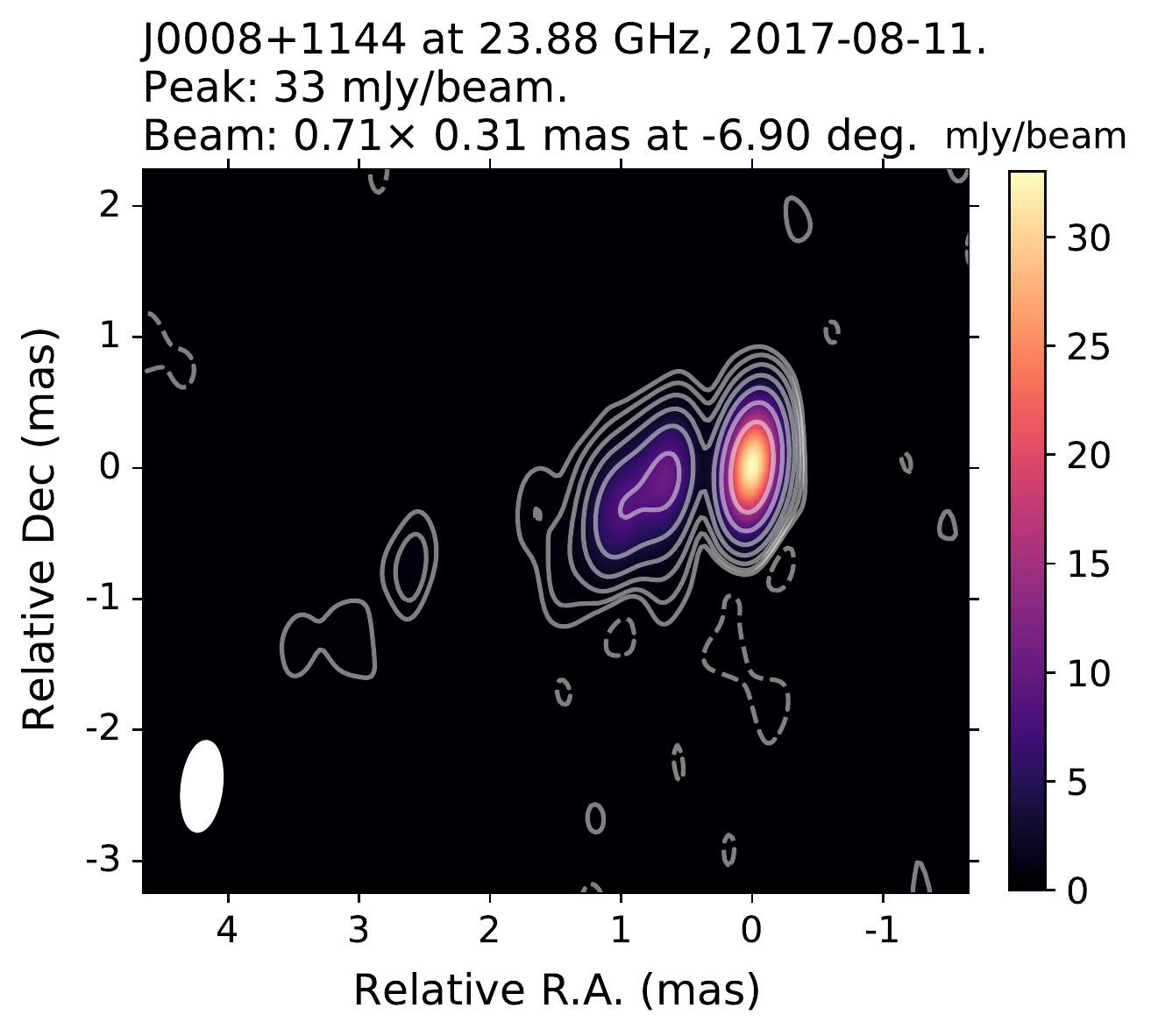}
        \label{}
    } 

    \caption{Self-calibrated images of J0008+1144. The images were obtained at a) 4.148\,GHz, b) 7.652\,GHz, c) 15.352\,GHz and d) 23.88\,GHz. The rms noise level from the lower to the higher frequency is 0.2, 0.1, 0.1, 0.1   mJy/beam. The interferometric beam (ellipse) is displayed on the bottom-left corner of each image. Contours represent -0.8\%, 0.8\%, 1.6\%, 3.2\%, 6.4\%, 12.8\%, 25.6\%, and 51.2\% of the peak intensity at each image.} 
    \label{fB3} 
\end{figure*}

\clearpage
\twocolumn
\section{Core spectrum of the calibrators}

The correlated flux density of the calibrators as a function of frequency is shown in Figures~C.1, C.2 and C.3. The black filled circles represent the core for all cases; the blue curves are the fits. The spectrum of J0006 is fitted with a function of the form $S_\nu \propto \nu^{i}(1-e^{-\nu^{j-i}})$. 
On the other hand, the spectra of J0007 and J0008 are fitted with a function of the form $S_\nu \propto \nu^{i}$. Both the core and the bright  component 5\,mas to North of the core in J0007 are fully optically thin regions. The spectrum of J0008 appears to be composed of both a rising and a declining part. Further observations are needed to determine the spectral index of the rising side for both J0006 and J0008.

\begin{figure}[h]
      \includegraphics[width=0.42\textwidth]{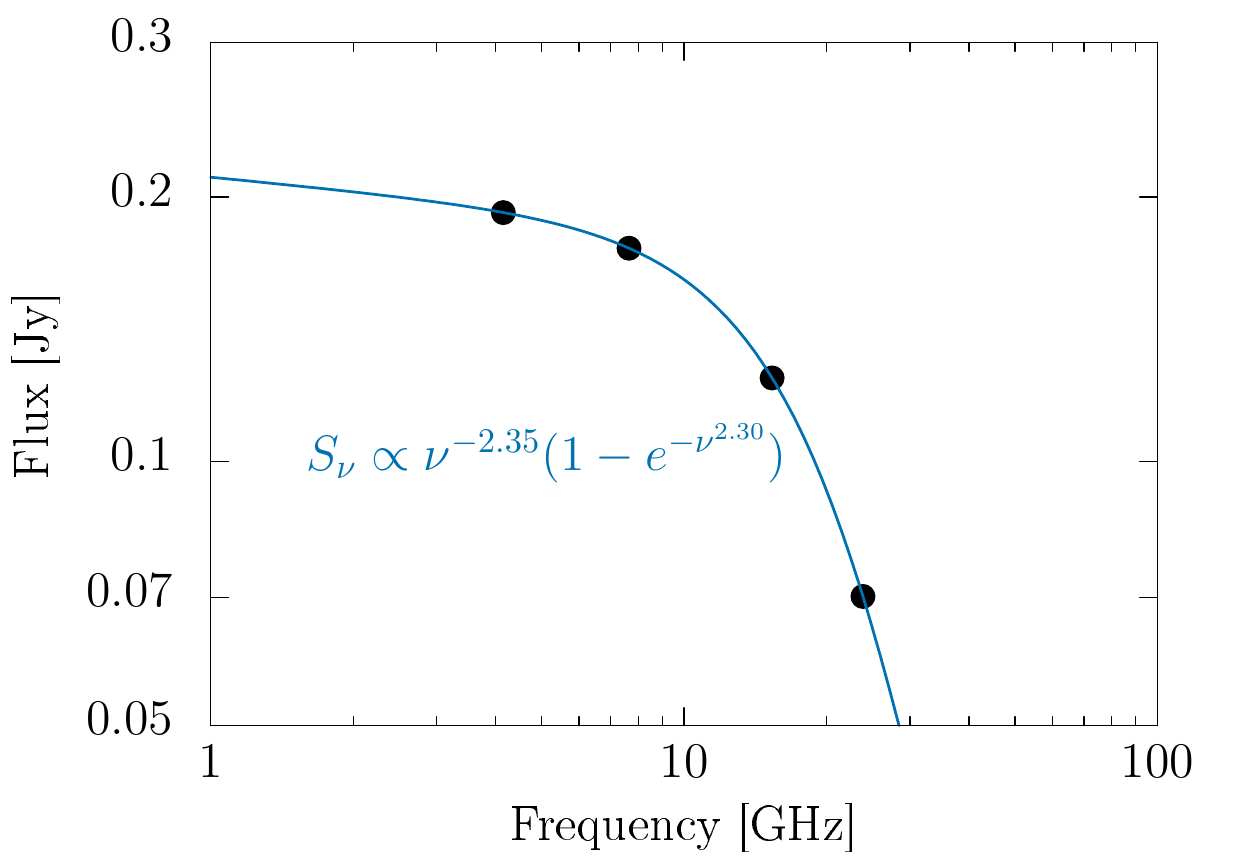}
        \caption{Core spectrum of J0006.}
    \label{fC1}
\end{figure}

\begin{figure}[h]
      \includegraphics[width=0.42\textwidth]{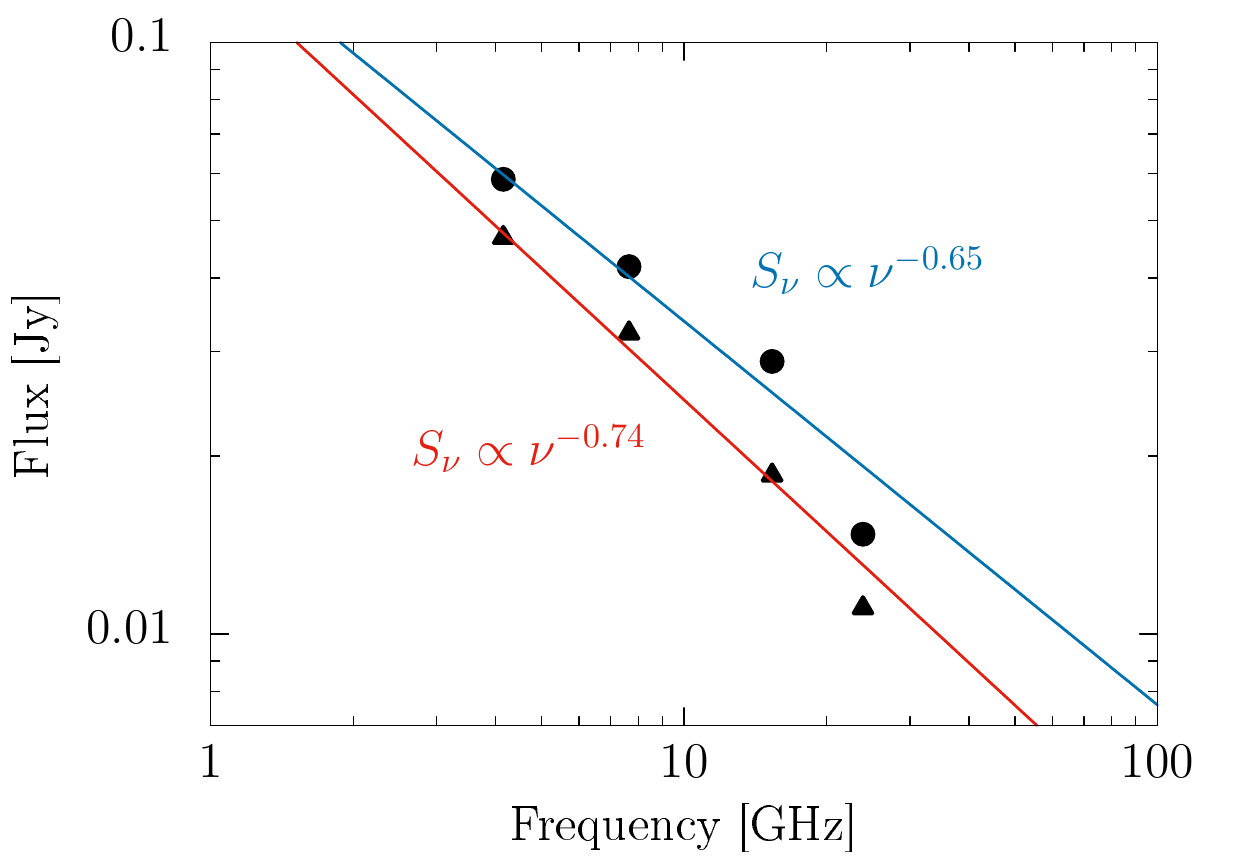}
        \caption{Core spectrum of J0007. The black triangle represents the bright knot 5\,mas to North from the core. }
    \label{fC2}
\end{figure}

\begin{figure}[h]
      \includegraphics[width=0.42\textwidth]{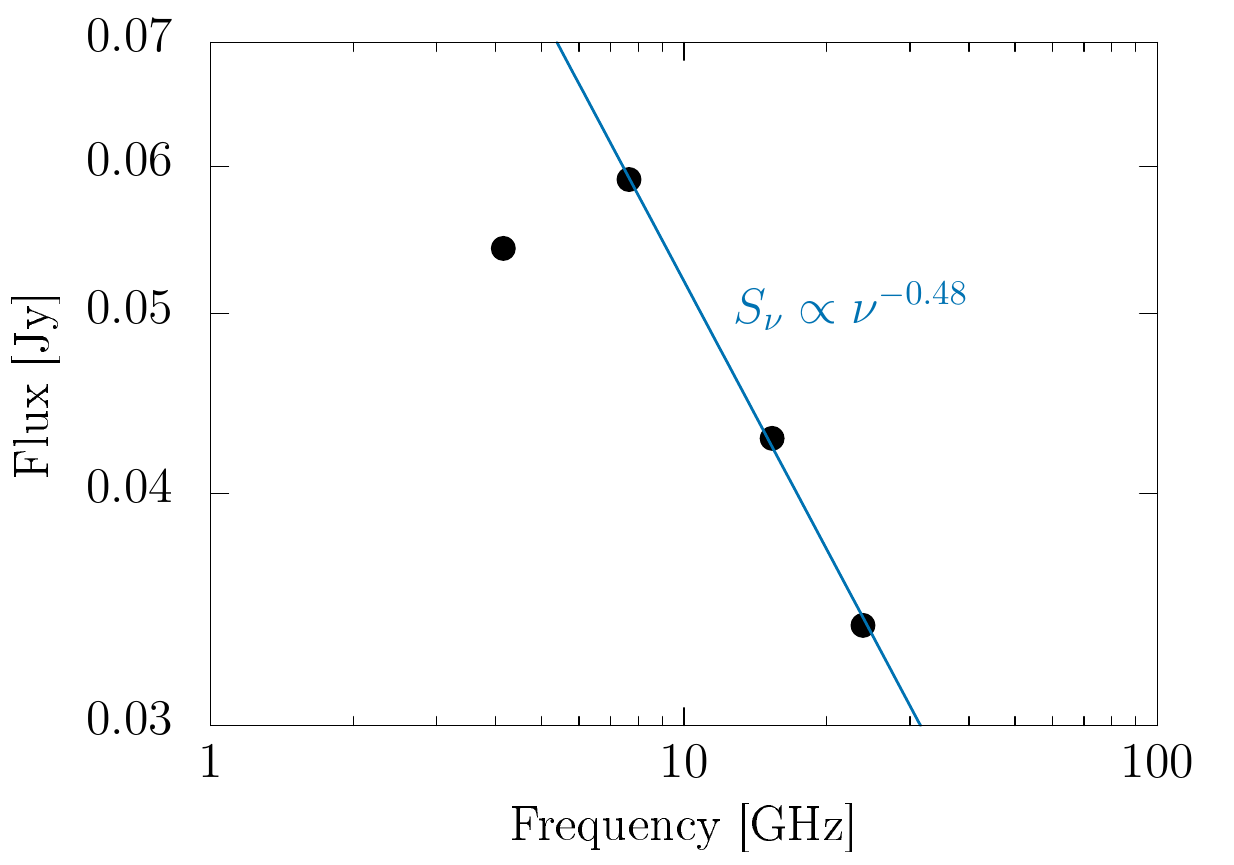}
        \caption{Core spectrum of J0008.}
    \label{fC3}
\end{figure}

\clearpage
\onecolumn

\section{Self- and phase-referencing errors} \label{app:errors}
\subsection{Self-referencing core-shift errors}

\begin{table*}[h]
\centering
\caption{Image shift errors}
\label{table:2D_err}
\begin{tabular}{ccccccc}
\hline \noalign{\smallskip}
\multirow{2}{*}{Frequency} & \multicolumn{2}{c}{J0006} & \multicolumn{2}{c}{J0007} & \multicolumn{2}{c}{J0008} \\
 & \begin{tabular}[c]{@{}c@{}}R.A.\\ (mas)\end{tabular} & \begin{tabular}[c]{@{}c@{}}Dec\\ (mas)\end{tabular} & \begin{tabular}[c]{@{}c@{}}R.A.\\ (mas)\end{tabular} & \begin{tabular}[c]{@{}c@{}}Dec\\ (mas)\end{tabular} & \begin{tabular}[c]{@{}c@{}}R.A.\\ (mas)\end{tabular} & \begin{tabular}[c]{@{}c@{}}Dec\\ (mas)\end{tabular} \\ \hline \noalign{\smallskip}
C-X & 0.029 & 0.026 & 0.026 & 0.026 & 0.031 & 0.028 \\
X-U & 0.014 & 0.013 & 0.013 & 0.014 & 0.015 & 0.014 \\
U-K & 0.008 & 0.008 & 0.008 & 0.008 & 0.009 & 0.009 \\ \hline
\end{tabular}
\end{table*}

\begin{table*}[h]
\centering
\caption{Core position errors.}
\label{table:core_pos_err1}
\begin{tabular}{ccccccc}
\hline \noalign{\smallskip}
\multirow{2}{*}{Frequency} & \multicolumn{2}{c}{J0006} & \multicolumn{2}{c}{J0007} & \multicolumn{2}{c}{J0008} \\
 & \begin{tabular}[c]{@{}c@{}}R.A.\\ (mas)\end{tabular} & \begin{tabular}[c]{@{}c@{}}Dec\\ (mas)\end{tabular} & \begin{tabular}[c]{@{}c@{}}R.A.\\ (mas)\end{tabular} & \begin{tabular}[c]{@{}c@{}}Dec\\ (mas)\end{tabular} & \begin{tabular}[c]{@{}c@{}}R.A.\\ (mas)\end{tabular} & \begin{tabular}[c]{@{}c@{}}Dec\\ (mas)\end{tabular} \\ \hline \noalign{\smallskip}
C & 0.003 & 0.006  & 0.017 & 0.028 & 0.075 &0.045  \\
X & 0.010 & 0.006  & 0.015 & 0.015 & 0.009 &0.009  \\
U & 0.006 & 0.003  & 0.006 & 0.012 & 0.003 &0.006  \\
K & 0.005 & 0.005  & 0.003 & 0.004 & 0.002 &0.005  \\ \hline
\end{tabular}
\end{table*}

\begin{table*}[h]
\centering
\caption{Core identification errors.}
\label{table:core_id_err1}
\begin{tabular}{ccccccc}
\hline \noalign{\smallskip}
\multirow{2}{*}{Frequency} & \multicolumn{2}{c}{J0006} & \multicolumn{2}{c}{J0007} & \multicolumn{2}{c}{J0008} \\
 & \begin{tabular}[c]{@{}c@{}}R.A.\\ (mas)\end{tabular} & \begin{tabular}[c]{@{}c@{}}Dec\\ (mas)\end{tabular} & \begin{tabular}[c]{@{}c@{}}R.A.\\ (mas)\end{tabular} & \begin{tabular}[c]{@{}c@{}}Dec\\ (mas)\end{tabular} & \begin{tabular}[c]{@{}c@{}}R.A.\\ (mas)\end{tabular} & \begin{tabular}[c]{@{}c@{}}Dec\\ (mas)\end{tabular} \\ \hline \noalign{\smallskip}
C & 0.044 & 0.037 & 0.016 & 0.042 & 0.050 & 0.011 \\
X & 0.012 & 0.014 & 0.009 & 0.022 & 0.021 & 0.019 \\
U & 0.015 & 0.017 & 0.000 & 0.004 & 0.010 & 0.007 \\
K & 0.002 & 0.014 & 0.006 & 0.005 & 0.002 & 0.004 \\ \hline
\end{tabular}
\end{table*}

\subsection{Phase-referencing core-shift errors}

\begin{table*}[h]
\centering
\caption{Thermal errors.}
\label{table:thermal_err}
\begin{tabular}{ccccccc}
\hline \noalign{\smallskip}
\multirow{2}{*}{Frequency} & \multicolumn{2}{c}{J0006} & \multicolumn{2}{c}{J0007} & \multicolumn{2}{c}{J0008} \\
 & \begin{tabular}[c]{@{}c@{}}R.A.\\ (mas)\end{tabular} & \begin{tabular}[c]{@{}c@{}}Dec\\ (mas)\end{tabular} & \begin{tabular}[c]{@{}c@{}}R.A.\\ (mas)\end{tabular} & \begin{tabular}[c]{@{}c@{}}Dec\\ (mas)\end{tabular} & \begin{tabular}[c]{@{}c@{}}R.A.\\ (mas)\end{tabular} & \begin{tabular}[c]{@{}c@{}}Dec\\ (mas)\end{tabular} \\ \hline \noalign{\smallskip}
C & 0.012 & 0.025 & 0.016 & 0.036 & 0.011 & 0.023 \\
X & 0.005 & 0.011 & 0.018 & 0.039 & 0.013 & 0.028 \\
U & 0.003 & 0.007 & 0.005 & 0.011 & 0.004 & 0.009 \\
K & 0.003 & 0.006 & 0.003 & 0.007 & 0.002 & 0.005 \\ \hline
\end{tabular}
\end{table*}

\begin{table*}[h]
\centering
\caption{Core identification errors. }
\label{table:core_id_err2}
\begin{tabular}{ccccccc}
\hline \noalign{\smallskip}
\multirow{2}{*}{Frequency} & \multicolumn{2}{c}{J0006} & \multicolumn{2}{c}{J0007} & \multicolumn{2}{c}{J0008} \\
 & \begin{tabular}[c]{@{}c@{}}R.A.\\ (mas)\end{tabular} & \begin{tabular}[c]{@{}c@{}}Dec\\ (mas)\end{tabular} & \begin{tabular}[c]{@{}c@{}}R.A.\\ (mas)\end{tabular} & \begin{tabular}[c]{@{}c@{}}Dec\\ (mas)\end{tabular} & \begin{tabular}[c]{@{}c@{}}R.A.\\ (mas)\end{tabular} & \begin{tabular}[c]{@{}c@{}}Dec\\ (mas)\end{tabular} \\ \hline \noalign{\smallskip}
C & 0.014 & 0.013 & 0.003 & 0.101 & 0.070 & 0.019 \\
X & 0.007 & 0.009 & 0.017 & 0.004 & 0.052 & 0.060 \\
U & 0.008 & 0.019 & 0.002 & 0.016 & 0.019 & 0.003 \\
K & 0.008 & 0.020 & 0.017 & 0.003 & 0.006 & 0.016 \\ \hline
\end{tabular}
\end{table*}

\begin{table*}[h]
\centering
\caption{Estimated errors for the time-variable part of the tropospheric delay gradient between the target and phase reference calibrator. The values are given per source and per frequency pair. Since C and X band observations were simultaneous, the non-dispersive tropospheric delay error vanishes when a difference between the frequencies is taken; hence, C-X error is zero.}
\label{table:trop_err}
\begin{tabular}{ccccccc}
\hline \noalign{\smallskip}
\multirow{2}{*}{Frequency} & \multicolumn{2}{c}{J0006} & \multicolumn{2}{c}{J0007} & \multicolumn{2}{c}{J0008} \\
 & \begin{tabular}[c]{@{}c@{}}R.A.\\ (mas)\end{tabular} & \begin{tabular}[c]{@{}c@{}}Dec\\ (mas)\end{tabular} & \begin{tabular}[c]{@{}c@{}}R.A.\\ (mas)\end{tabular} & \begin{tabular}[c]{@{}c@{}}Dec\\ (mas)\end{tabular} & \begin{tabular}[c]{@{}c@{}}R.A.\\ (mas)\end{tabular} & \begin{tabular}[c]{@{}c@{}}Dec\\ (mas)\end{tabular} \\ \hline \noalign{\smallskip}
C-X & 0 & 0 & 0 & 0 & 0 & 0 \\
X-U & 0.004 & 0.007 & 0.002 & 0.003 & 0.003 & 0.004 \\
U-K & 0.010 & 0.016 & 0.004 & 0.007 & 0.005 & 0.008 \\ \hline
\end{tabular}
\end{table*}

\begin{table*}[h]
\centering
\caption{Estimated errors due to residual ionospheric delay gradient between the target source and phase reference calibrator. Values are given per source and per frequency.}
\label{table:ion_err}
\begin{tabular}{ccccccc}
\hline \noalign{\smallskip}
\multirow{2}{*}{Frequency} & \multicolumn{2}{c}{J0006} & \multicolumn{2}{c}{J0007} & \multicolumn{2}{c}{J0008} \\
 & \begin{tabular}[c]{@{}c@{}}R.A.\\ (mas)\end{tabular} & \begin{tabular}[c]{@{}c@{}}Dec\\ (mas)\end{tabular} & \begin{tabular}[c]{@{}c@{}}R.A.\\ (mas)\end{tabular} & \begin{tabular}[c]{@{}c@{}}Dec\\ (mas)\end{tabular} & \begin{tabular}[c]{@{}c@{}}R.A.\\ (mas)\end{tabular} & \begin{tabular}[c]{@{}c@{}}Dec\\ (mas)\end{tabular} \\ \hline \noalign{\smallskip}
C & 0.038 & 0.082 & 0.016 & 0.035 & 0.020 & 0.042 \\
X & 0.012 & 0.025 & 0.005 & 0.011 & 0.006 & 0.013 \\
U & 0.003 & 0.006 & 0.001 & 0.003 & 0.001 & 0.003 \\ K & 0.001 & 0.003 & 0.001 & 0.001 & 0.001 & 0.001 \\ \hline
\end{tabular}
\end{table*}

\clearpage

\section{Spectral index maps after alignment}

\begin{figure*}[h]
\centering
    \subfigure[]
    {
\includegraphics[width=0.44\textwidth]{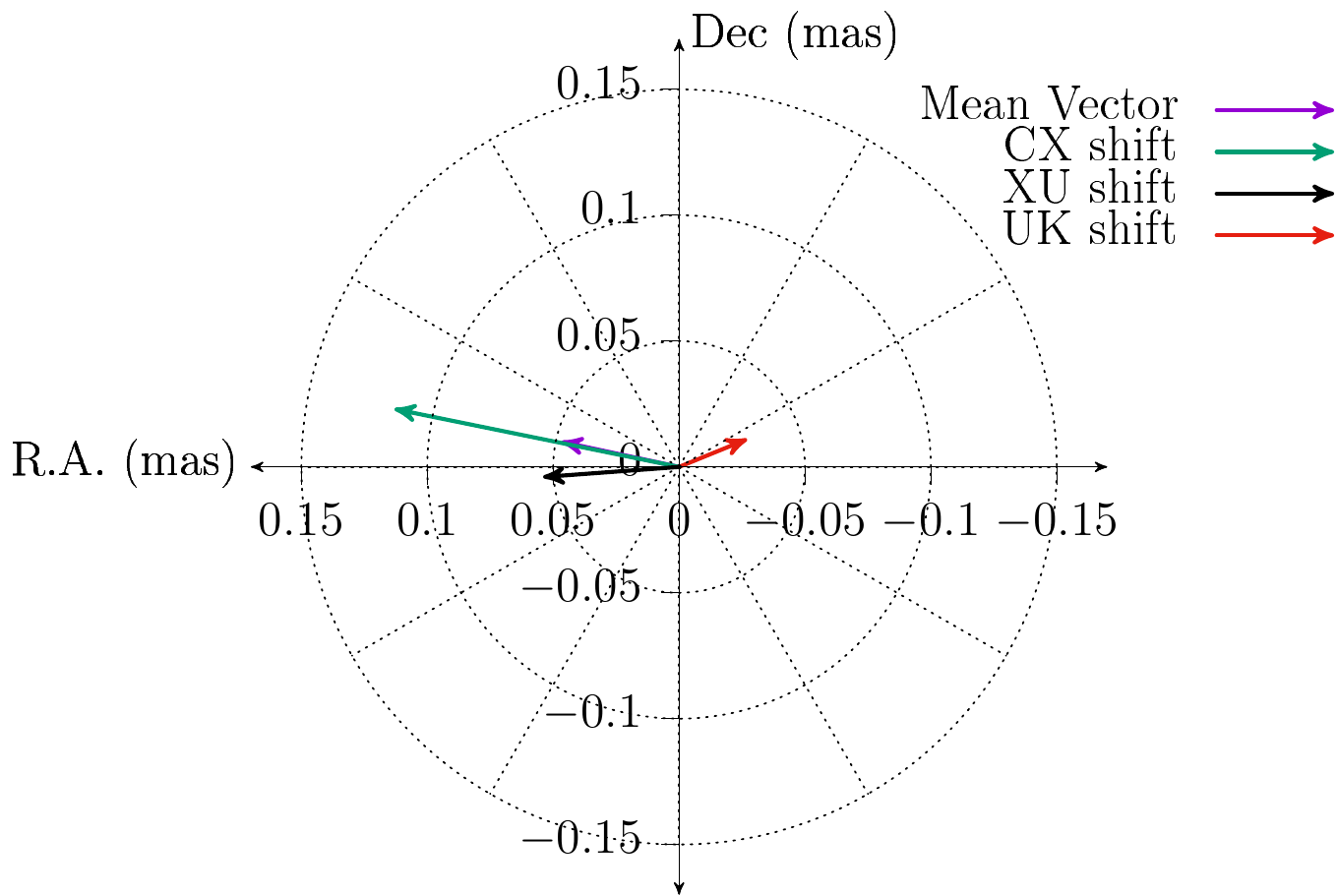}
        \label{}
    }
    \hspace{1cm}
    \subfigure[]
    {
        \includegraphics[width=0.44\textwidth]{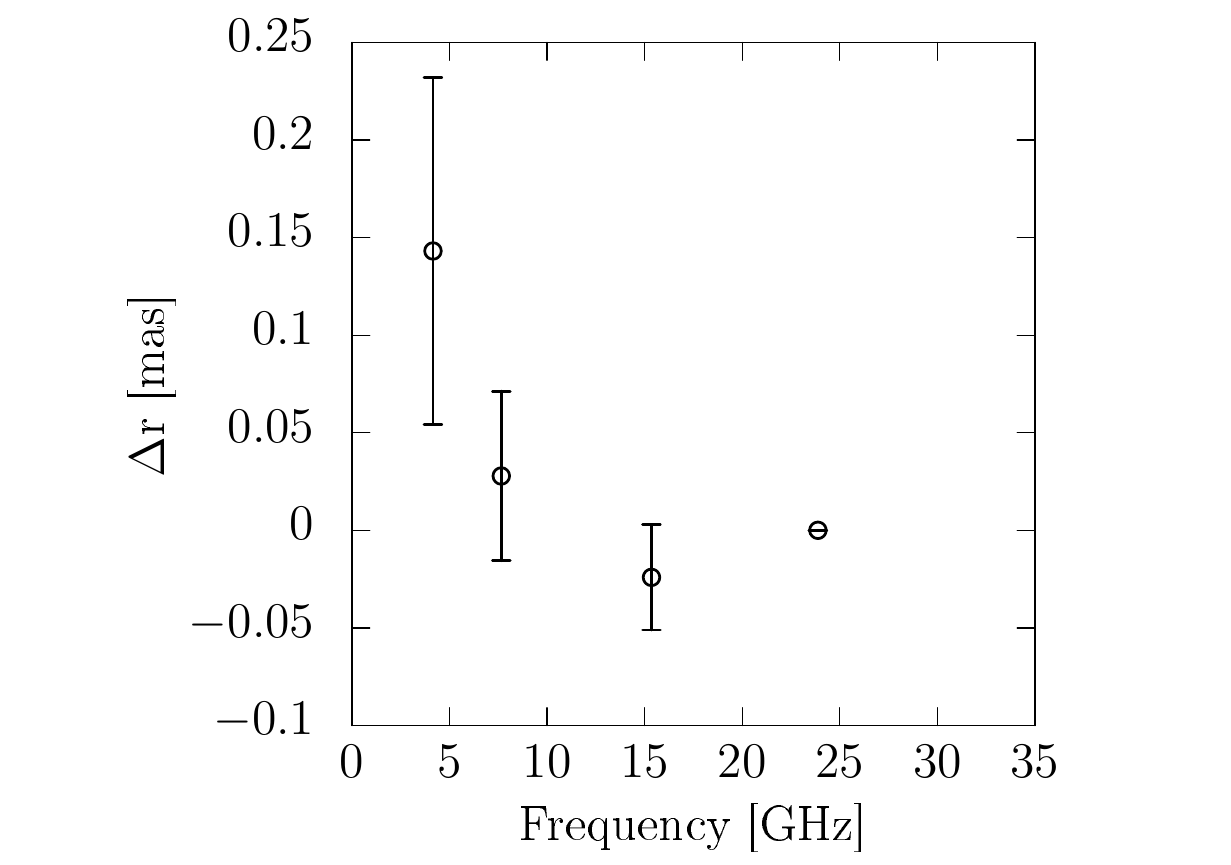}
        \label{}
    }
    \caption{(a)  Core-shift vectors of III\,Zw\,2 with calibrator J0006, (b) Core-shift values with K-band as the reference frequency.}
    \label{fF1}
\end{figure*}

\begin{figure*}[h]
\centering
    \subfigure[]
    {
\includegraphics[width=0.44\textwidth]{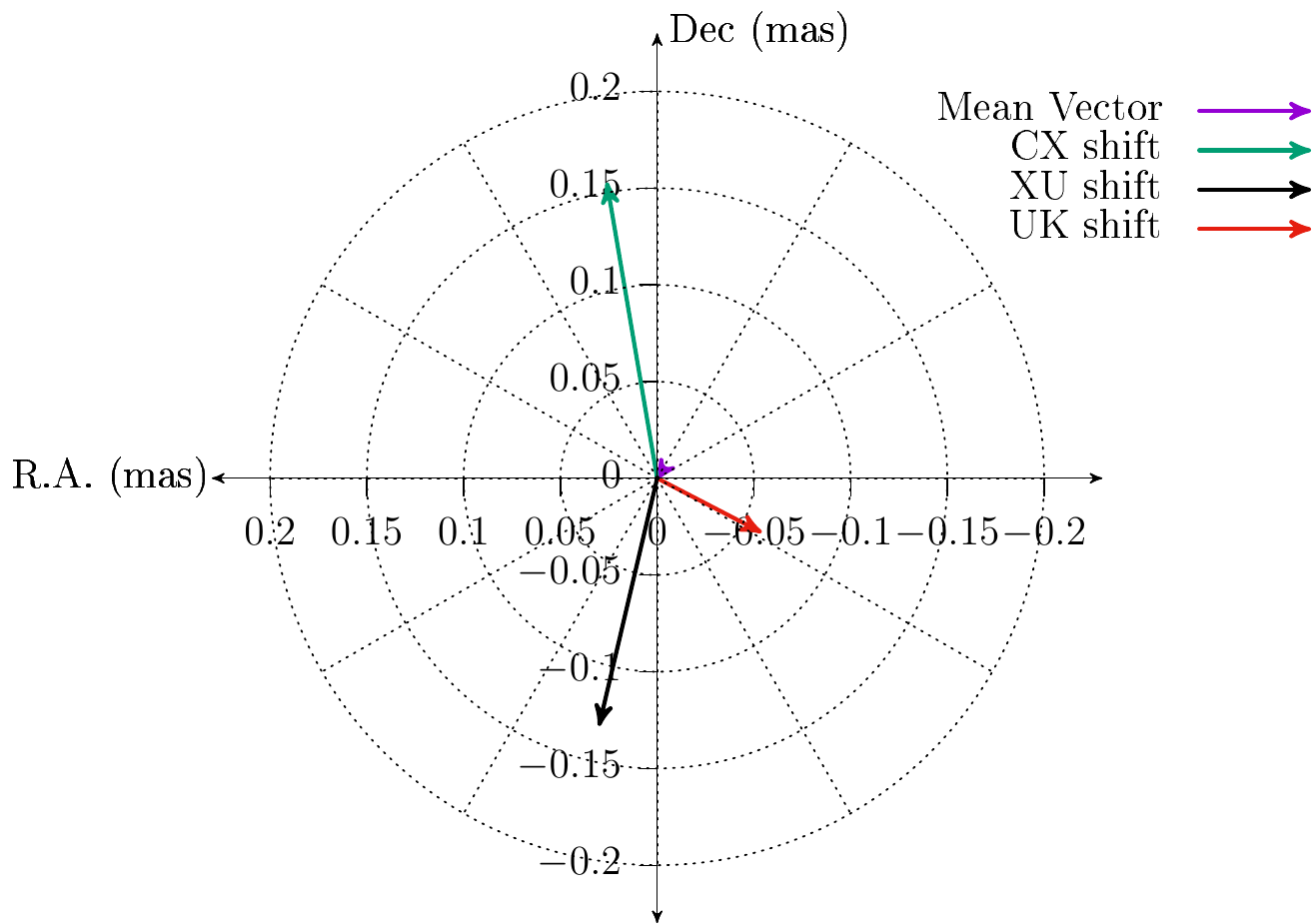}
        \label{}
    }
    \hspace{1cm}
    \subfigure[]
    {
        \includegraphics[width=0.44\textwidth]{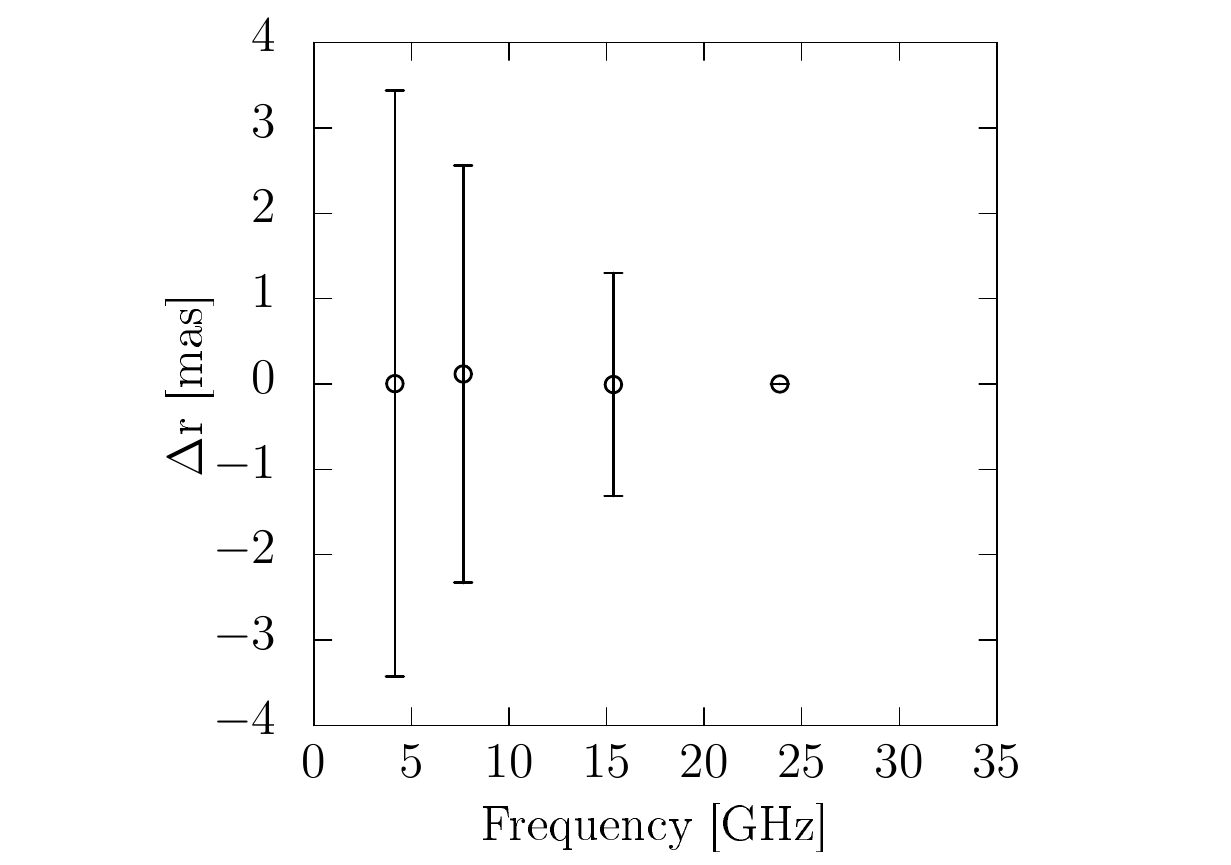}
        \label{}
    }
    \caption{(a) Core-shift vectors of III\,Zw\,2 with calibrator J0007, (b) Core-shift values with K-band as the reference frequency. }
    \label{fF2}
\end{figure*}

\begin{figure*}[h]
\centering
    \subfigure[]
    {
\includegraphics[width=0.44\textwidth]{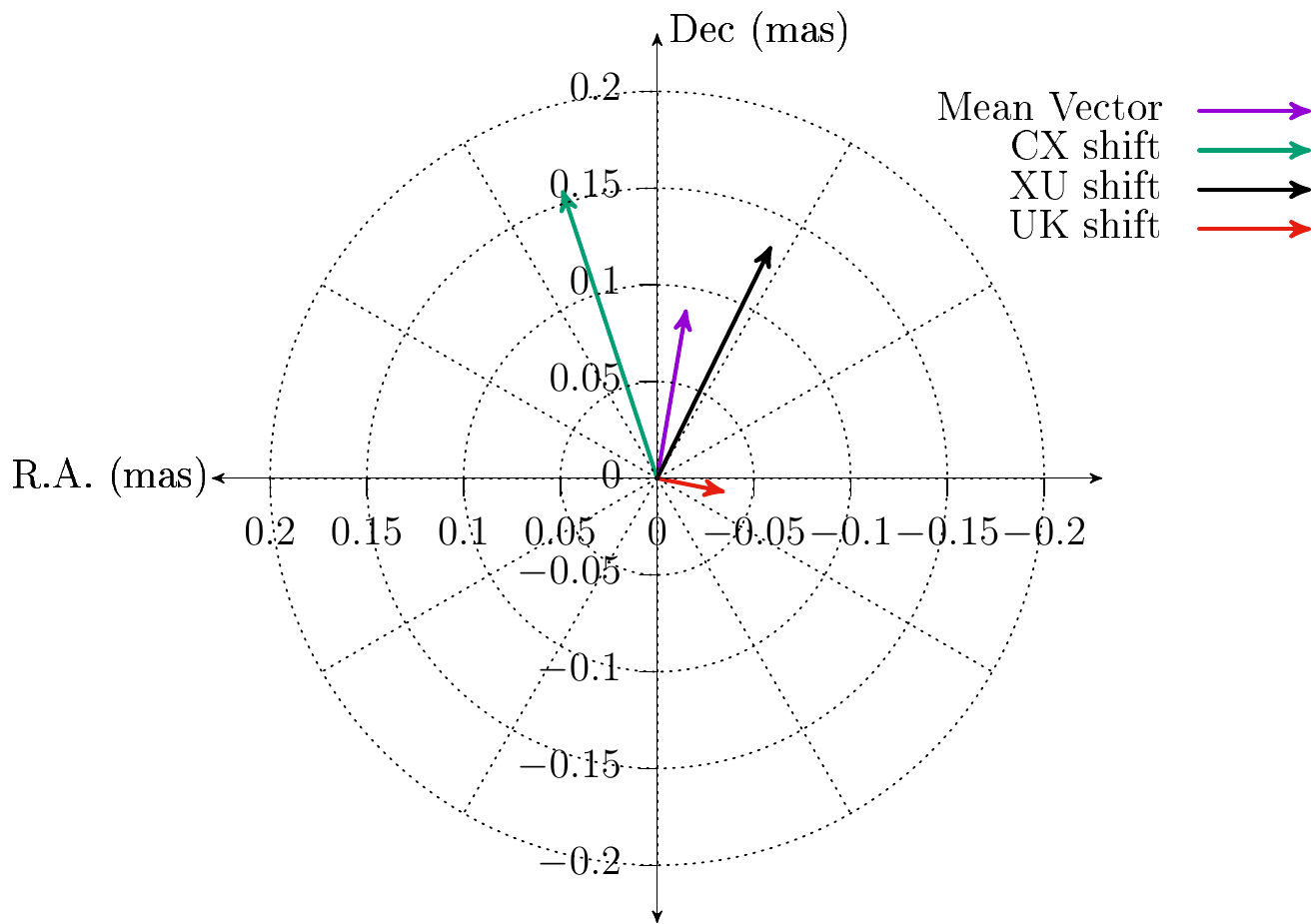}
        \label{}
    }
    \hspace{1cm}
    \subfigure[]
    {
        \includegraphics[width=0.44\textwidth]{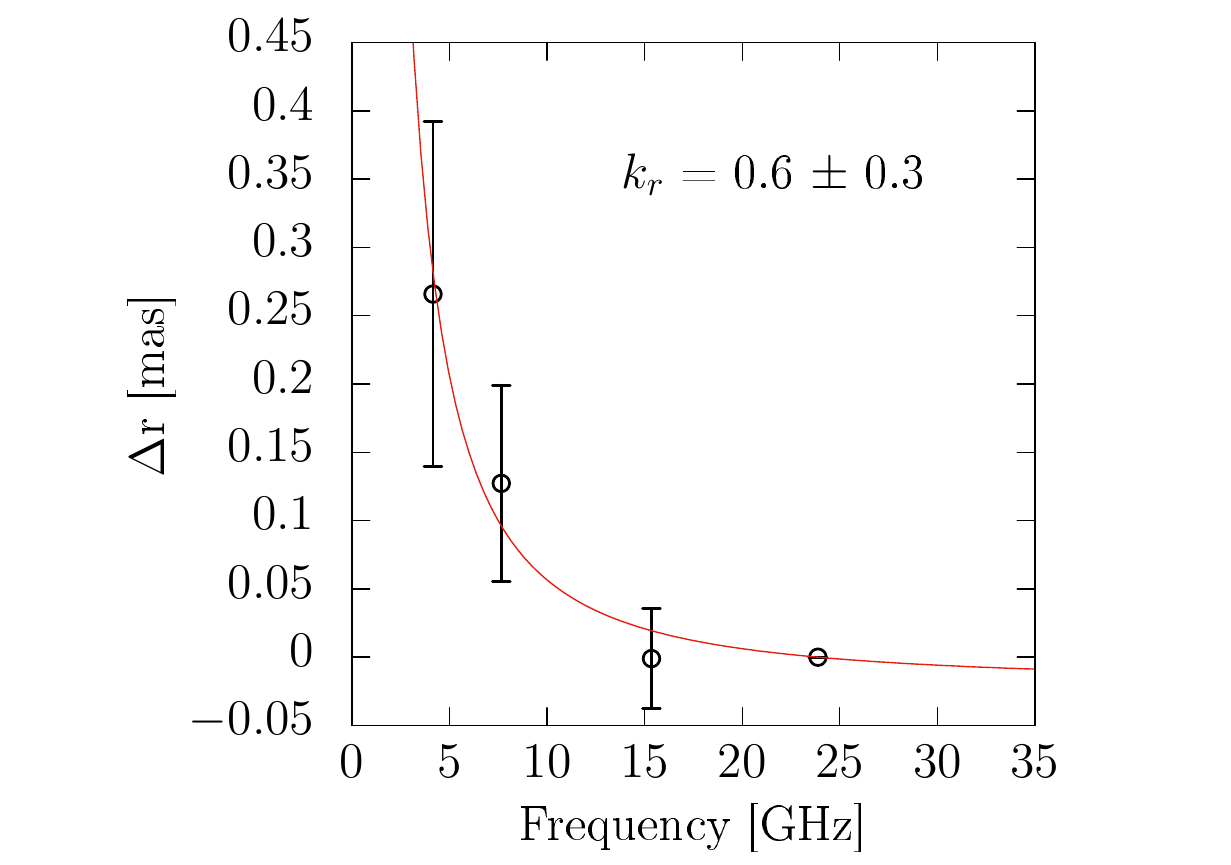}
        \label{}
    }
    \caption{(a) Core-shift vectors of III\,Zw\,2 with calibrator J0008, (b) Core-shift values with K-band as the reference frequency. The red curve indicates the best fit of the form: $\Delta r = a(\nu_{GHz}^{-1/k_r} - 23.88^{-1/k_r}$). The fitting parameters are $a = 2.8 \pm 2.9$ and $k_r = 0.6 \pm 0.3$.}
    \label{fF3}
\end{figure*}

\begin{itemize}
    \item J0006+1235: smooth spectral index gradients along the jet are seen in all frequency pairs (Figure~\ref{fF1}). The core is located near the optically thick region for the 4.15$-$7.65\,GHz section with highest spectral index in the most upstream region. At higher frequencies the core becomes first flat and then slightly optically thin. This result agrees with the core spectrum index shown in Figure~\ref{fC1}
    \item J0007+1027: interestingly the core is optically thin in all images (Figure~\ref{fF2}). This result agrees with the steep spectrum seen in both the core and the northern knot in Figure~ \ref{fC2}.
    \item J0008+1144: smooth spectral index gradients along the jet are detected in all frequency pairs (Figure~\ref{fF3}). The core is found to be the optically thick region only for the 4.15$-$7.65\,GHz section. At higher frequencies, the maps show flat or optically thin spectral index. These results also agree with the core spectrum shown in Figure~\ref{fC3}, which comprises of a rising (C-X bands) and a declining part (X-K bands).
\end{itemize}

\begin{figure*}[h]
\centering
   \subfigure[]
   {
        \includegraphics[width=0.45\textwidth]{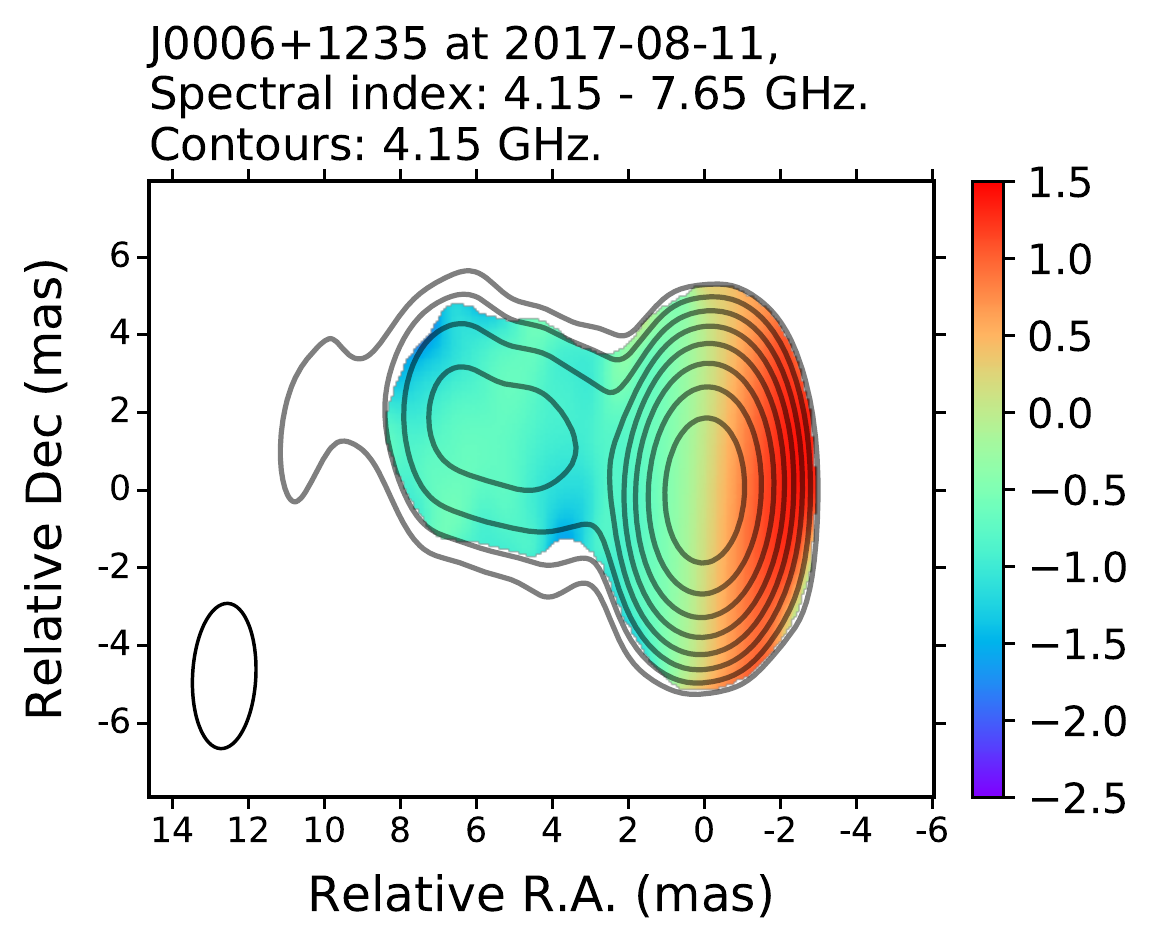}
        \label{}
    }
     \subfigure[]
    {
        \includegraphics[width=0.4\textwidth]{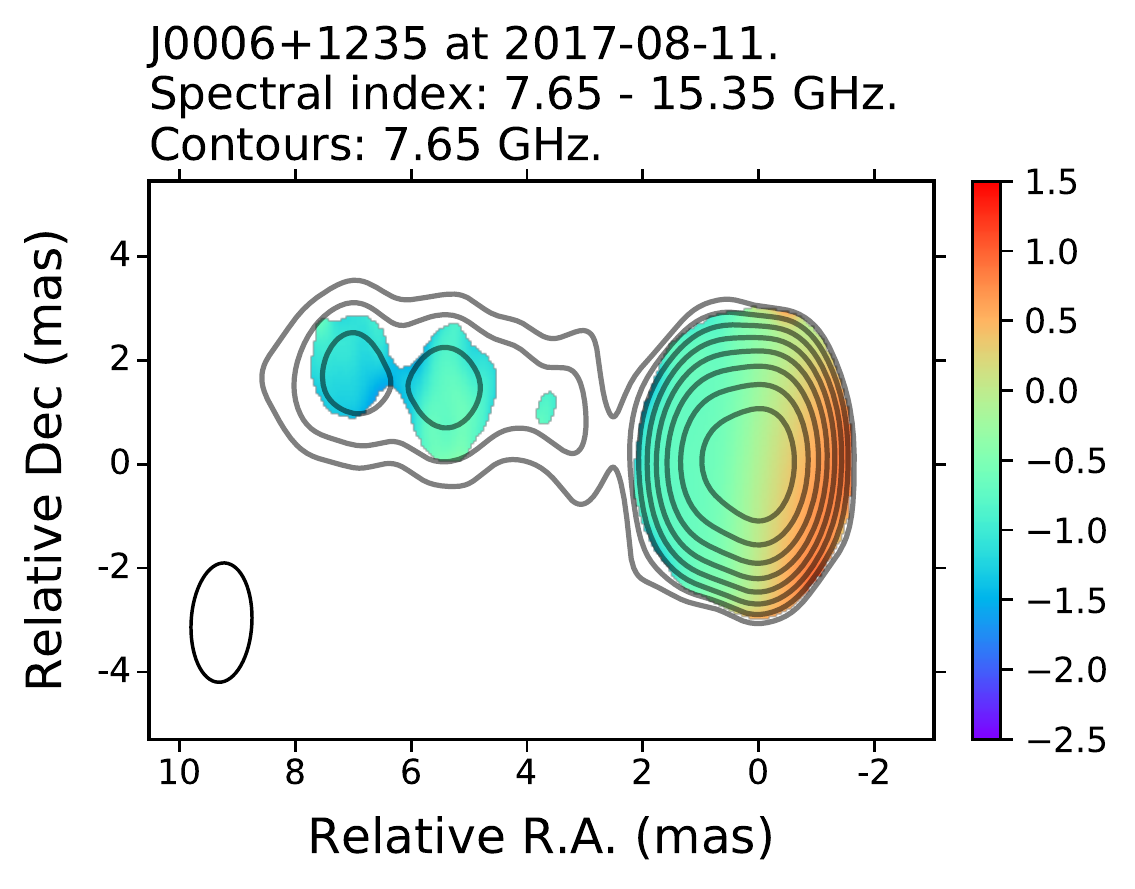}
        \label{}
    }
    \subfigure[]
    {
        \includegraphics[width=0.4\textwidth]{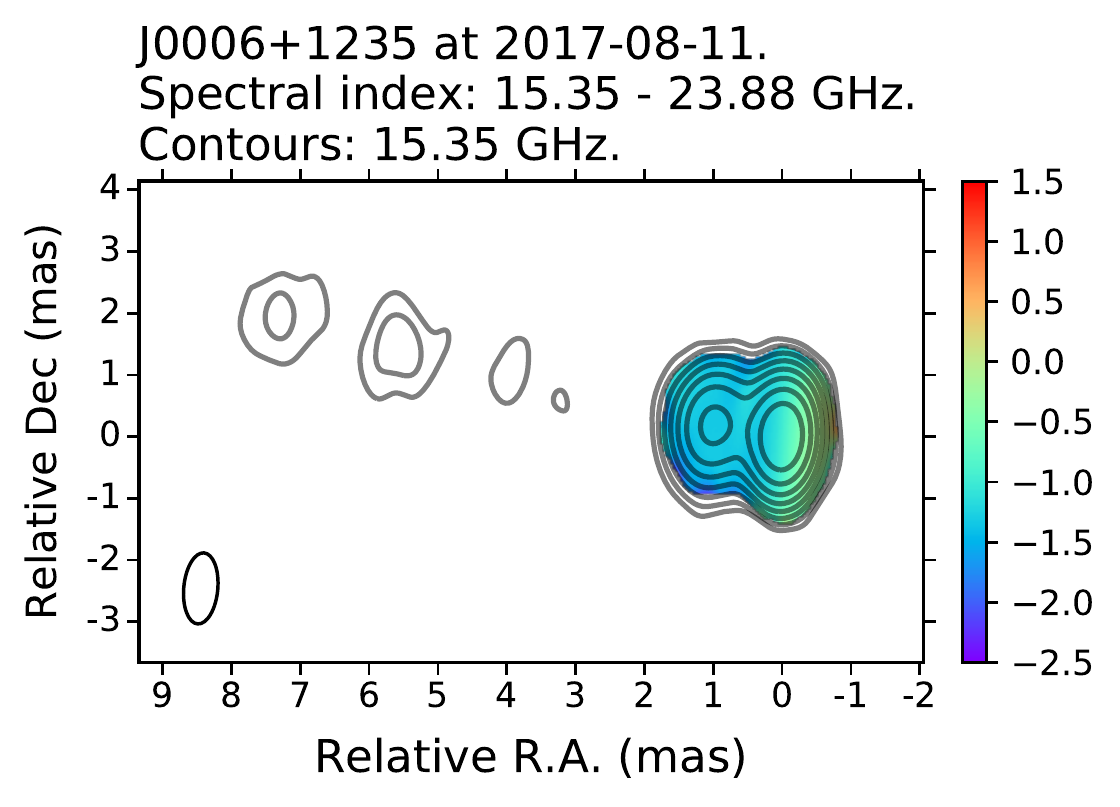}
        \label{}
    }
        \caption{Spectral index maps for J0006 per frequency pair. The spectral index is indicated in the colourbar. The aligned images correspond to 4.15$-$7.65\,GHz, 7.65$-$15.35\,GHz and 15.35$-$23.88\,GHz, respectively. The interferometric beam (ellipse) is displayed on the bottom-left corner of each image.}
    \label{fE1}
\end{figure*}

\begin{figure*}[h]
\centering
     \subfigure[]
    {
        \includegraphics[width=0.4\textwidth]{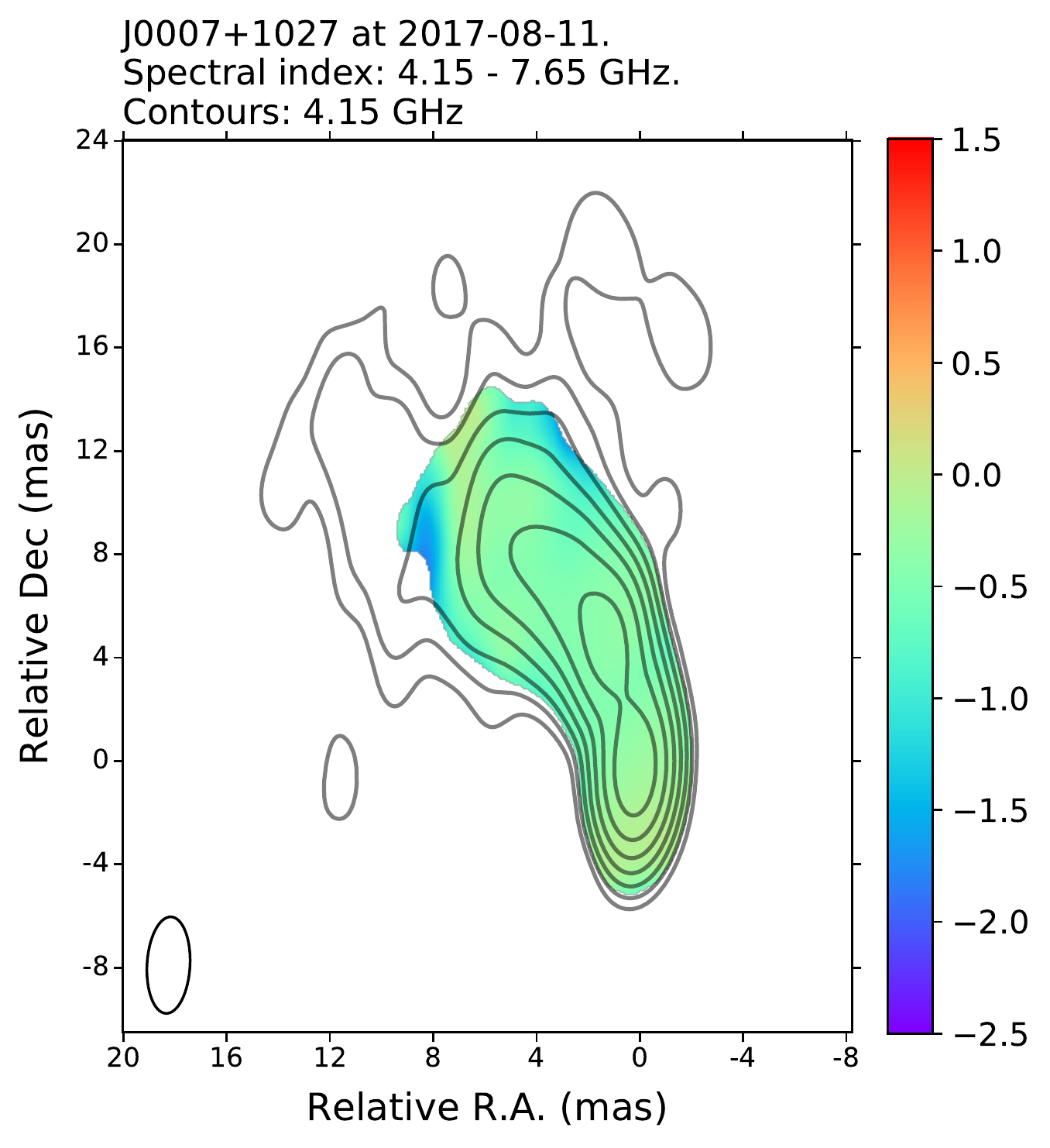}
        \label{}
    } 
       \subfigure[]
    {
        \includegraphics[width=0.4\textwidth]{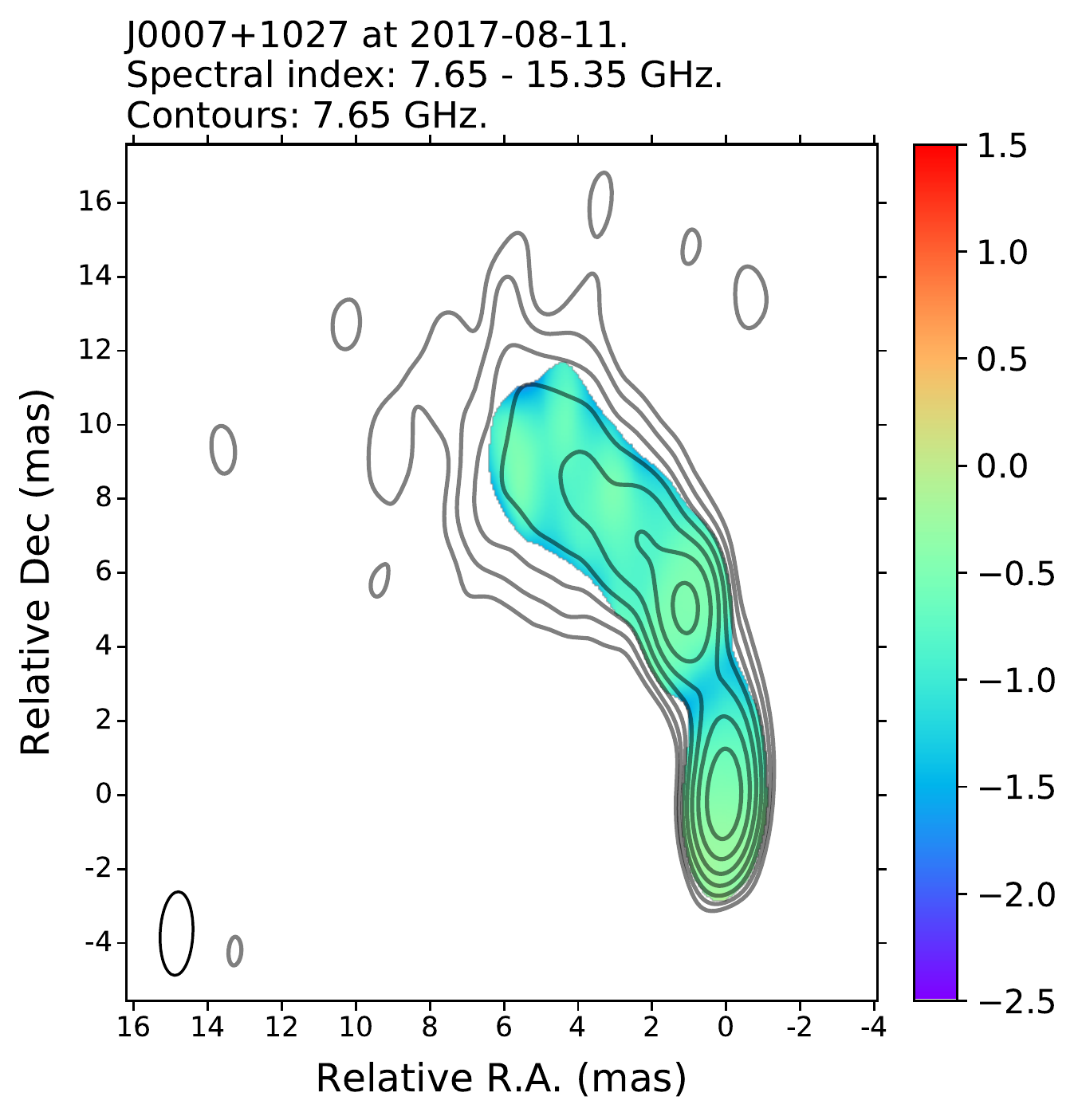}
        \label{}
    } 
   \subfigure[]
    {
        \includegraphics[width=0.4\textwidth]{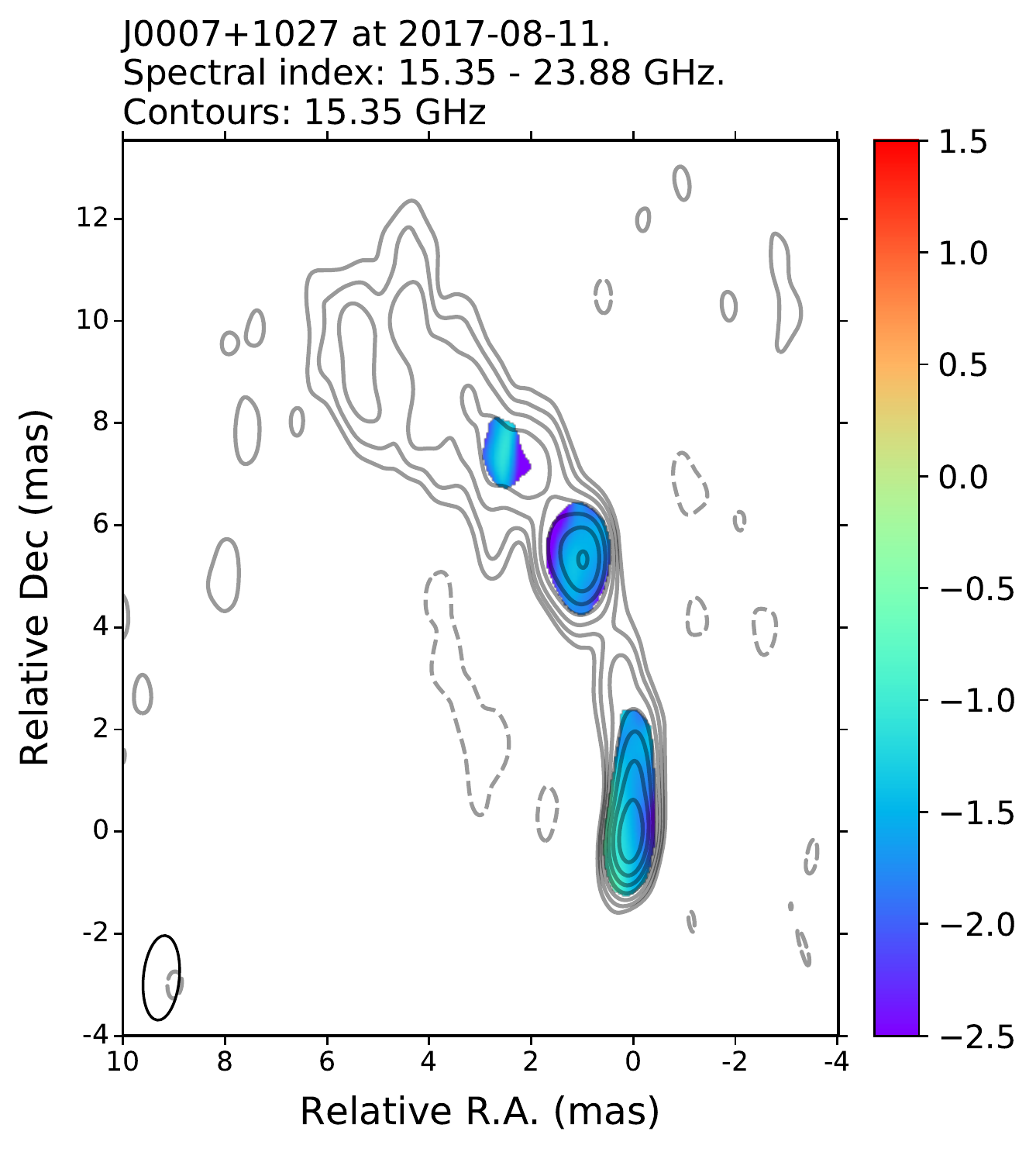}
        \label{}
    } 
        \caption{Spectral index maps for J0007 per frequency pair. The spectral index is indicated in the colourbar. The aligned images correspond to 4.15$-$7.65\,GHz, 7.65$-$15.35\,GHz and 15.35$-$23.88\,GHz, respectively. The interferometric beam (ellipse) is displayed on the bottom-left corner of each image.}
    \label{fE2}
    
\end{figure*}

\begin{figure*}[h]   
\centering
    \subfigure[]
   {
        \includegraphics[width=0.4\textwidth]{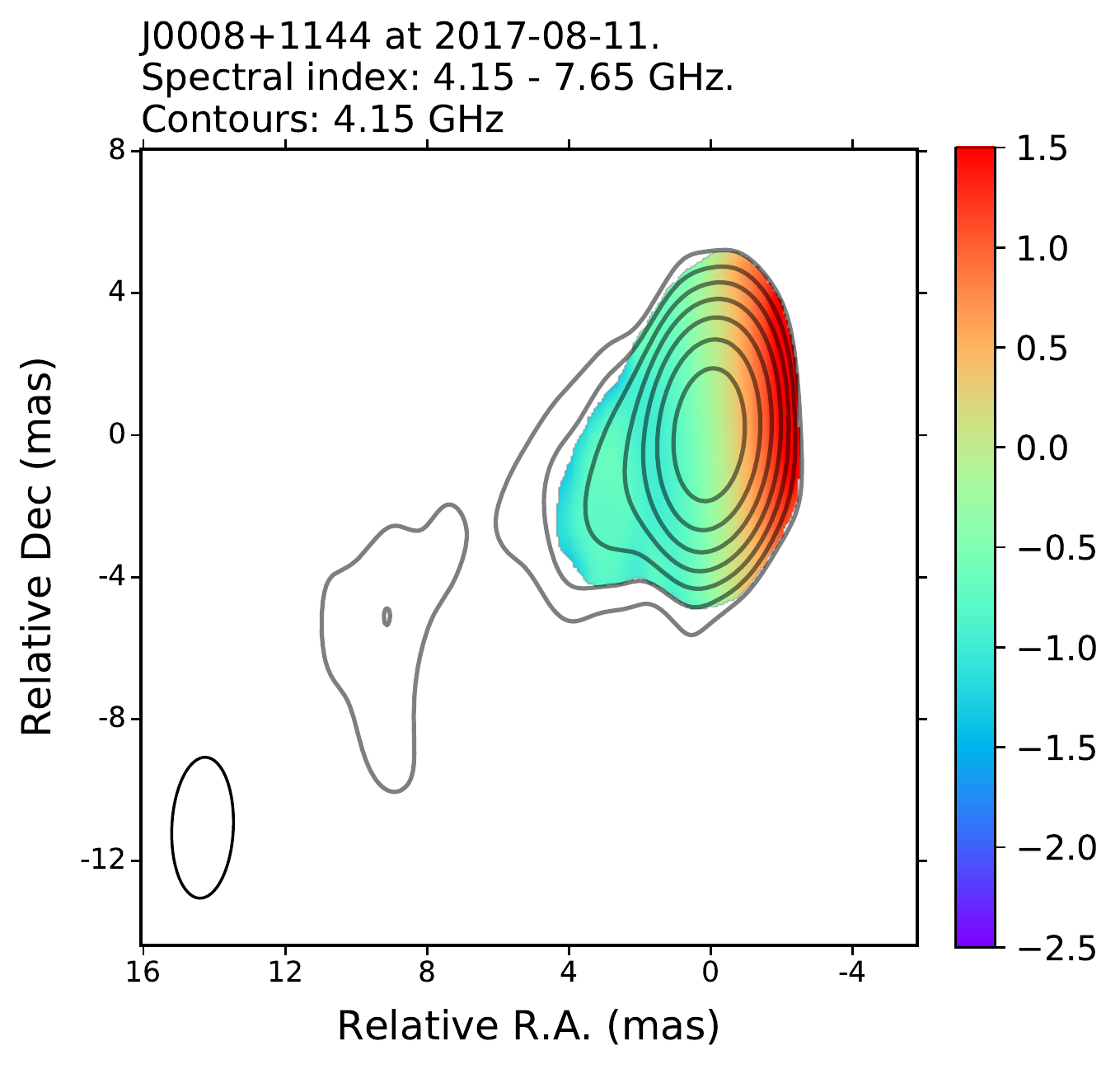}
        \label{}
    }
     \subfigure[]
    {
        \includegraphics[width=0.4\textwidth]{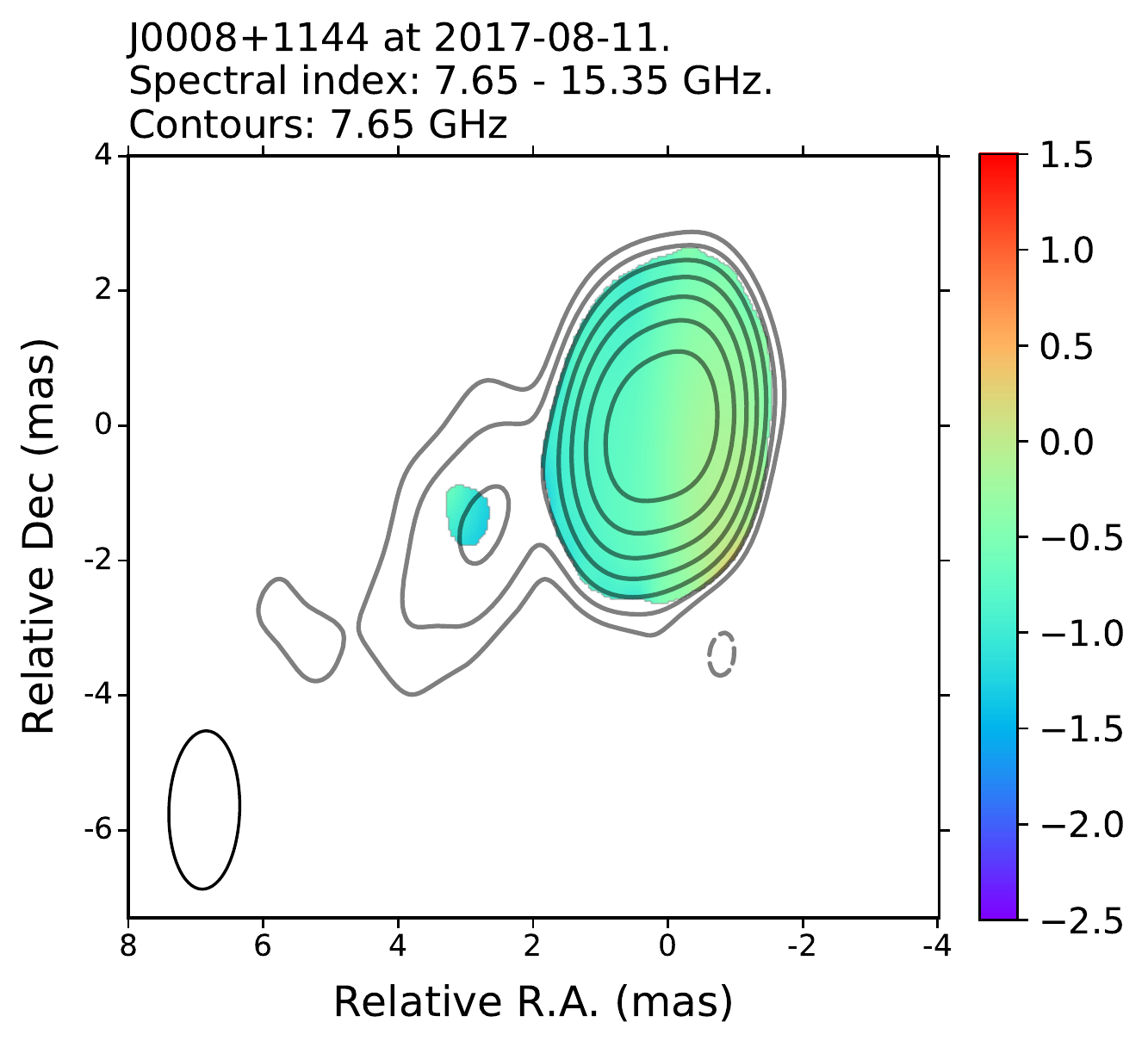}
        \label{}
    }
      \subfigure[]
    {
        \includegraphics[width=0.4\textwidth]{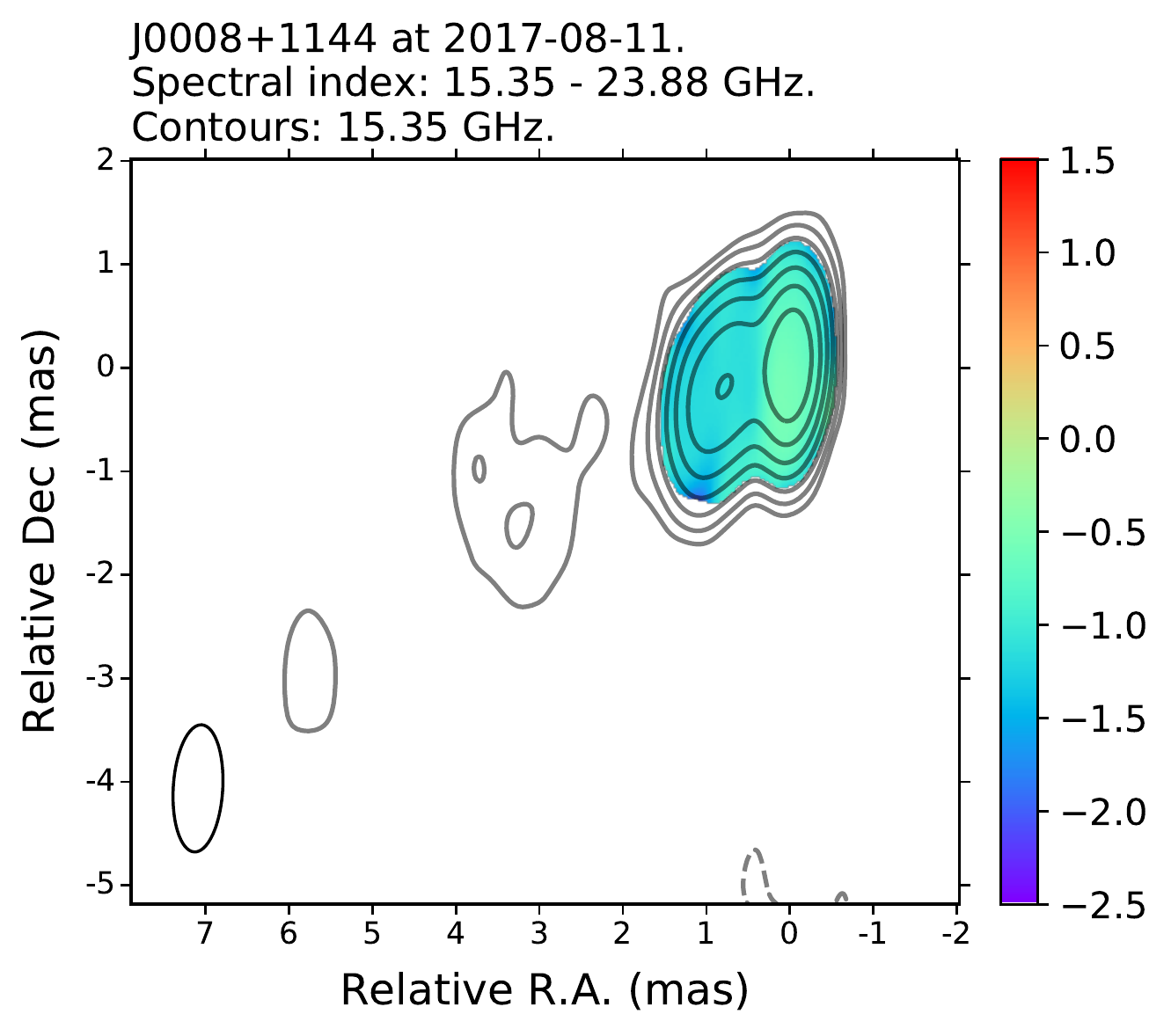}
        \label{}
    }
        \caption{Spectral index maps for J0008 per frequency pair. The spectral index is indicated in the colourbar. The aligned images correspond to 4.15$-$7.65\,GHz, 7.65$-$15.35\,GHz and 15.35$-$23.88\,GHz, respectively. The interferometric beam (ellipse) is displayed on the bottom-left corner of each image.}
    \label{fE3}
\end{figure*}

\clearpage

\section{Projected core-shifts of III\,Zw\,2 with each calibrator}

\end{appendix}

\end{document}